\documentclass{aa}

\usepackage{graphicx}
\usepackage{txfonts}
\usepackage{natbib}
\usepackage{longtable,lscape}
\usepackage{rotating}
\defcitealias{A2005013}{W\&D}
\defcitealias{P2005002}{Paper~I}
\defcitealias{P2006002}{Paper~II}

\def\teff{$T_\mathrm{eff}$}

\newcommand{\BP}{$G_{\rm BP}$}
\newcommand{\RP}{$G_{\rm RP}$}
\newcommand{\RVS}{$G_{\rm RVS}$}
\newcommand{\Gaia}{\emph{Gaia}}
\newcommand{\Hipparcos}{\emph{Hipparcos}}
\newcommand{\Tycho}{\emph{Tycho}}
\def\Av{$A_{\lambda=\mathrm 550}$}
\usepackage{verbatim}
\usepackage{color}

\bibpunct{(}{)}{;}{a}{}{,} 
\begin{document}
 \title{{\Gaia} broad band photometry\thanks{Tables 11 to 13 are only available in electronic format.}}
\author{C. Jordi\inst{1,2}, M. Gebran\inst{1,2}, J.M. Carrasco\inst{1,2}, J. de Bruijne\inst{3}, H. Voss\inst{1,2}, C. Fabricius\inst{4,2}, J. Knude\inst{5}, A. Vallenari\inst{6}, R. Kohley\inst{7}, A. Mora\inst{7}}

\offprints{C. Jordi}

\institute{
  Departament d'Astronomia i Meteorologia, Universitat de Barcelona, c/ Mart\'{\i} i Franqu\`es, 1, 08028 Barcelona, Spain\\
\email{carme,carrasco@am.ub.es}
\and
Institut de Ci\`encies del Cosmos, ICC-UB, c/ Mart\'{\i} i Franqu\`es, 1, 08028 Barcelona, Spain
\and
Research and Scientific Support Department of the European Space Agency, European Space Research and Technology Centre, Keplerlaan 1, 2201 AZ, Noordwijk, The Netherlands
\and
Institut d'Estudis Espacials de Catalunya (IEEC), Edif. Nexus, C/ Gran Capit\`a, 2-4, 08034 Barcelona, Spain
\and
Niels Bohr Institute, Copenhagen University Juliane Maries Vej 30, DK-2100 Copenhagen \O
\and
INAF, Padova Observatory, Vicolo dell'Osservatorio 5, 35122 Padova, Italy
\and
Science Operations Department of the European Space Agency, European Space Astronomy Centre, Villanueva de la Ca\~nada, 28692 Madrid, Spain
}
 
 \date{Received / Accepted}

 \abstract
 {}
 {The scientific community needs to be prepared to analyse the data from {\Gaia}, one of the most ambitious ESA space missions, which is to be launched in 2012. The purpose of this paper is to provide data and tools to predict how {\Gaia} photometry is expected to be. To do so, we provide relationships among colours involving {\Gaia} magnitudes (white light $G$, blue {\BP}, red {\RP} and {\RVS} bands) and colours from other commonly used photometric systems (Johnson-Cousins, Sloan Digital Sky Survey, {\Hipparcos} and {\Tycho}). }
 {The most up-to-date information from industrial partners has been used to define the nominal passbands, and based on the BaSeL3.1 stellar spectral energy distribution library, relationships were obtained for stars with different reddening values, ranges of temperatures, surface gravities and metallicities.}
{The transformations involving {\Gaia} and Johnson-Cousins $V-I_C$ and Sloan DSS $g-z$ colours have the lowest residuals. A polynomial expression for the relation between the effective temperature and the colour \BP$-$\RP \ was derived for stars with $T_{\rm{eff}}\geq$ 4500~K. For stars with $T_{\rm{eff}}<$ 4500~K, dispersions exist in gravity and metallicity for each absorption value in $g-r$ and $r-i$. Transformations involving two Johnson or two Sloan DSS colours yield lower residuals than using only one colour. We also computed several ratios of total-to-selective absorption including absorption $A_G$ in the $G$ band and colour excess $E($\BP--\RP$)$ for our sample stars. A relationship involving $A_{G}/A_{V}$ and the intrinsic $(V-I_{C})$ colour is provided. The derived {\Gaia} passbands have been used to compute tracks and isochrones using the Padova and BASTI models, and the passbands have been included in both web sites. Finally, the performances of the predicted {\Gaia} magnitudes have been estimated according to the magnitude and the celestial coordinates of the star.}
{The provided dependencies among colours can be used for planning scientific exploitation of {\Gaia} data, performing simulations of the {\Gaia}-like sky, planning ground-based complementary observations and for building catalogues with auxiliary data for the {\Gaia} data processing and validation.}

\authorrunning{Jordi et al.}
 \maketitle

\section{Introduction}
\label{sec:intro}

\begin{figure*}[t!]
 \centering
 \includegraphics[scale=0.3]{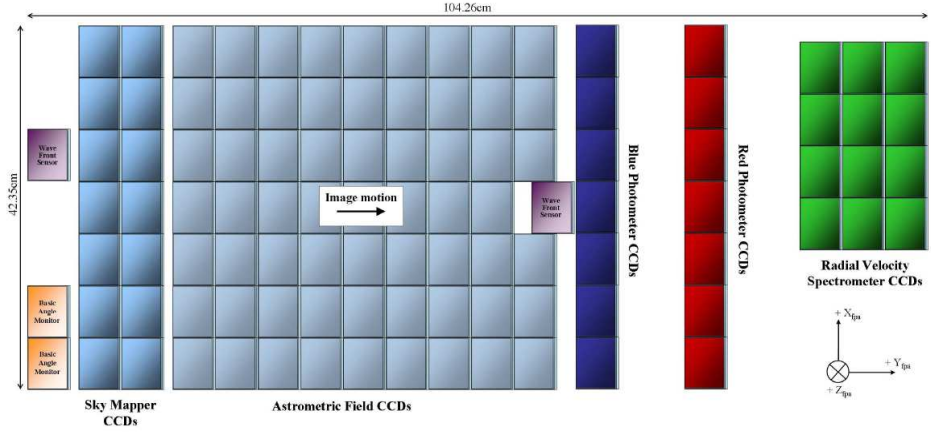}
  \caption{{\Gaia} focal plane. The viewing directions of both telescopes are superimposed on this common focal plane which features 7 CCD rows, 17 CCD strips, and 106 large-format CCDs, each with 4500 TDI lines, 1966 pixel columns, and pixels of size 10 $\mu$m along scan  by 30 $\mu$m across scan (59 mas $\times$ 177 mas). Star images cross the focal plane in the direction indicated by the arrow. Picture courtesy of ESA - A. Short.}
 \label{fig:focalplane}
\end{figure*}

{\Gaia} is an ESA mission that will chart a three-dimensional map of our Galaxy, the Milky Way. The main goal is to provide data to study the formation, dynamical, chemical, and star-formation evolution of the Milky Way. \cite{2001A&A...369..339P} and \cite{2000ESA} presented the mission as it was approved in 2000. While the instrumental design has undergone some changes during the study and design-development phases, the science case remains fully valid. {\Gaia} is scheduled for a launch in 2012 and over its 5-year mission will measure positions, parallaxes, and proper motions for every object in the sky brighter than about magnitude 20, i.e. about 1 billion objects in our Galaxy and throughout the Local Group, which means about 1\% of the Milky Way stellar content. 

Besides the positional and kinematical information (position, parallax, proper motion, and radial velocity), {\Gaia} will provide the spectral energy distribution of every object sampled by a dedicated spectrophotometric instrument that will provide low-resolution spectra in the blue and red. In this way, the observed objects will be classified, parametrized (for instance, determination of effective temperature, surface gravity, metallicity, and interstellar reddening, for stars) and monitored for variability. Radial velocities will also be acquired for more than 100 million stars brighter than 17 magnitude through Doppler-shift measurements from high-resolution spectra by the Radial Velocity Spectrometer (RVS)\footnote{The resolution for bright stars up to $V \sim 11$ is $R \sim 11500$ and for faint stars up to $V \sim 17$ is R$ \sim 5000$.} with a precision ranging from 1 to 15 km s$^{-1}$ depending on the magnitude and the spectral type of the stars \citep{2004MNRAS.354.1223K,2005MNRAS.359.1306W}. These high-resolution spectra will also provide astrophysical information, such as interstellar reddening, atmospheric parameters, elemental abundances for different chemical species, and rotational velocities for stars brighter than $V\simeq$13 mag. 

With such a deep and full-sky coverage and end-of-mission parallax precisions of about 9--11~$\mu$as at $V=10$, 10--27~$\mu$as at $V=15$ and up to 100--350~$\mu$as at $V=20$, {\Gaia} will revolutionise the view of our Galaxy and its stellar content. And not only this, because {\Gaia} will also observe about 300\,000 solar system objects, some 500\,000 QSOs, several million external galaxies, and thousands of exoplanets.

From measurements of unfiltered (white) light from about 350 to 1000~nm {\Gaia} will yield $G$-magnitudes that will be monitored through the mission for variability. The integrated flux of the low-resolution BP (blue photometer) and RP (red photometer) spectra will yield {\BP}-- and {\RP}--magnitudes as two broad passbands in the ranges 330--680~nm and 640--1000~nm, respectively. In addition, the radial velocity instrument will disperse light in the range 847--874~nm (region of the CaII triplet) and the integrated flux of the resulting spectrum can be seen as measured with a photometric narrow band yielding $G_{\rm RVS}$ magnitudes.

The goal of the BP/RP photometric instrument is to measure the spectral energy distribution of all observed objects to allow on-ground corrections of image centroids measured in the main astrometric field for systematic chromatic shifts caused by aberrations. In addition, these photometric observations will allow the classification of the sources by deriving the astrophysical characteristics, such as effective temperature, gravity, and chemical composition for all stars. Once the astrophysical parameters are determined, age and mass will enable the chemical and dynamical evolution of the Galaxy over a wide range of distances to be described. 

\begin{figure}[h!]
 \centering
\includegraphics[scale=1.5]{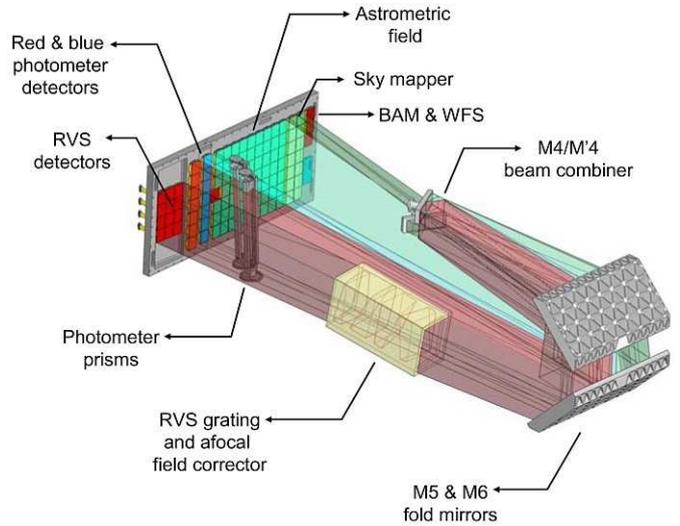}
  \caption{On their way to the BP/RP and RVS sections of the focal plane, light from the two {\Gaia} telescopes is dispersed in wavelength. Picture courtesy of EADS-Astrium.}
 \label{fig:rays}
\end{figure}

There is a huge expectation of the broader scientific community in general, from solar system to extragalactic fields through stellar astrophysics and galactic astronomy, for the highly precise, large and deep survey of {\Gaia}. 
There is an ongoing effort by the scientific community to prepare proper modellings and simulations to help in the scientific analysis and interpretation of {\Gaia} data. To cover some of the needs of this scientific exploitation preparation, this paper aims to provide characterization of the {\Gaia} passbands ($G$, {\BP}, {\RP} and $G_{\rm RVS}$), {\Gaia} colours for a stellar library of spectral energy distributions, and isochrones in this {\Gaia} system. Finally, colour--colour transformations with the most commonly used photometric systems (Johnson-Cousins, {\it Hipparcos}, {\it Tycho} and Sloan) are also included. Altogether, this will allow users to predict how the {\Gaia} sky will look, how a specific object will be observed and with which precision. Although this paper is mainly dedicated to broadband photometry, we have included the $G_{RVS}$ passband in order to predict which stars will have {\Gaia} radial velocity measurements.

Sections~\ref{sec:G-phot} and \ref{sec:phot-inst} are dedicated to the description of the photometric instrument and the photometric bands used for the transformation. Section~\ref{sec:lib} describes the library used to derive the magnitudes and colour-colour transformations shown in Sect.~\ref{sec:transf}. 
Sections~\ref{sec:Mbol} and \ref{sec:Av} describe the computations of the bolometric corrections and the extinction factors in the {\Gaia} passbands. Colours derived from isochrones with different metallicities are given in Sect.~\ref{sec:tracks}. The computation of the magnitude errors and the expected performances are discussed in Sect.~\ref{sec:performances}. Finally, the conclusions are presented in Sect.~\ref{sec:conclusions}.


\section{$G$, $G_{BP}$, $G_{RP}$, and $G_{RVS}$ passbands}
\label{sec:G-phot}

In \cite{2006MNRAS.367..290J}, the {\Gaia} photometric instrument was introduced. 
With the selection of EADS-Astrium as prime contractor, the photometric and
spectroscopic instruments and the focal plane designs were changed. A major
change was the integration of astrometry, photometry, and spectroscopy in the
two main telescopes and only one focal plane as explained in \cite{2010IAUS..261..296L} and shown in Fig.~\ref{fig:focalplane}.

\begin{figure}[h]
 \centering
 \includegraphics[scale=0.32]{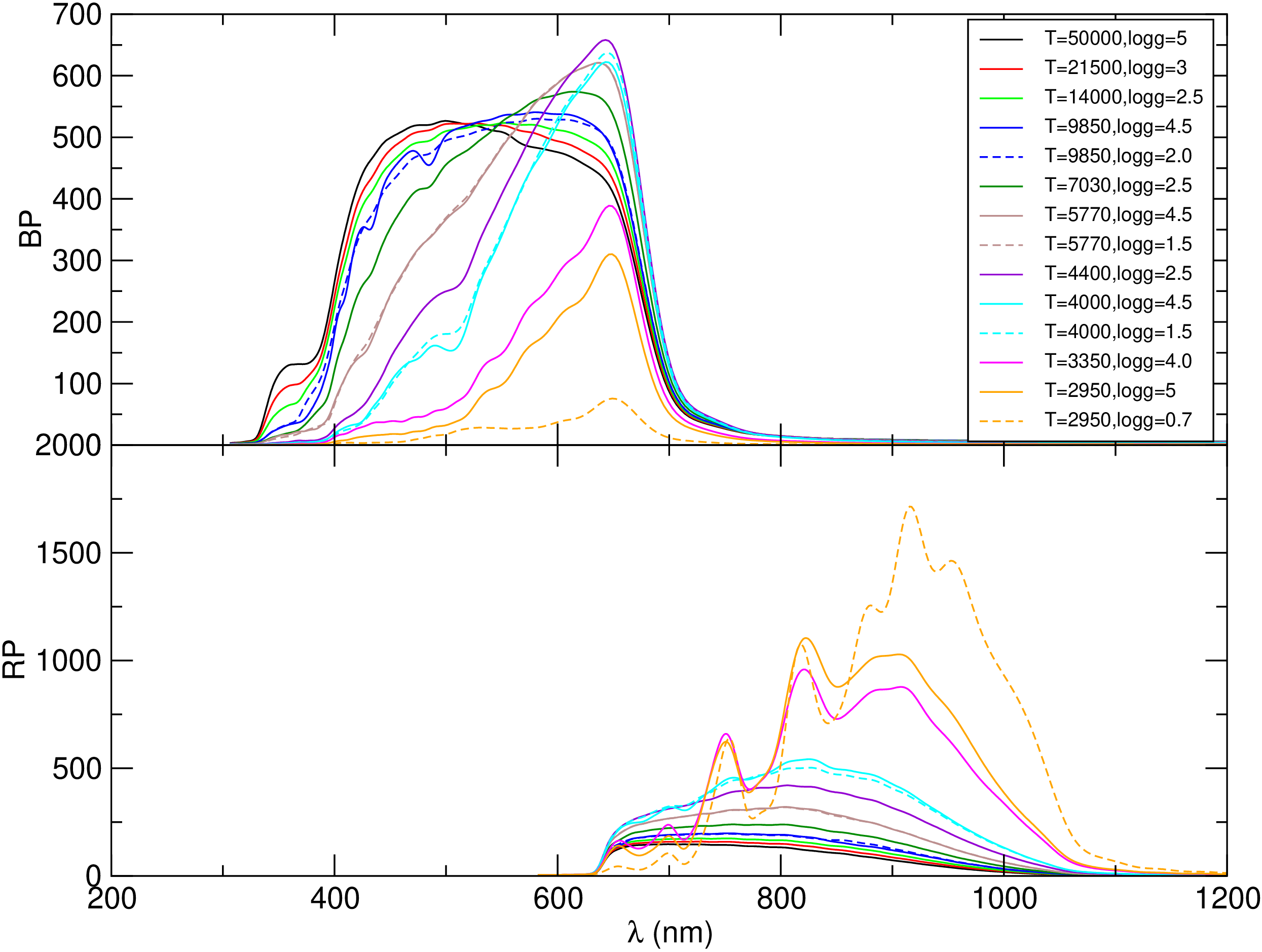}
  \caption{BP/RP low-resolution spectra for a sample of 14 stars with solar metallicity and $G$=15 mag. The flux is in photon$\cdot$s$^{-1}\cdot$pixel$^{-1}$.}
 \label{fig:BPRP_spectra}
\end{figure}

The {\Gaia} photometry is obtained for every source by means of two low-resolution 
dispersion optics located in the common path of the two telescopes (see Fig.~\ref{fig:rays}): one for the blue wavelengths (BP) and one for the red wavelengths (RP). These two low-resolution
prisms substitute the previous set of medium and narrow passbands described in \cite{2006MNRAS.367..290J}. The spectral dispersion of the BP and RP spectra has been chosen to allow the synthetic
production of measurements as if they were made with the old passbands. The spectral resolution is a function of wavelength and varies in BP from 4 to 
32~nm pixel$^{-1}$ covering the wavelength range 330--680~nm. In RP, the 
wavelength range is 640--1000~nm with a resolution of 7 to 15~nm pixel$^{-1}$. We 
display in Fig.~\ref{fig:BPRP_spectra} a sample of 14 BP/RP spectra for stars 
with effective temperatures ranging from 2950 to 50000 K. These noiseless spectra 
were computed with the {\Gaia} Object Generator\footnote{The {\Gaia} Object Generator has been developed by Y. Isasi et al. within the 'Simulations coordination unit' in the Data Processing and Analysis Consortium.}.

The $G$ passband described in \cite{2006MNRAS.367..290J} from the 
unfiltered light in the Astrometric Field (AF) measurements has not undergone any
conceptual change. Since 2006, nearly all CCD devices have been built and some mirrors have already been coated, thus the measurements of the transmission curves provide updated values
for the $G$ passband.

Sixty-two charge-coupled devices (CCDs) are used in AF, while BP and RP spectra are recorded in strips of 7 CCDs each. Twelve CCDs are used in the RVS instrument. Every CCD will have its own QE curve and there will be pixel-to-pixel sensitivity variations. In addition, the reflectivity of the mirrors and prisms will change through their surfaces. {\Gaia} will observe each object several times in each of the two fields of view at different positions in the focal plane (in different CCD), and each observation will have its own characteristics (dispersion, PSF, geometry, overall transmission, etc). The comparison of several observations of a large set of reference sources will allow an internal calibration that will smooth out the differences and will refer all the observations onto a mean instrument. This internal calibration will yield epoch and combined spectra and integrated photometry for all sources with the mean instrument configuration. The transmission of the optics and the QEs used in this paper have to be understood as corresponding to this averaged {\Gaia} instrument.

\begin{figure}[h]
 \centering
 \includegraphics[scale=0.35]{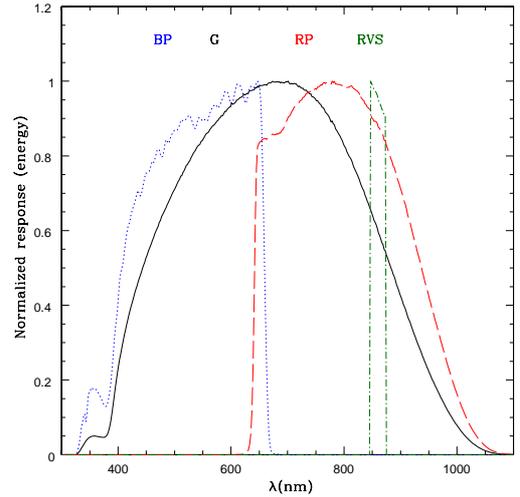}
  \caption{{\Gaia} $G$ (solid line), {\BP} (dotted line), {\RP} (dashed line) and $G_{RVS}$ (dot-dashed line) normalised passbands.}
 \label{fig:Gaia_trans}
\end{figure}

The passbands are derived by the convolution of the response curves of the optics and the QE curves of the CCDs and are shown in Fig.~\ref{fig:Gaia_trans}. The mirrors are coated with Ag and are the same for all instruments, while the coatings of the prisms act as low-pass and high-pass bands for BP/RP. Three different QE curves are in place: one 'yellow' CCD for the astrometric field, an enhanced 'blue' sensitive CCD for the BP spectrometer and a 'red' sensitive CCD for the RP and RVS spectrometers. We have 
used the most up-to-date information from {\Gaia} partners to compute the passbands. Some of the data, however, are still (sometimes ad-hoc) model predictions and not yet real measurements of flight hardware.

The zero  magnitudes have been fixed through the precise energy-flux measurement of Vega. 
\cite{1995A&A...296..771M} gives a monochromatic measured flux of $3.46\cdot 10^{-11}$ 
W m$^{-2}$ nm$^{-1}$ at 555.6~nm, equivalent to $3.56 \cdot 10^{-11}$ W m$^{-2}$ 
nm$^{-1}$ at 550~nm, being $V=0.03$ the apparent visual magnitude of 
Vega\footnote{\cite{2004AJ....127.3508B} give a value of $V=0.026\pm0.008$ for the same 
flux at 555.6~nm, and \cite{2007ASPC..364..315B} gives a revised value 
$V=0.023\pm0.008$ mag.}. Thus, for a star with $m_{\rm 550nm}= 0.0$ we will measure a 
flux of $3.66\cdot 10^{-11}$ W m$^{-2}$ nm$^{-1}$. Vega's spectral energy distribution 
has been modelled according to \cite{Bessel98}, who parameterizes it using Kurucz ATLAS9 
models with $T_{\rm eff}=9550$~K, log$g=3.95$~dex, $[Fe/H]=-0.5$~dex and $\xi_{t}= 
2$~km s$^{-1}$.

The integrated synthetic flux for a Vega-like star has been computed for the $G$, {\BP}, {\RP}, and $G_{\rm RVS}$ passbands. A magnitude equal to 0.03 has been assumed for each synthetic flux. In 
that way, $G =$ {\BP} $=$ {\RP} $=$ {\RVS} = V = 0.03~mag for a Vega-like star. The derivation 
of the magnitudes in $G$, {\BP}, {\RP} and \RVS\ is given as follows:

\tiny
\begin{equation}
\label{Magnitude_eq}  
G_{X}=-2.5\cdot\log\left( \frac{\int_{\lambda_{min}}^{\lambda_{max}}  d\lambda \ F(\lambda) \ 10^{-0.4A_{\lambda}} \ T(\lambda) \  P_{X}(\lambda) \ \lambda Q_{X}(\lambda) }{\int_{\lambda_{min}}^{\lambda_{max}} d\lambda \ F^{Vega}(\lambda) \ T(\lambda) \ P_{X}(\lambda) \  \lambda Q_{X}(\lambda) } \right) + G^{Vega}_{X},
\end{equation}
\normalsize
\noindent where $G_{X}$ stands for $G$, {\BP}, {\RP} and $G_{\rm RVS}$. 
$F(\lambda)$ is the flux of 
the source and $F^{vega}(\lambda)$ is the flux of Vega (A0V spectral type) used as the 
zero point. Both these fluxes are in energy per wavelength and above the Earth's atmosphere. 
$ G^{Vega}_{X}$ is the 
apparent magnitude of Vega in the $G_{X}$ passband. $A_{\lambda}$ is the extinction.
$T(\lambda)$ denotes the telescope transmission, $P_{X}(\lambda)$ is the prism transmission
($P_{X}(\lambda)=1$ is assumed for the $G$ passband) and finally,
$Q_{X}(\lambda)$ is the detector response (CCD quantum efficiency). Therefore, the 
$G$, {\BP}, {\RP} and $G_{\rm RVS}$ passbands are defined by 
$S_{X}(\lambda)=T(\lambda) P_{X}(\lambda) \lambda Q_{X}(\lambda)$.


\section{Other photometric systems}
\label{sec:phot-inst}

As discussed in the introduction, relationships between {\Gaia}'s magnitudes and other photometric systems are provided. In this section we briefly introduce the most commonly used broadband photometric systems, which are used in Sect.~\ref{sec:transf} to derive their relationships with {\Gaia} bands. The photometric systems considered are the following: a) the Johnson-Cousins photometric system, which is one of the oldest systems used in astronomy \citep{1963bad..book..204J}, b) the Sloan Digital Sky Survey photometric passbands \citep{1996AJ....111.1748F} are and will be used in several large surveys such as UVEX, VPHAS, SSS, LSST, SkyMapper, PanSTARRS... and c) finally, as {\Gaia} is the successor of {\it Hipparcos}, and all its objects fainter than $V\sim 6$~mag will be observed with {\Gaia}, we establish the correspondence between the very broadbands of the two missions. 
For completeness, we include {\it Tycho} passbands as well.

\begin{table*}[!htb]
  \caption{{\small Central wavelength and FWHM for the {\Gaia}, Johnson, Sloan and {\it Hipparcos-Tycho} passbands}.}
\label{tabsist}
\begin{center}
\begin{tabular}{l|cccc|ccccc|ccccc|ccc}
\hline
&\multicolumn{4}{c}{\textbf{{\Gaia}}}&\multicolumn{5}{c}{\textbf{Johnson-Cousins}}&\multicolumn{5}{c}{\textbf{SDSS}}&\multicolumn{3}{c}{\textbf{\it Hipparcos}}\\
\hline
Band	   &$G$&\BP&\RP&$G_{RVS}$&	$U$&$B$&$V$&$R_{C}$&$I_{C}$&$u$&$g$&$r$&$i$&z&$H_p$& $B_T$ & $V_T$  \\
\hline
$\lambda_{\rm o}$ (nm)&673 &532&797&860& 361  &   441	&   551  &   647  &   806  &  357  &   475   &   620  &    752  &   899 &528&420 &532\\
${\Delta\lambda}$ (nm)&440&253&296&28&   64  &   95   &    85  &   157  &   154  &   57  &   118   &   113  &    68  &   100 &222&71 &98\\
\hline
\multicolumn{12}{l}{\footnotesize{$\lambda_{\rm o}$: Mean wavelength}}\\
\multicolumn{12}{l}{\footnotesize{${\Delta\lambda}$: Full Width at Half Maximum (FWHM)}}\\
\end{tabular}
\end{center}
\end{table*}

\subsection{Johnson-Cousins $UBVRI$ photometric system} 

\begin{figure}[h]
 \centering
 \includegraphics[scale=0.35]{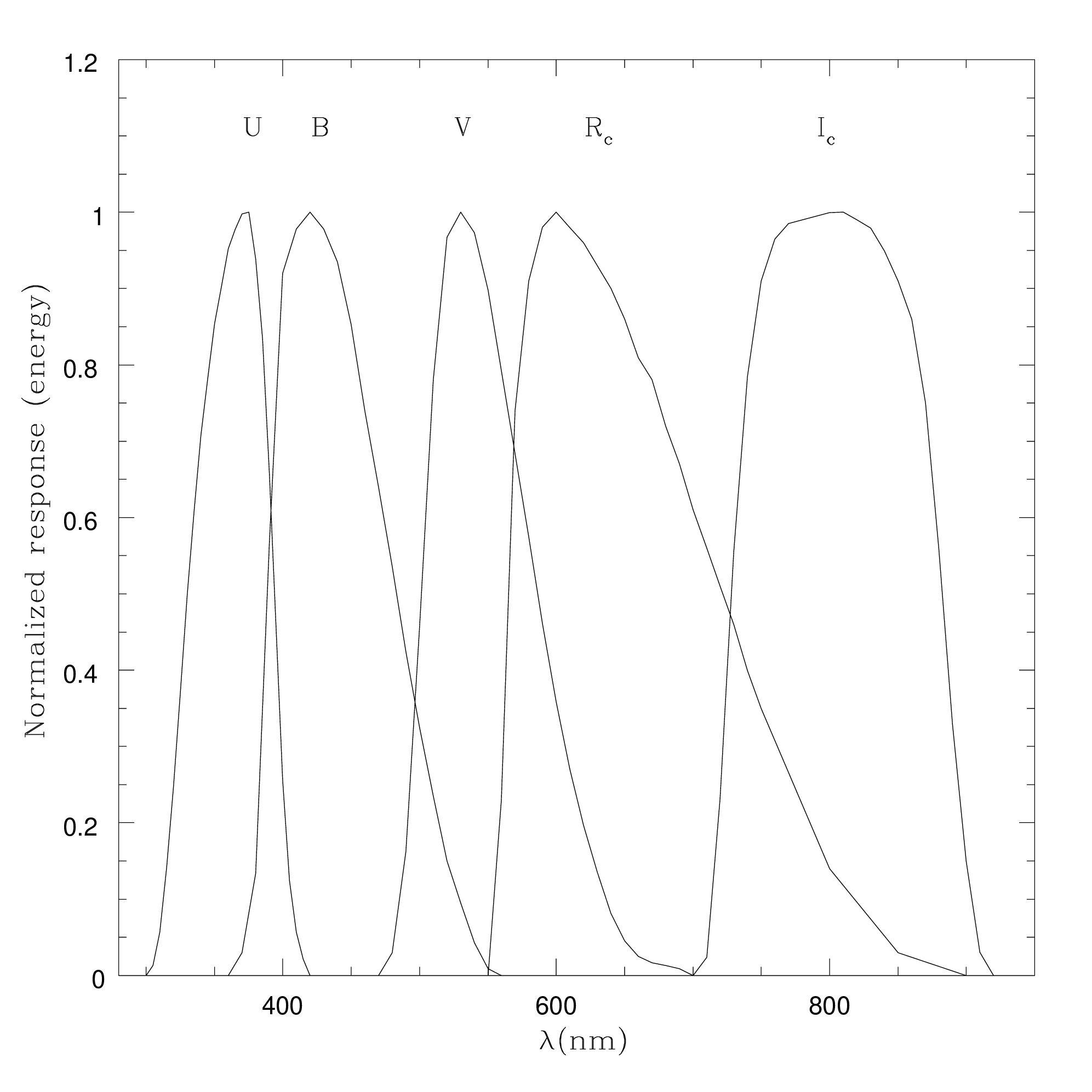}
  \caption{Johnson-Cousins normalised passbands \citep{1990PASP..102.1181B}.}
 \label{fig:JOHNSON_PB}
\end{figure}

The $UBVRI$ system consists of five passbands which stretch from the blue end of the visible spectrum to beyond the red end. The $UBV$ magnitudes and colour indices have always been based on the original system by \cite{1963bad..book..204J} while we can find several RI passbands in the literature. Here we adopt the passband curves in \cite{1990PASP..102.1181B}, which include the RI passbands based on the work of \cite{1976MmRAS..81...25C}. The mean wavelengths of the bands and their FWHM are displayed in Table~\ref{tabsist}. For Johnson-Cousins magnitudes, the zero magnitudes are defined through Vega and they are $U=0.024$~mag, $B=0.028$~mag, $V=0.030$~mag, $R_C=0.037$~mag and $I_C=0.033$~mag \citep{Bessel98}. Figure \ref{fig:JOHNSON_PB} displays the five Johnson-Cousins passbands.

\subsection{SDSS photometric system}

\begin{figure}[h]
 \centering
 \includegraphics[scale=0.35]{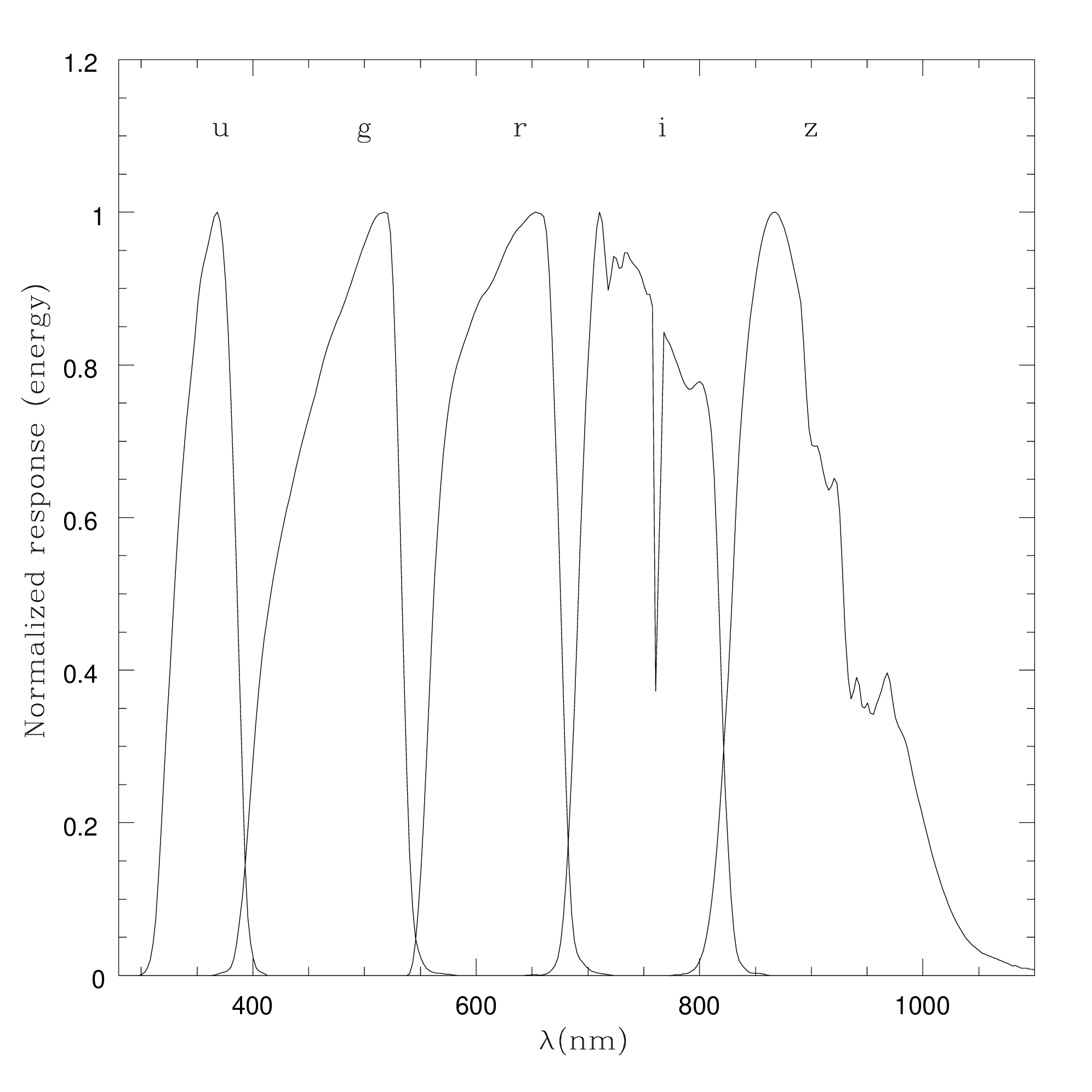}
  \caption{SDSS normalised passbands (http://www.sdss.org/).}
 \label{fig:SDSS_PB}
\end{figure}

The Sloan Digital Sky Survey (SDSS) photometric system comprises five CCD-based wide-bands with wavelength coverage from 300 to 1100~nm \ \citep{1996AJ....111.1748F}. The five filters are called $u, g, r, i$, and $z$ and their mean 
wavelengths and their widths are displayed in Table~\ref{tabsist}. This photometric system includes extinction through an airmass of 1.3 at Apache Point Observatory and "$ugriz$" refers to the magnitudes in the SDSS 2.5m system\footnote{Other systems exist as the $u'g'r'i'z'$ magnitudes which are in the USNO 40-in system.}.  The zero point of this photometric system is the $AB$ system of \cite{1983ApJ...266..713O} and thus 
$m_{\nu}=0$ corresponds to a source with a flat spectrum of 3.631$\times$10$^{-23}$ W 
m$^{-2}$ Hz$^{-1}$. Figure \ref{fig:SDSS_PB} displays the SDSS passbands. All the data concerning the passbands can be found in \cite{2007AJ....134..973I}. These passbands are based on the QEs provided on the SDSS web site\footnote{Available at http://www.sdss.org/.}, where one can also find the conversion between the various SDSS magnitude systems. 

\subsection{{\it Hipparcos} photometric system}

\begin{figure}[h]
 \centering
 \includegraphics[scale=0.35]{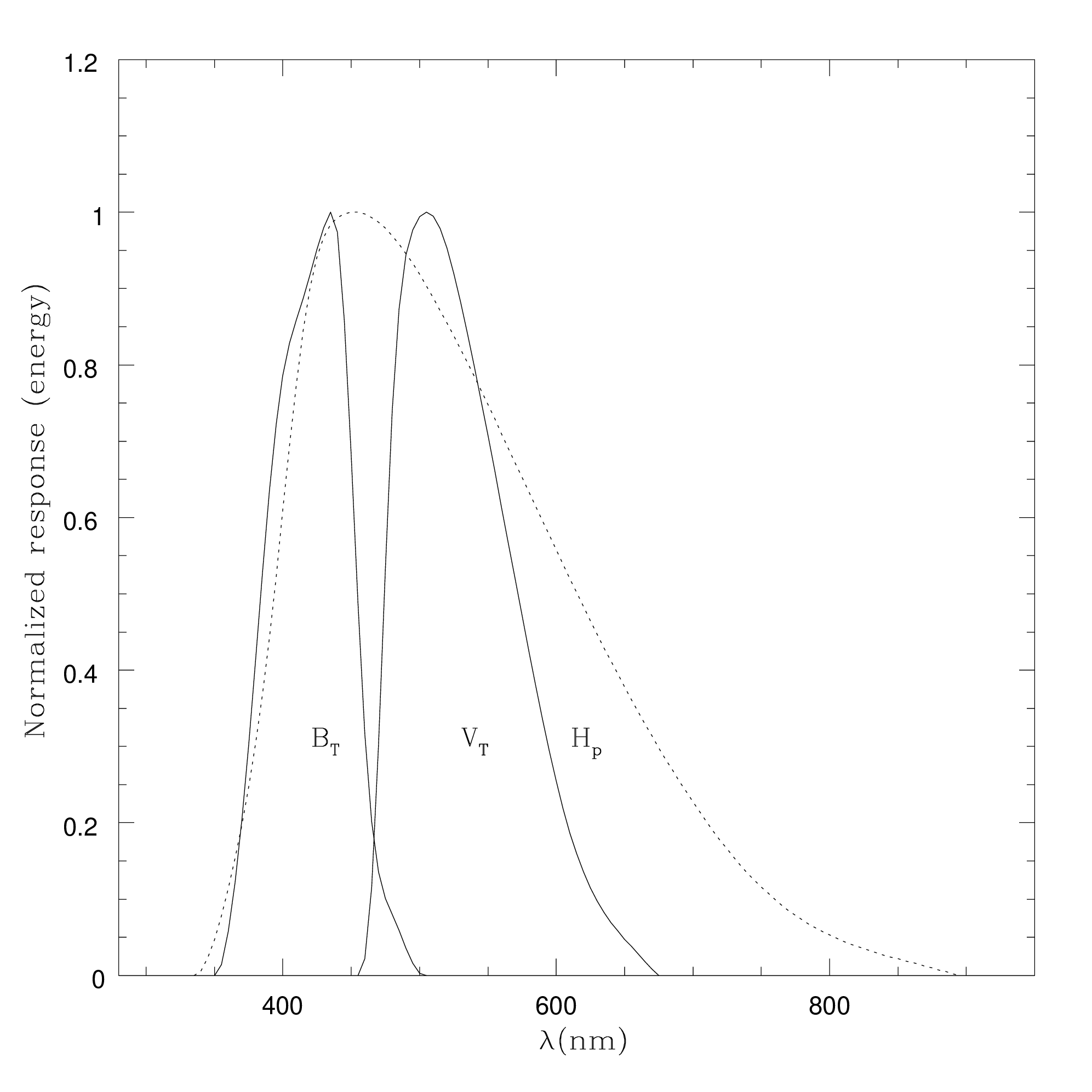}
  \caption{{\it Hipparcos} and {\it Tycho} normalised passbands \citep{1997yCat.1239....0E}.}
 \label{fig:HIP_PB}
\end{figure}

ESA's {\it Hipparcos} space astrometry mission produced a highly precise
astrometric and photometric catalogue for about 120\,000 stars
\citep{1997A&A...323L..49P}. The unfiltered image provided the $H_{p}$ magnitude, and 
the white light of the Sky Mapper was divided by a dichroic beam splitter onto 
two photomultiplier tubes providing the two {\it Tycho} magnitudes $B_{T}$ and $V_{T}$. The 
{\it Hipparcos} ($H_{p}$) and {\it Tycho} ($V_{T}$ and $B_{T}$) passbands are displayed in Fig.~\ref{fig:HIP_PB} (see Table~\ref{tabsist} to check the values for the mean wavelengths and the widths of 
these passbands). The zero points of the {\it Hipparcos/Tycho} photometry are chosen to match 
the Johnson system in a way that $H_{p}=V_{T}=V$ and $B_{T}=B$ for $B-V=0$ \citep{1997A&A...323L..61V}. Hence this is a Vega-like system, with $H_{p}^{Vega}=V^{Vega}_{T}=B^{Vega}_{T}=0.03$~mag. {\it Tycho} passbands ($B_{T}$, $V_{T}$) are very similar to the Johnson ($B$, $V$) passbands and relations already exist between these two systems \citep{1997yCat.1239....0E}. Additional discussion of the {\it Hipparcos-Tycho} passbands and their relationship with the Johnson system can be found in \cite{2000PASP..112..961B}.

\section{The choice of stellar libraries}
\label{sec:lib}
\begin{figure}[th!]
 \begin{center}
 \includegraphics[scale=0.4]{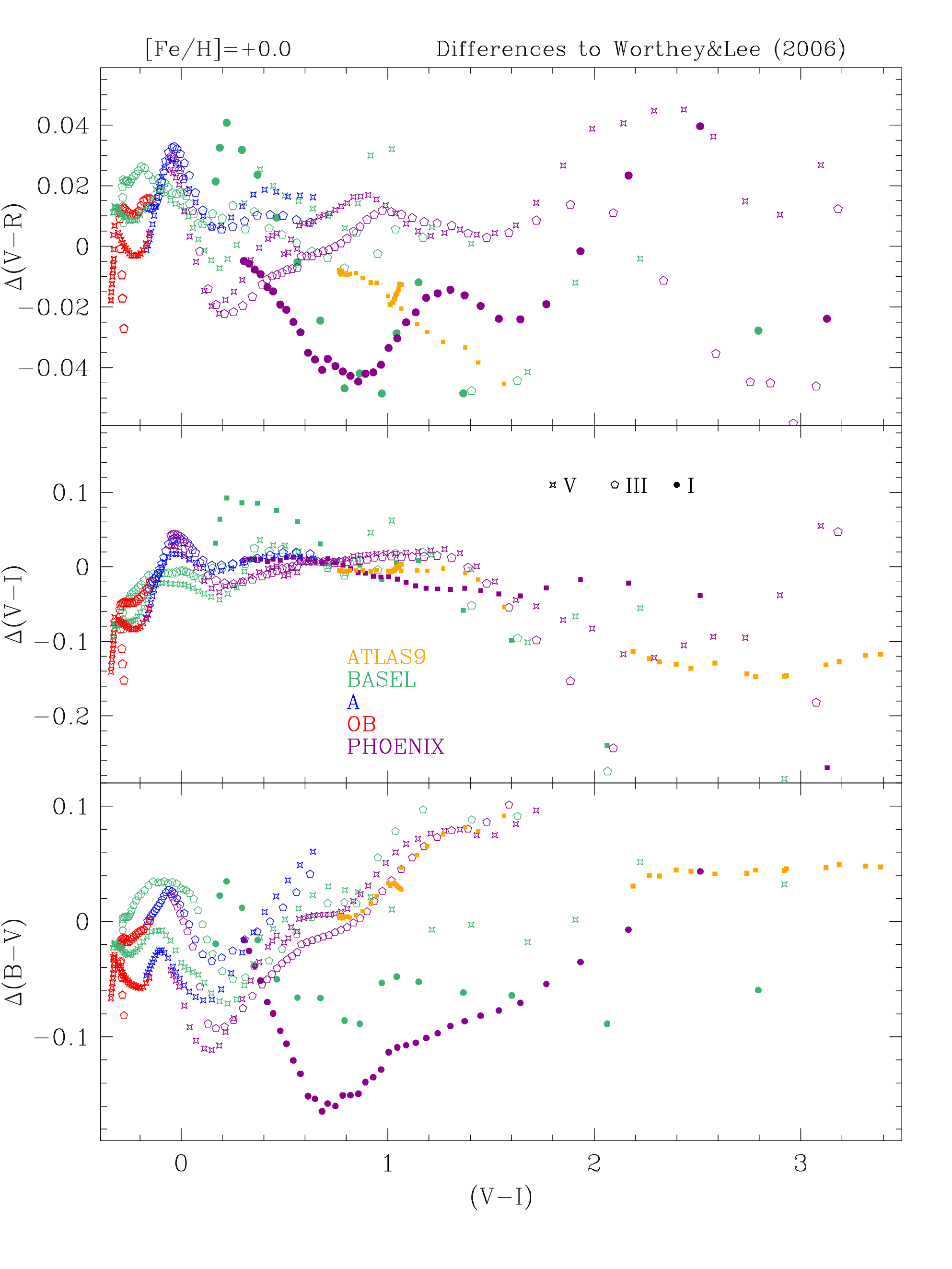}
  \caption{ Residuals in the colour-colour diagram of all available high-resolution libraries with the empirical calibration of \cite{Worthey06} for solar metallicity stars.}
\label{A_VR}
\end{center}
\end{figure}

A large community of scientists has agreed to produce state-of-the-art libraries of synthetic spectra, with a homogeneous and complete coverage of the astrophysical-parameters space at the two resolutions required to produce {\Gaia} simulations: 0.1 nm for the low-dispersion (300--1100 nm) and 0.001 nm for the high resolution mode (840--890 nm).
The capability of reproducing real spectra is improving, and each code producing synthetic spectra is tuned for a given type of stars.  These libraries, summarized in Table~\ref{sinopt}, span a large range in atmospheric parameters, from super-metal-rich to very metal-poor stars, from cool stars to hot stars, from dwarfs to giant stars, with small steps in all parameters, typically $\Delta$\teff=250~K (for cool stars), $\Delta \log g$=0.5 dex, $\Delta$[Fe/H]=0.5 dex.  Depending on \teff, these libraries rely mostly on MARCS (F,G,K stars), PHOENIX (cool and C stars), KURUCZ ATLAS9 and TLUSTY (A, B, O stars) models. Those models are based on different assumptions: KURUCZ are LTE plane-parallel models, MARCS implements also spherical symmetry, while PHOENIX and TLUSTY (hot stars) can calculate NLTE models both in plane-parallel mode and spherical symmetry \citep[for a more detailed discussion see][]{Gust08}. 
MARCS spectra are also calculated including a global [$\alpha$/Fe] enhancement (from -0.2 to 0.4~dex with a step of 0.2~dex). Moreover, enhancements of individual $\alpha$ elements (O, Mg, Si, Ca) are considered.
Hot-star spectra take into account the effects of magnetic fields, peculiar abundances,  mass loss, and circumstellar envelopes (Be).
The  impact of the underlying assumptions, of the different input physics (i.e. atomic and molecular line lists, convection treatment) or of the inclusion of NLTE effects can be seen when comparing the broadband colours ($B-V$, $V-R$, $V-I$) of the different libraries. As an example we show in Fig.~\ref{A_VR} the comparison between the colours derived for solar metallicities from the empirical calibration of \cite{Worthey06} and those derived from the libraries described in Table~\ref{sinopt}.
They show a similar behaviour and are a good reproduction of the empirical relations in the diagram $\Delta(V-R)-(V-I)$, where the residuals are $< 0.07 $ mag, except for very red colours, as expected. The agreement is  worse in the $\Delta(B-V)-(V-I)$ diagram, where the residuals are of the order of $\pm 0.1$. 

The work of \cite{Lejeune3} is based on that of \cite{Lejeune1} who presented the first hybrid library of synthetic stellar spectra, BaSeL, using three original grids of model atmospheres (\cite{1992IAUS..149..225K}, \cite{Fluks} and  \cite{Bessell1,Bessell2}, respectively) in order to cover the largest possible ranges in stellar parameters ($T_{\mathrm{eff}}$, $\log g$, and [M/H]).
The important point in the BaSeL library is that it includes correction functions that have been applied to a (theoretical) solar-abundance model flux spectrum in order to yield synthetic $UBVRIJHKL$ colours that match the (empirical) colour-temperature calibrations derived from observations. In this way, the discontinuity in the flux level provided by the match of several libraries is taken into account. \cite{Lejeune2} extended this library to M dwarfs by using the models of \cite{1999ApJ...512..377H} and to non-solar metallicities, down to [M/H]$\sim$-5.0 dex (BaSeL2.2). The version 3.1 of BaSeL (BaSeL3.1, \citealt{Lejeune3}) differs from the preceding 2.2 library by the colour-calibration at all metallicities using Galactic globular cluster photometric data. The BaSeL3.1 library was constructed to improve the calibration models, especially at low metallicities. The M-giants used in the BaSeL2.2 were replaced in the version 3.1 with \cite{Scholz1997} models. 

Given the level of differences in Fig.\ref{A_VR}, we decided to use the latest version of the BaSeL library (BaSeL3.1, \citealt{Lejeune3}) below. The last row of Table~\ref{sinopt} represents the grid coverage of the BaSeL library used in this work.

\begin{table*}[htbp!]
\caption{Synthetic stellar libraries.}
\label{sinopt} 
\begin{center}
\begin{tabular}{l c c c c}
\\[-6pt]
\hline 
\\[-6pt]
Name &  $T_{\mathrm {eff}}$  (K)& $\log g$ (dex) & [Fe/H] (dex) &Notes \\[+2pt]
\hline
\\[-6pt]

O,  B stars $^1$  &15000 -- 50000 &~1.0 -- 5.0&-5.0 -- +1.0&TLUSTY code, NLTE, wind, mass loss		\\
Ap/Bp stars	$^2$      &~7000 -- 16000 &~4.0&+0.0        &LLmodels code, non-solar abundance\\
B-F stars	$^3$      &~6000 -- 16000 &2.5 -- 4.5&+0.0        &LLmodels	\\
MARCS	$^4$      & 4000 -- 8000  &-0.5 -- 5.5&-5.0 -- +1.0& Variations in individual $\alpha$-elements abundances\\
C stars	$^{5,6}$  & 4000 -- 8000  &~0.0 -- 5.0&-5.0 -- +0.0& $\Delta$\teff= 500 K; [C/Fe]=0,1,2,3; [$\alpha$/Fe]=+0.0, +0.4		\\
PHOENIX	$^7$      &~3000 -- 10000 &-0.5 -- 5.5&-3.5 -- +0.5& $\Delta$\teff= 100 K; [$\alpha$/Fe]=-0.2--+0.8, $\Delta$[$\alpha$/Fe]=0.2\\
Ultra-cool dwarfs$^8$ &~~100 -- 6000  &~0.0 -- 6.0& +0.0       & Different dust model implemented;\\[+2pt]
\hline
\\[-6pt]
BaSeL3.1$^{9}$ &2000 -- 50000& -1.0 --5.5 & -2.0 -- 0.5 & Based on \cite{1992IAUS..149..225K}, \cite{Fluks}, \cite{Bessell1,Bessell2}, \\
&&&&\cite{1999ApJ...512..377H} and \cite{Scholz1997}\\
\hline
\end{tabular}
\tablefoot{  References:  $^1$\cite{Bouret08}, $^2$\cite{Koch08},$^3$\cite{2004A&A...428..993S},  $^4$\cite{Gust08}, $^5$\cite{Alva98}, $^6$\cite{Brott05}, $^7$\cite{Allard00}, $^8$\cite{Martayan08}, $^{9}$\cite{Lejeune3}.}
\end{center}
\end{table*}
\normalsize


\section{{\Gaia} magnitudes and colour-colour transformations}
\label{sec:transf}

In this section, we introduce the transformations 
between the {\Gaia} system and the other photometric systems introduced in Sect.~\ref{sec:phot-inst}.
\cite{2006MNRAS.367..290J} presented the relationship among the $G-V$ and $V-I_{C}$ 
colours. The $G$ is now slightly different from the one used in that paper because the 
QEs of the CCDs, the properties of the prism coatings, and the mirror reflectivities 
have been updated since. 

The SEDs of the BaSeL3.1 library described in Sect.~\ref{sec:lib}  have been reddened by several amounts 
(\Av $=0,1,3,5$ mag) 
following the \cite{1989ApJ...345..245C} reddening law and assuming $R_{\rm v}=3.1$ (see Sect.~\ref{sec:Av} for a discussion of the extinction law). Colours have been derived from synthetic photometry on all created SEDs and can be found in Table\footnote{Table 11 is only available in electronic form.}~. The number of figures/relationships that could be done with these data is numerous. Therefore, we have only computed and only display those relations which we believe are most useful to potential users. The transformations are only valid for the astrophysical parameters of the BaSeL3.1 library and no extrapolation is possible. 
Usually, the dispersion is found to increase for $T_{\rm{eff}}<4500$~K owing to gravity and metallicity. We will not analyse each case in detail. 
The better residuals are in the range 0.02--0.10 mag depending on the colours involved. 
For many applications this accuracy is sufficient. Anyway, 
the online tables are provided to allow readers to compute the desired relationship according to their needs or to look for specific stars.

\subsection{\BP$-$\RP\ as indicator of $T_{\rm{eff}}$}

\begin{figure}[h!]
 \centering
 \includegraphics[scale=0.35]{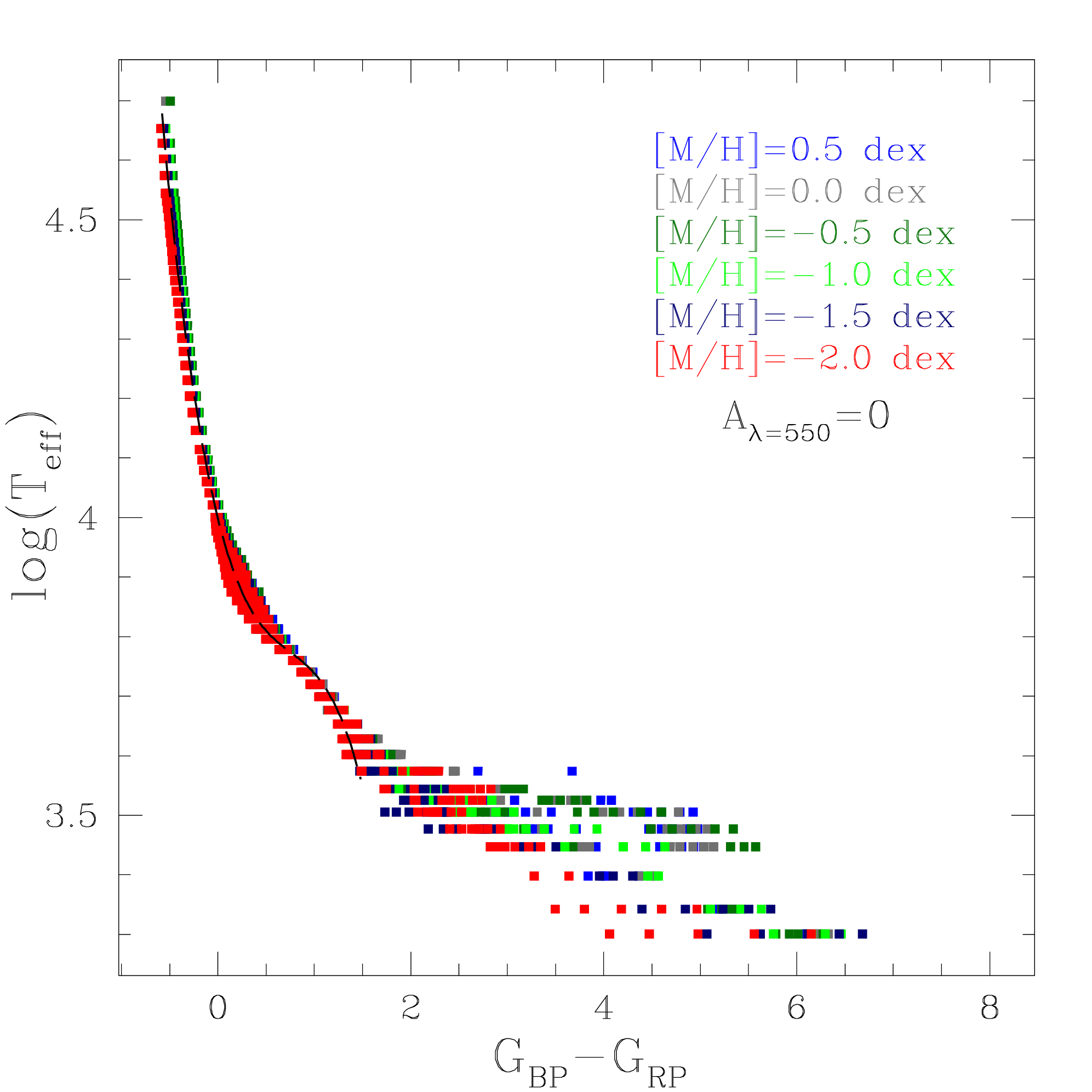}
  \caption{Effective temperatures versus the colour \BP$-$\RP \ for  all SEDs in
BaSeL3.1 library. No reddening has been considered. The dashed line corresponds to the polynomial expression of Eq.~\ref{EQ:Teff_BP_RP}.}
 \label{fig:Teff-BP-RP}
\end{figure}

We discuss here the relation between the effective temperature and the colour \BP$-$\RP. This relation, displayed in Fig. \ref{fig:Teff-BP-RP}, is almost equivalent to the well known relation of 
$T_{\rm{eff}}=f(V-I_{C})$ because, as we will see in Fig.~\ref{fig:transformation-J-C}, 
\BP$-$\RP\ plays the same role as $V-I_{C}$. There is a scatter owing to metallicity and surface gravity, but it is a rather tight relationship for \BP$-$\RP$<$1.5 ($T_{\rm{eff}}\geq 4500$~K). For a fixed metallicity and effective temperature, the horizontal scatter is due to the gravity because a redder \BP$-$\RP\ colour corresponds to lower gravity.  We have derived a polynomial expression for the relation between the effective temperature and the colour \BP$-$\RP\ for stars with \BP$-$\RP$<1.5$ and without reddening. For cooler stars, the dispersion increases drastically and a mean relation is useless. The polynomial fitting is displayed in Fig.~\ref{fig:Teff-BP-RP} and the expression is
\begin{equation}
\log (T_{\rm{eff}})=3.999-0.654(C_{XP})+0.709(C_{XP})^{2}-0.316(C_{XP})^{3},
\label{EQ:Teff_BP_RP}
\end{equation}
where $C_{XP} \equiv G_{BP}-G_{RP}$ and the residual of the fit is equal to 0.02 dex, which is equivalent to a relative error $\Delta T_{\rm{eff}}/T_{\rm{eff}}$ of $\sim$4.6~\%.

\subsection{Colour-colour transformations}

Polynomial expressions of the form
$$C_1=a+b\cdot C_2+c\cdot C_2^2+ d\cdot C_2^3$$
have been fitted to colours $C_1$ and $C_2$, with $C_1$ a colour involving at least one 
{\Gaia} magnitude and $C_2$ a Johnson-Cousins, {\it Hipparcos} or SDSS colour. In many cases, the reddening vector runs almost parallel to the colour-colour relationships and consequently to a unique fit to the set of spectra (BaSeL spectra 
with four different {\Av} are considered) has been computed. The results of these fittings are 
shown in Tables~\ref{table:coeficients_Johnson} to \ref{table:coeficients_SDSS}. The standard deviations of the residuals of the fittings for Johnson-Cousins, {\it Hipparcos/Tycho} and Sloan systems are in the last columns of these tables. They are of the order of a few hundredths of a magnitude in almost all cases and of a few tenths in the others (like in case of $B-V$). The fits show the dependencies among colours and their scatter, which mainly depend on the reddening and range of colours and in second order on luminosity class and metallicity.

\begin{table}[h]
\begin{center}
\caption{Coefficients of the colour-colour polynomial fittings using Johnson-Cousins passbands.}
\label{table:coeficients_Johnson}
\tiny
\begin{tabular}{lrrrrcc}
\\ \hline
& & $(V-I_{C})$& $(V-I_{C})^2$&$(V-I_{C})^3$ &$\sigma$\\
$G-V$&  -0.0257 &  -0.0924 &  -0.1623 &   0.0090 & 0.05\\
 $G-G_{\rm RVS}$&  -0.0138 &   1.1168 &  -0.1811 &   0.0085 & 0.07\\
   $G-$\BP&  0.0387 &  -0.4191 &  -0.0736 &   0.0040 & 0.05\\
$G-$\RP&  -0.0274 &   0.7870 &  -0.1350 &   0.0082 & 0.03\\
  $V-G_{\rm RVS}$&   0.0119 &   1.2092 &  -0.0188 &  -0.0005 & 0.07\\
  $V-$\BP&   0.0643 &  -0.3266 &   0.0887 &  -0.0050 & 0.05\\
$V-$\RP&   -0.0017 &   0.8794 &   0.0273 &  -0.0008 & 0.06\\
\BP$-$\RP&   -0.0660 &   1.2061 &  -0.0614 &   0.0041 & 0.08\\
\hline

& & $(V-R_{C})$& $(V-R_{C})^2$&$(V-R_{C})^3$ &$\sigma$\\
  $G-V$&  -0.0120 &  -0.3502 &  -0.6105 &   0.0852 & 0.10\\
 $G-G_{\rm RVS}$&  0.0267 &   2.3157 &  -0.7953 &   0.0809 & 0.10\\
   $G-$\BP&   0.0344 &  -0.9703 &  -0.2723 &   0.0466 & 0.10\\
 $G-$\RP&   0.0059 &   1.5748 &  -0.5192 &   0.0558 & 0.05\\
 $V-G_{\rm RVS}$&   0.0388 &   2.6659 &  -0.1847 &  -0.0043 & 0.15\\
 $V-$\BP&   0.0464 &  -0.6200 &   0.3382 &  -0.0386 & 0.05\\
  $V-$\RP&    0.0180 &   1.9250 &   0.0913 &  -0.0294 & 0.13\\
   \BP$-$\RP&  -0.0284 &   2.5450 &  -0.2469 &   0.0092 & 0.14\\
\hline
& & $(R_{C}-I_{C})$& $(R_{C}-I_{C})^2$&$(R_{C}-I_{C})^3$ &$\sigma$\\
$G-V$&  -0.0056 &  -0.4124 &  -0.2039 &  -0.0777 & 0.13\\
 $G-G_{\rm RVS}$&   -0.0279 &   2.0224 &  -0.5153 &   0.0176 & 0.06\\
  $G-$\BP&    0.0682 &  -1.0505 &   0.1169 &  -0.1052 & 0.10\\
  $G-$\RP&  -0.0479 &   1.5523 &  -0.5574 &   0.0776 & 0.03\\
 $V-G_{\rm RVS}$&  -0.0223 &   2.4347 &  -0.3113 &   0.0953 & 0.14\\
 $V-$\BP&   0.0738 &  -0.6381 &   0.3208 &  -0.0276 & 0.06\\
 $V-$\RP&    -0.0423 &   1.9646 &  -0.3535 &   0.1553 & 0.14\\
  \BP$-$\RP& -0.1161 &   2.6028 &  -0.6743 &   0.1829 & 0.12\\
\hline
& & $(B-V)$& $(B-V)^2$&$(B-V)^3$ &$\sigma$\\
  $G-V$&  -0.0424 &  -0.0851 &  -0.3348 &   0.0205 & 0.38\\
   $G-G_{\rm RVS}$&   0.1494 &   1.2742 &  -0.2341 &   0.0080 & 0.15\\
 $G-$\BP&  -0.0160 &  -0.4995 &  -0.1749 &   0.0101 & 0.35\\
  $G-$\RP&   0.0821 &   0.9295 &  -0.2018 &   0.0161 & 0.09\\
   $V-G_{\rm RVS}$&  0.1918 &   1.3593 &   0.1006 &  -0.0125 & 0.45\\
  $V-$\BP&   0.0264 &  -0.4144 &   0.1599 &  -0.0105 & 0.05\\
  $V-$\RP&   0.1245 &   1.0147 &   0.1329 &  -0.0044 & 0.46\\
 \BP$-$\RP&   0.0981 &   1.4290 &  -0.0269 &   0.0061 & 0.43\\
\hline
\end{tabular}
\tablefoot{Data computed with four values of extinction (\Av$=$ 0, 1, 3 and 5 mag).}
\end{center}
\end{table}

\begin{figure}[htbp!]
 \centering
 \includegraphics[scale=0.35]{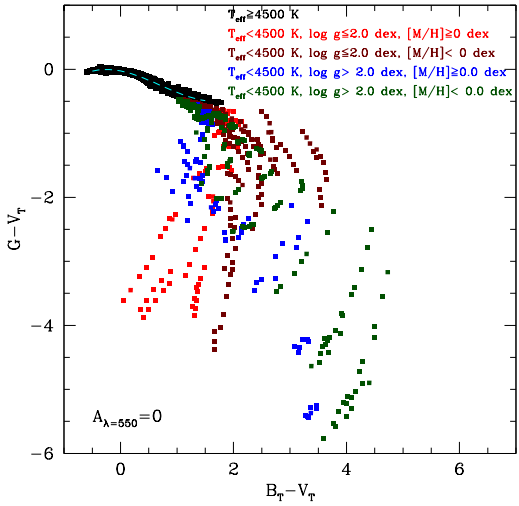}
  \caption{Colour-colour diagrams involving {\Gaia} $G$ and {\it Tycho} colour. No extinction has been considered. The stars are separated in effective temperature, surface gravity, and metallicity. The plot is very similar to the one displaying $(G-V)-(B-V)$. Dashed line corresponds to the fitting in Table \ref{table:coeficients_Hippa}.}
 \label{fig:Tycho_Teff}
\end{figure}

\begin{figure}[htbp!]
 \centering
 \includegraphics[scale=0.2]{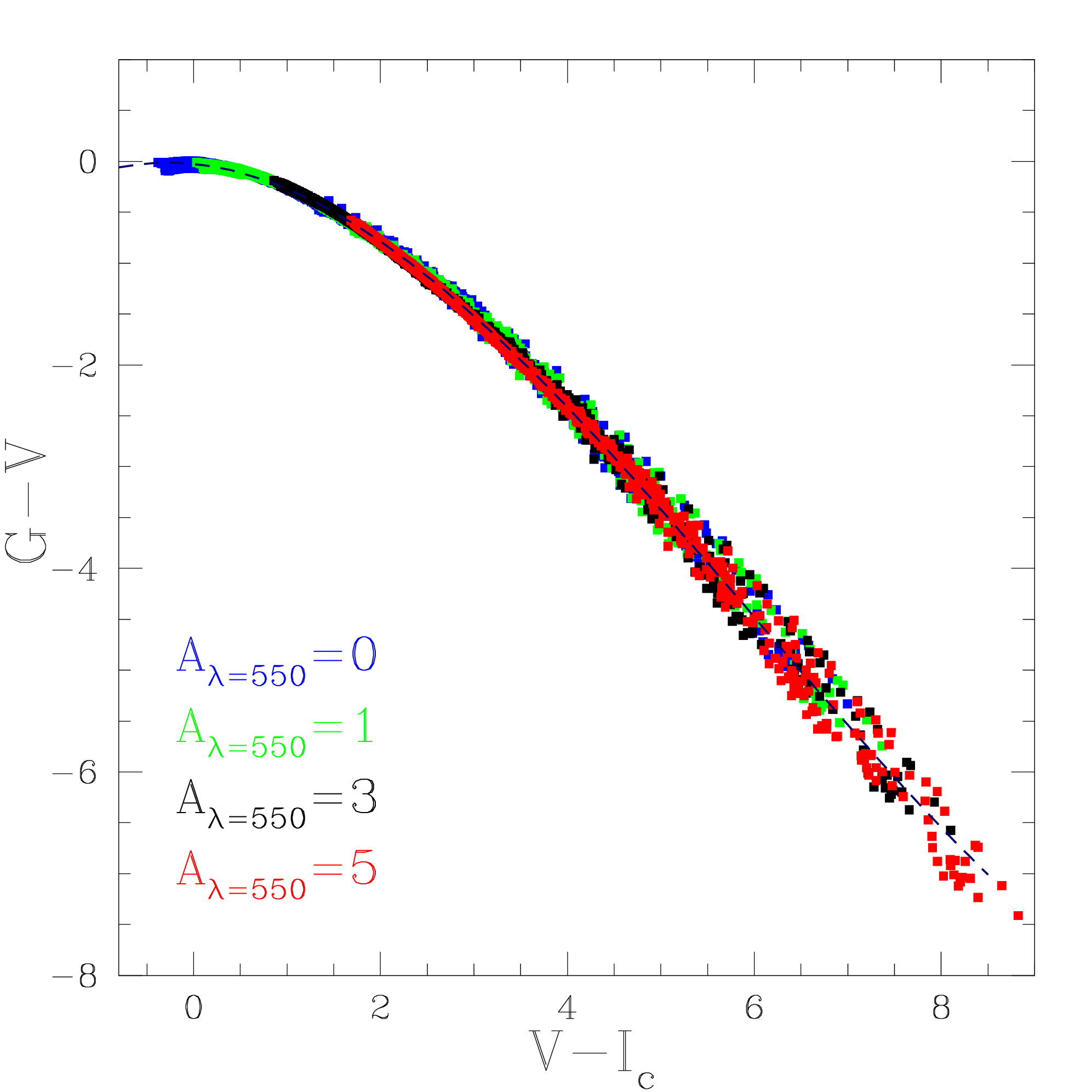}
  \includegraphics[scale=0.2]{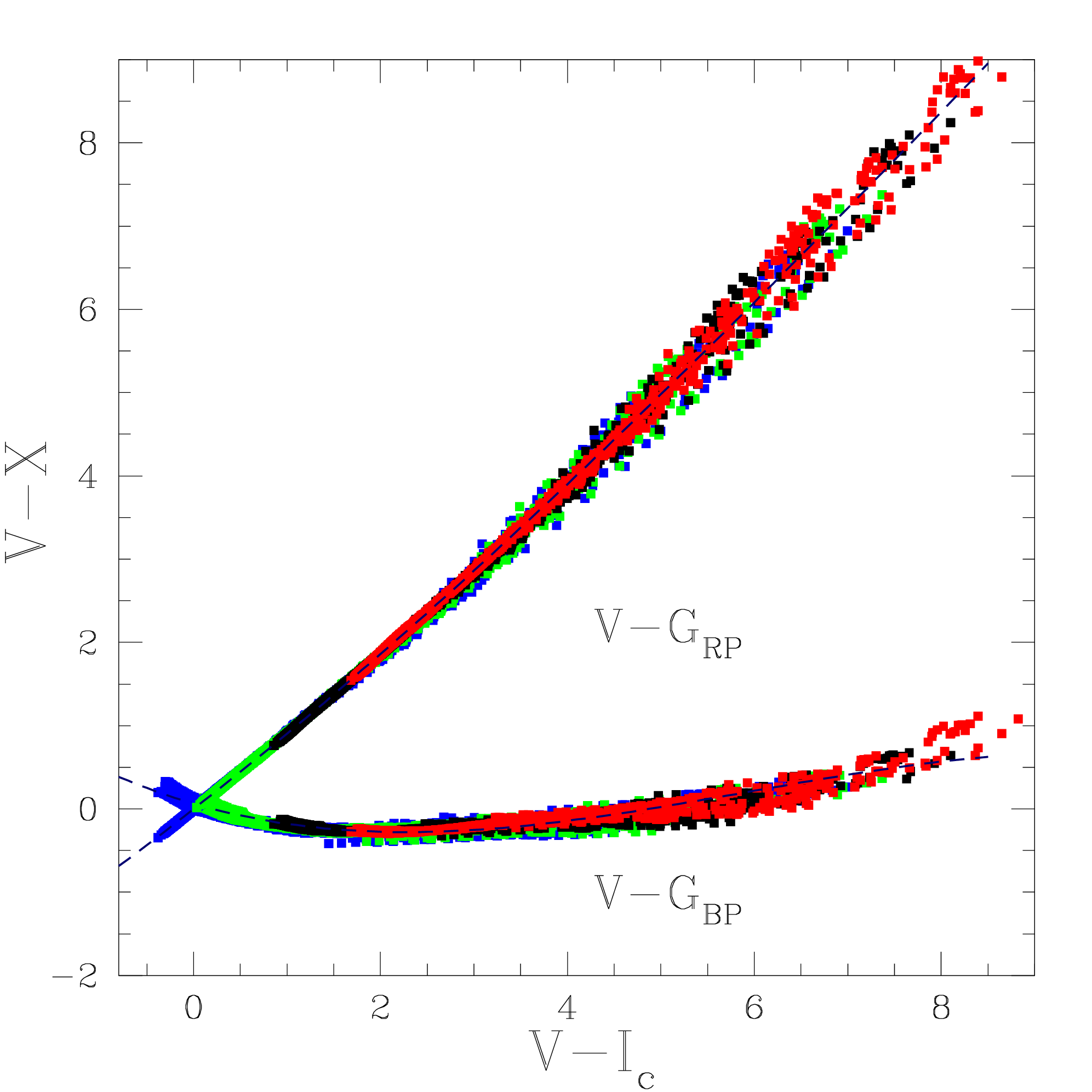}\\
 \includegraphics[scale=0.2]{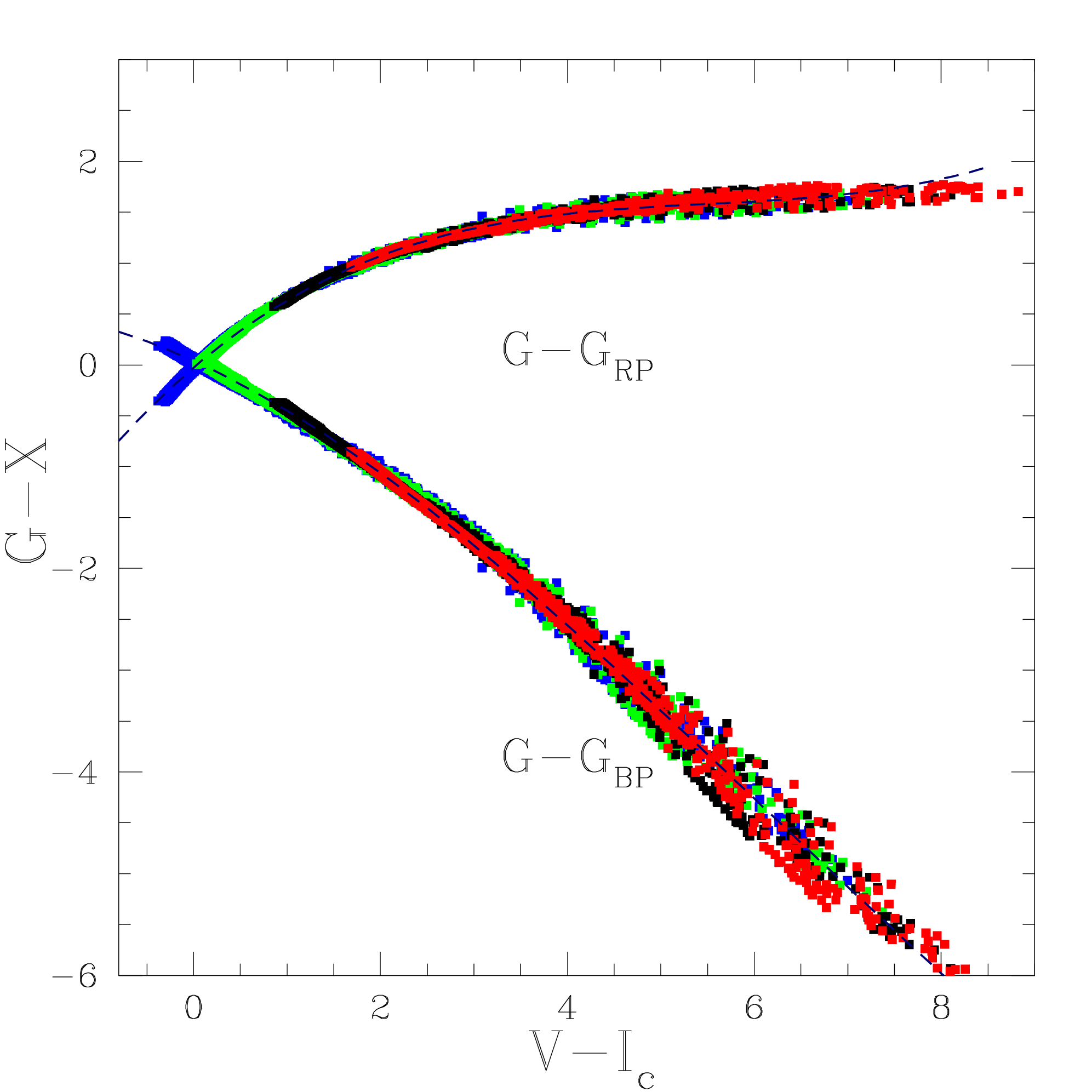}
 \includegraphics[scale=0.2]{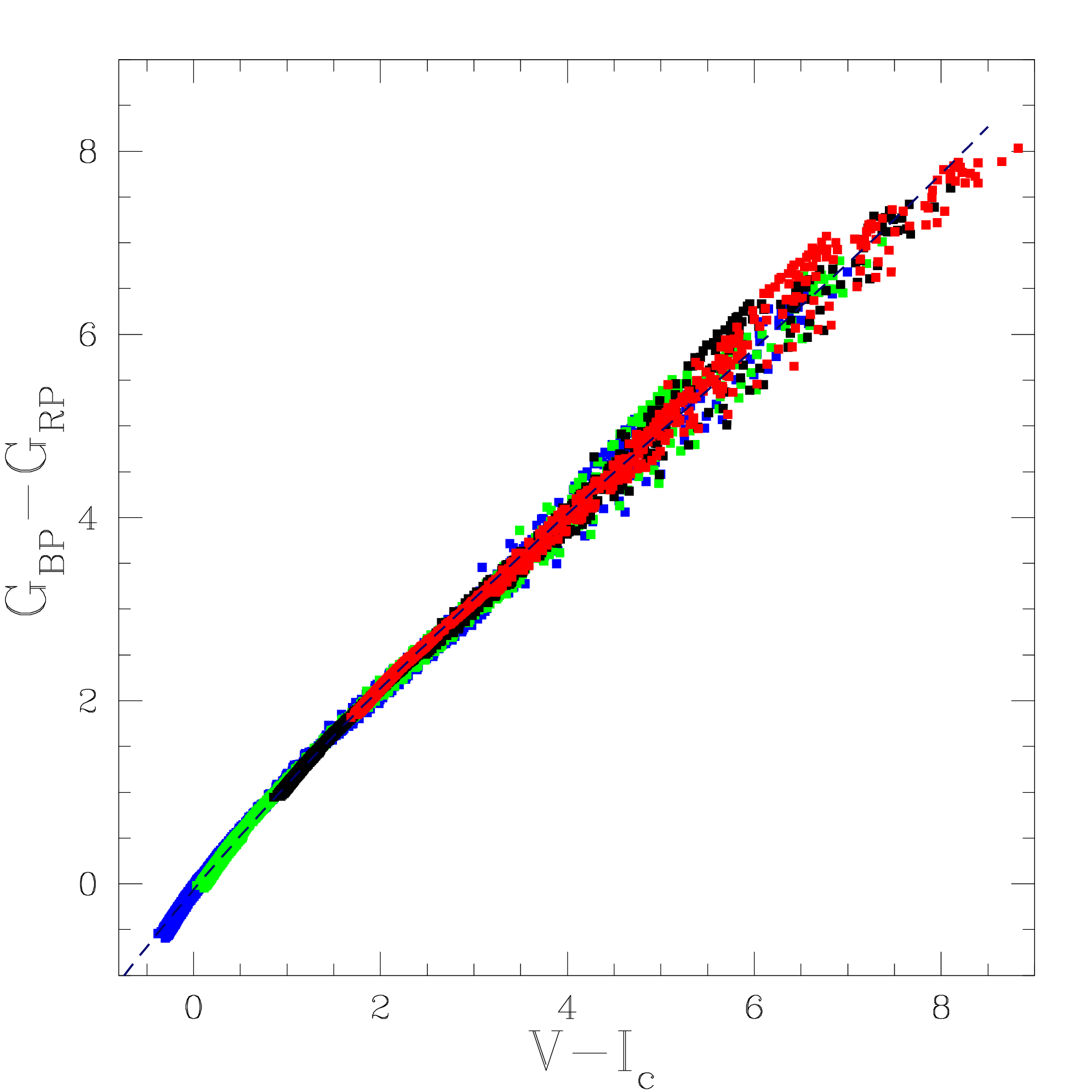}
  \caption{Colour-colour diagrams involving {\Gaia} passbands and $V-I_{C}$ Johnson-Cousins passbands. Different colours are used for different {\Av} values. Plots with $V-R_C$ or $R_C-I_C$ show very similar behaviour. Dashed lines correspond to the fitting in Table \ref{table:coeficients_Johnson}.}
 \label{fig:transformation-J-C}
 \end{figure}

\begin{figure}[htbp!]
 \centering
 \includegraphics[scale=0.2]{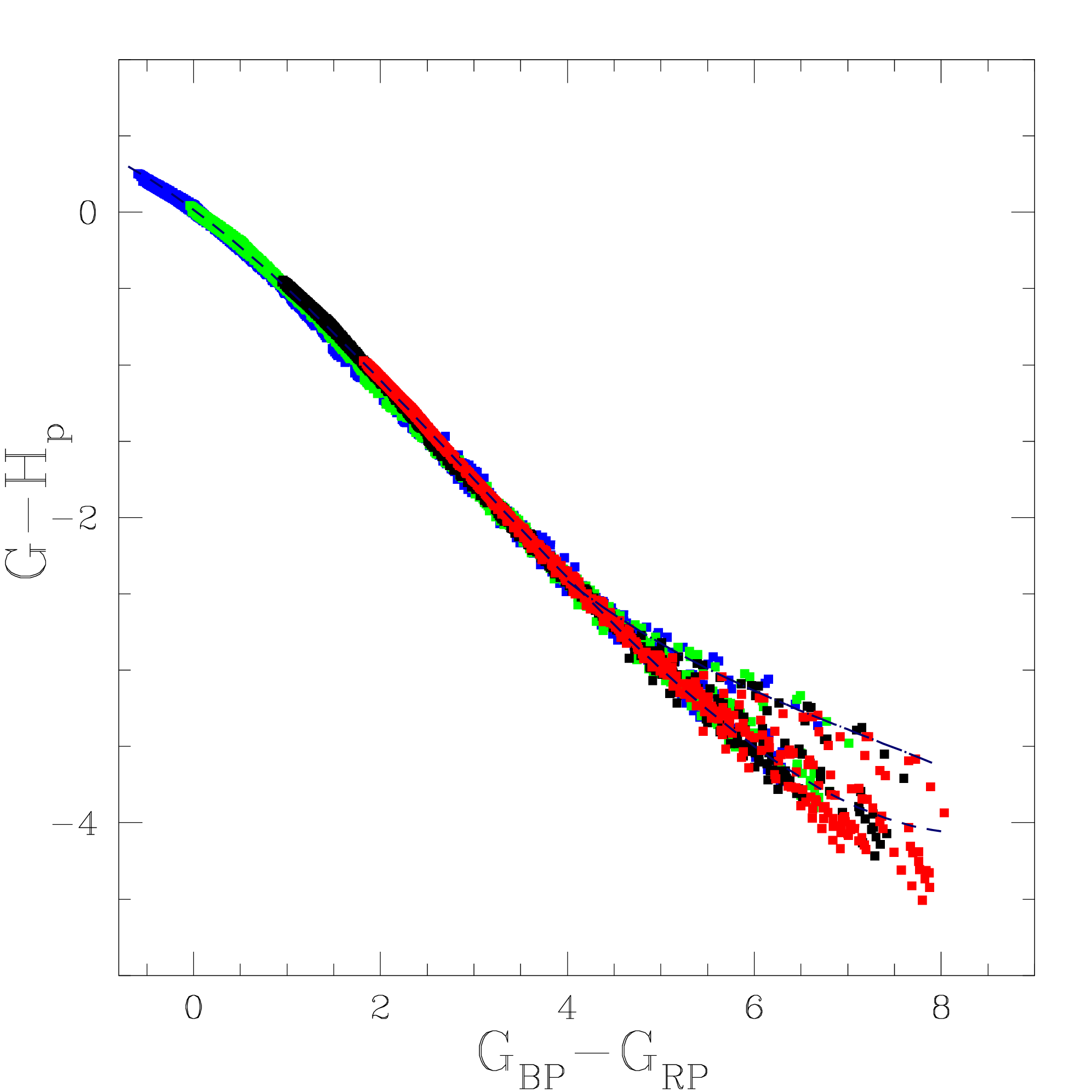}
 \includegraphics[scale=0.2]{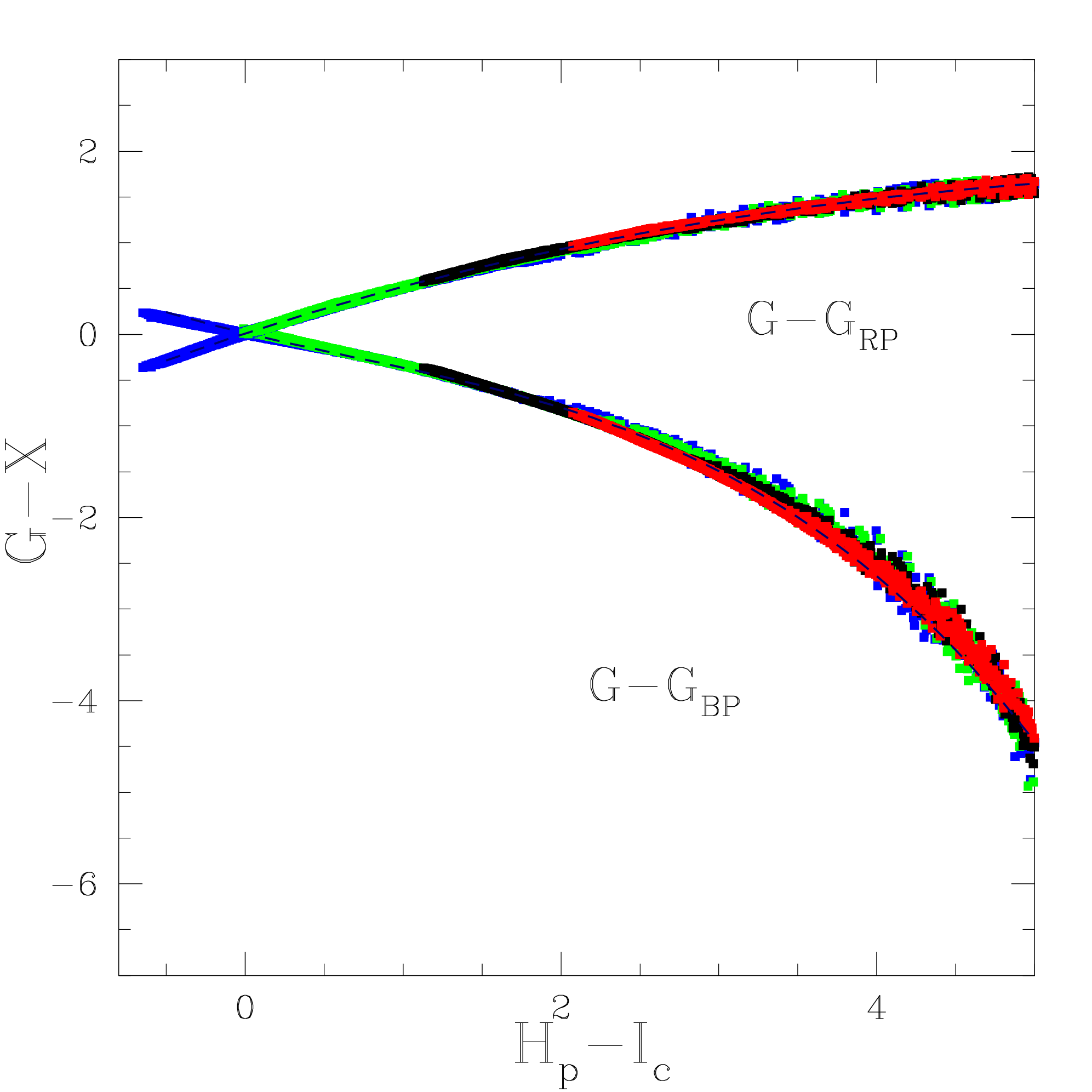}
  \caption{Colour-colour diagrams involving the three broad {\Gaia} passbands and {\it Hipparcos} $H_{p}$. Different colours are used for different absorption values as in Fig.~\ref{fig:transformation-J-C}. Dashed lines correspond to the fitting in Table \ref{table:coeficients_Hippa}.}
 \label{fig:Hipp-diag}
\end{figure}

\begin{figure}[htbp!]
 \centering
 \includegraphics[scale=0.2]{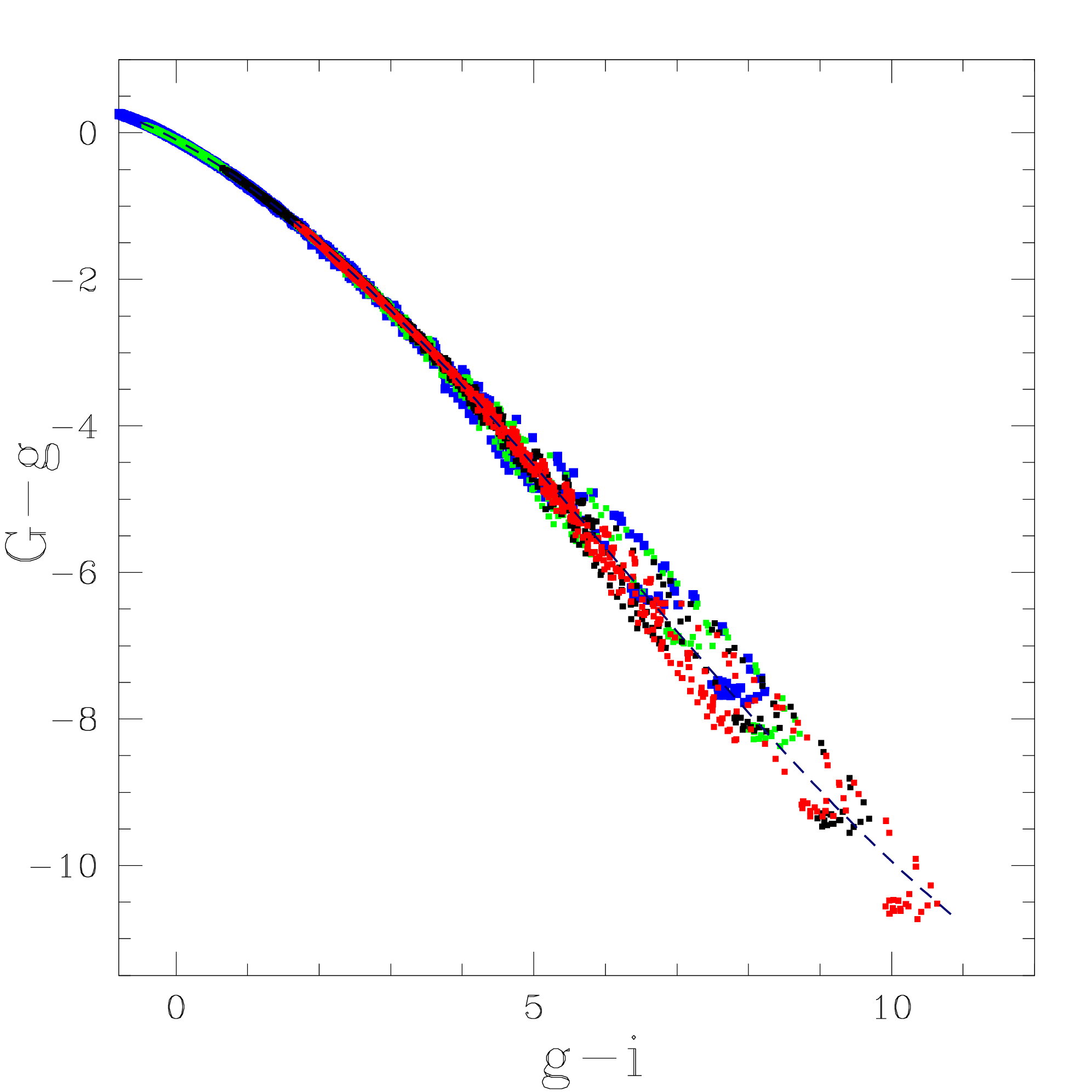}
 \includegraphics[scale=0.2]{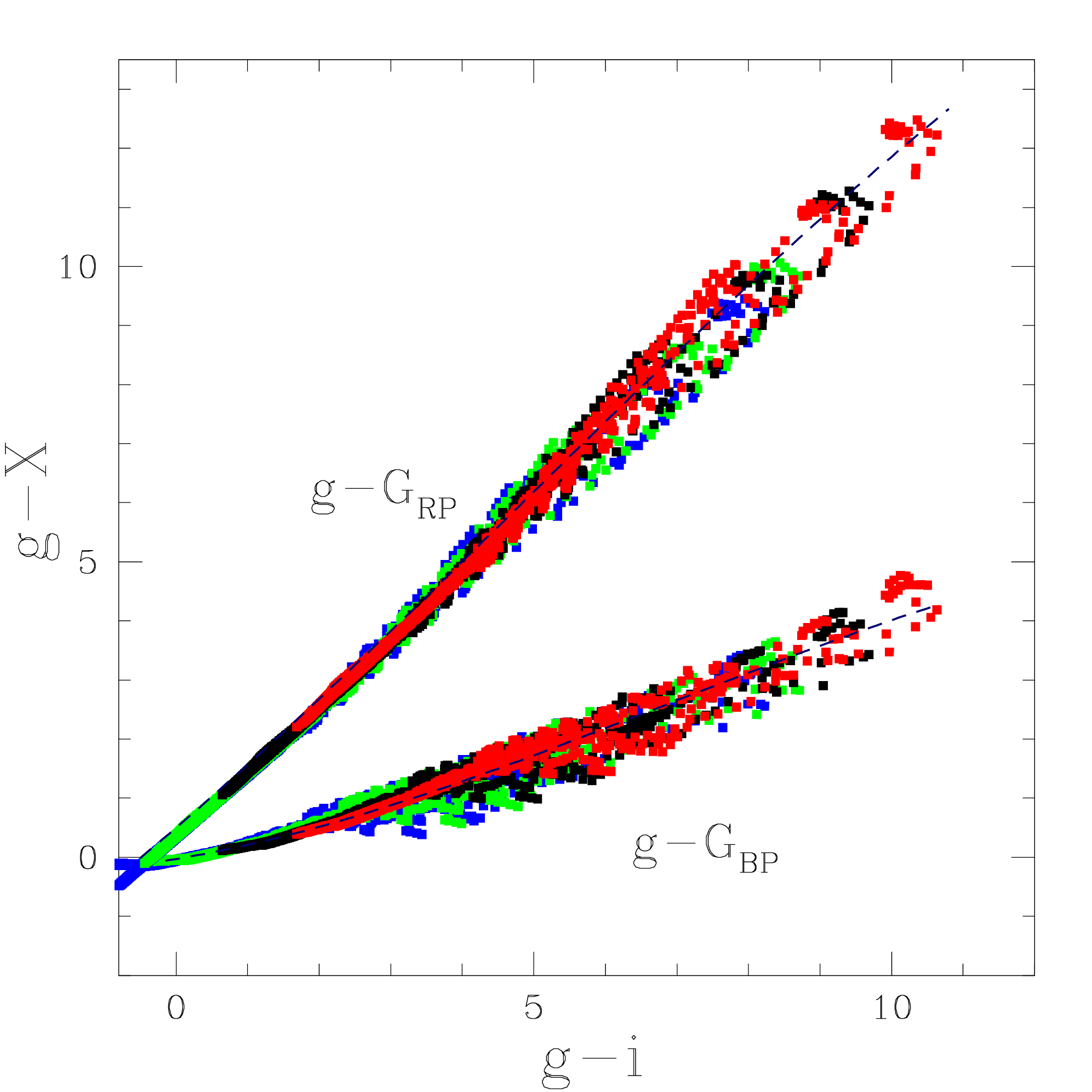}\\
 \includegraphics[scale=0.2]{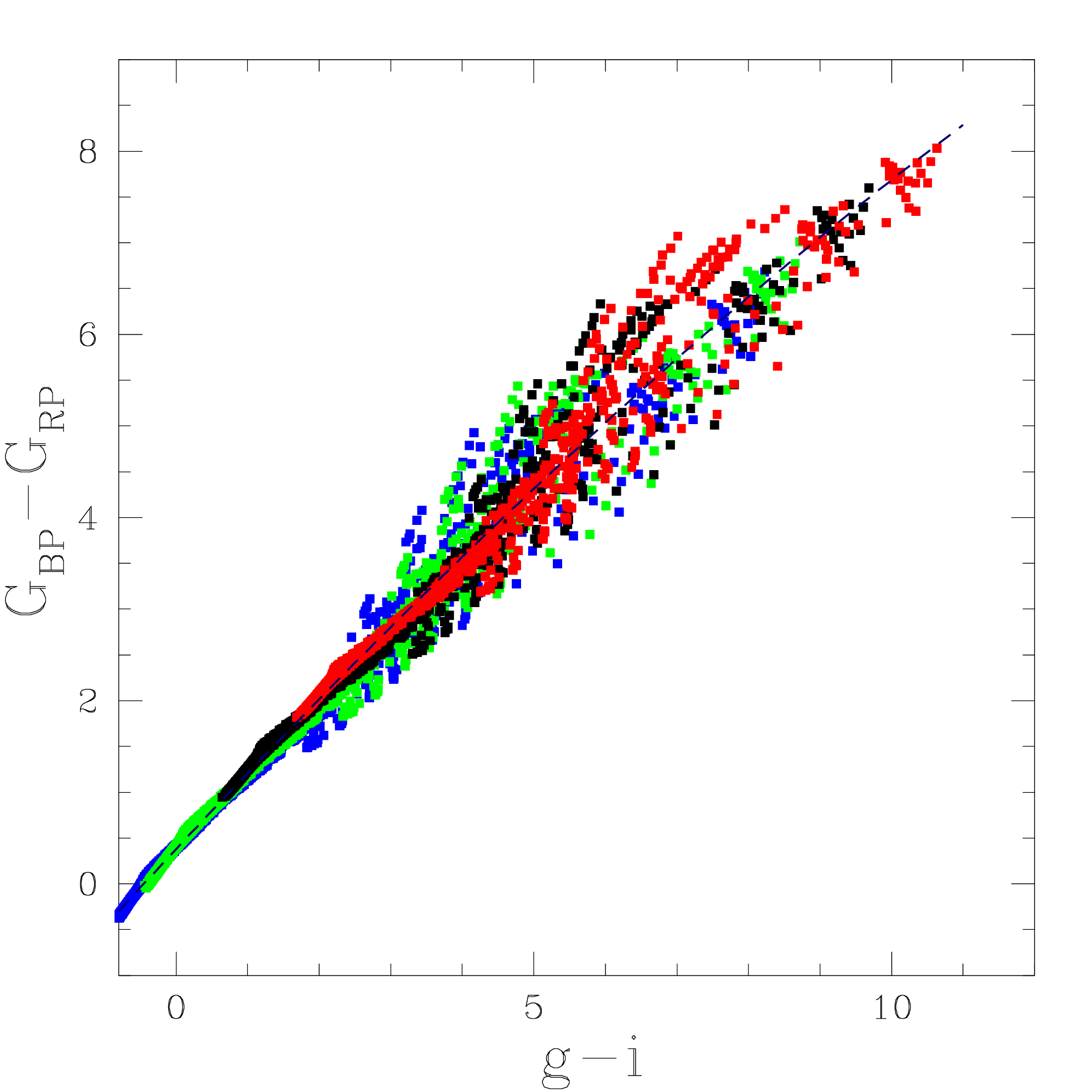}
 \includegraphics[scale=0.2]{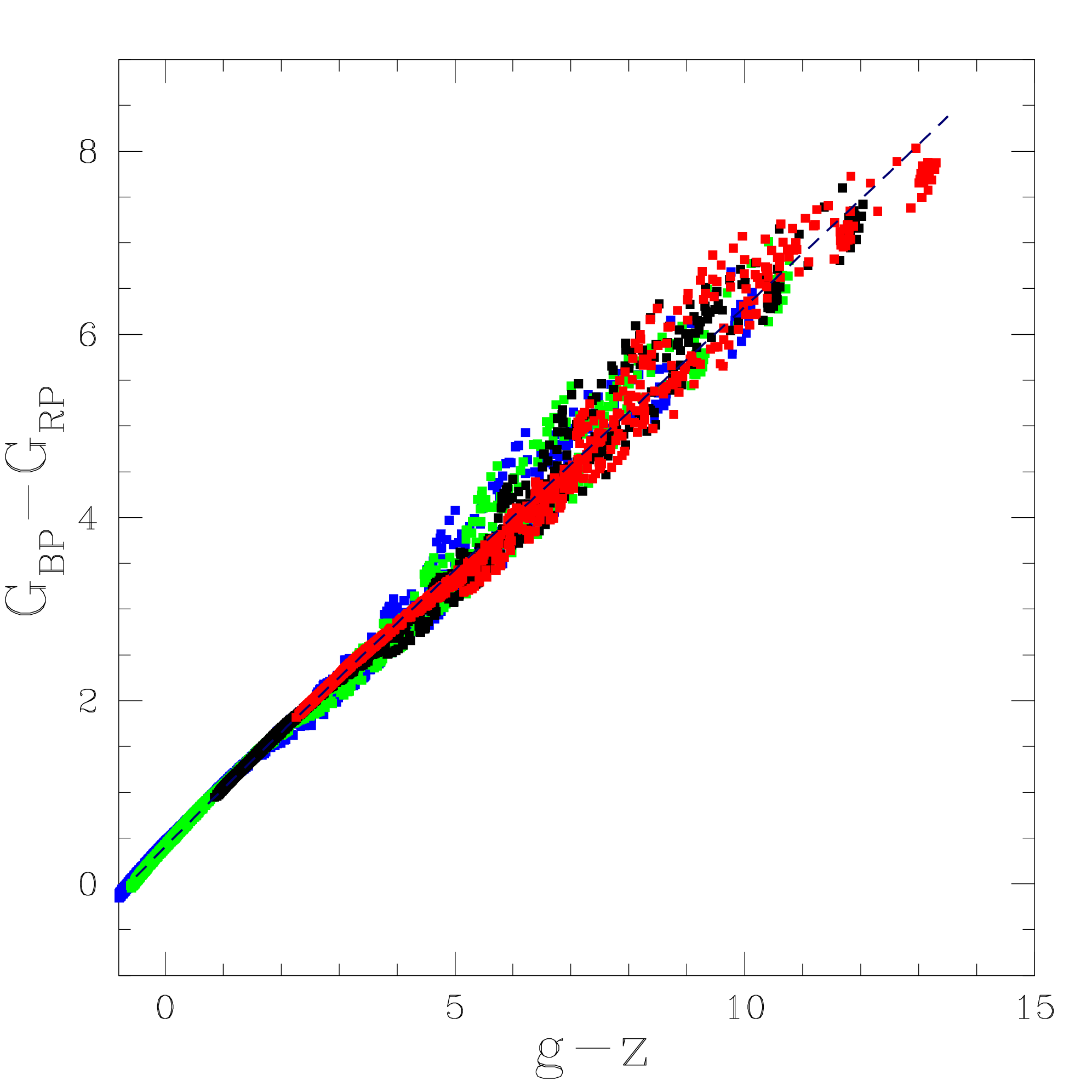}\\
 \includegraphics[scale=0.2]{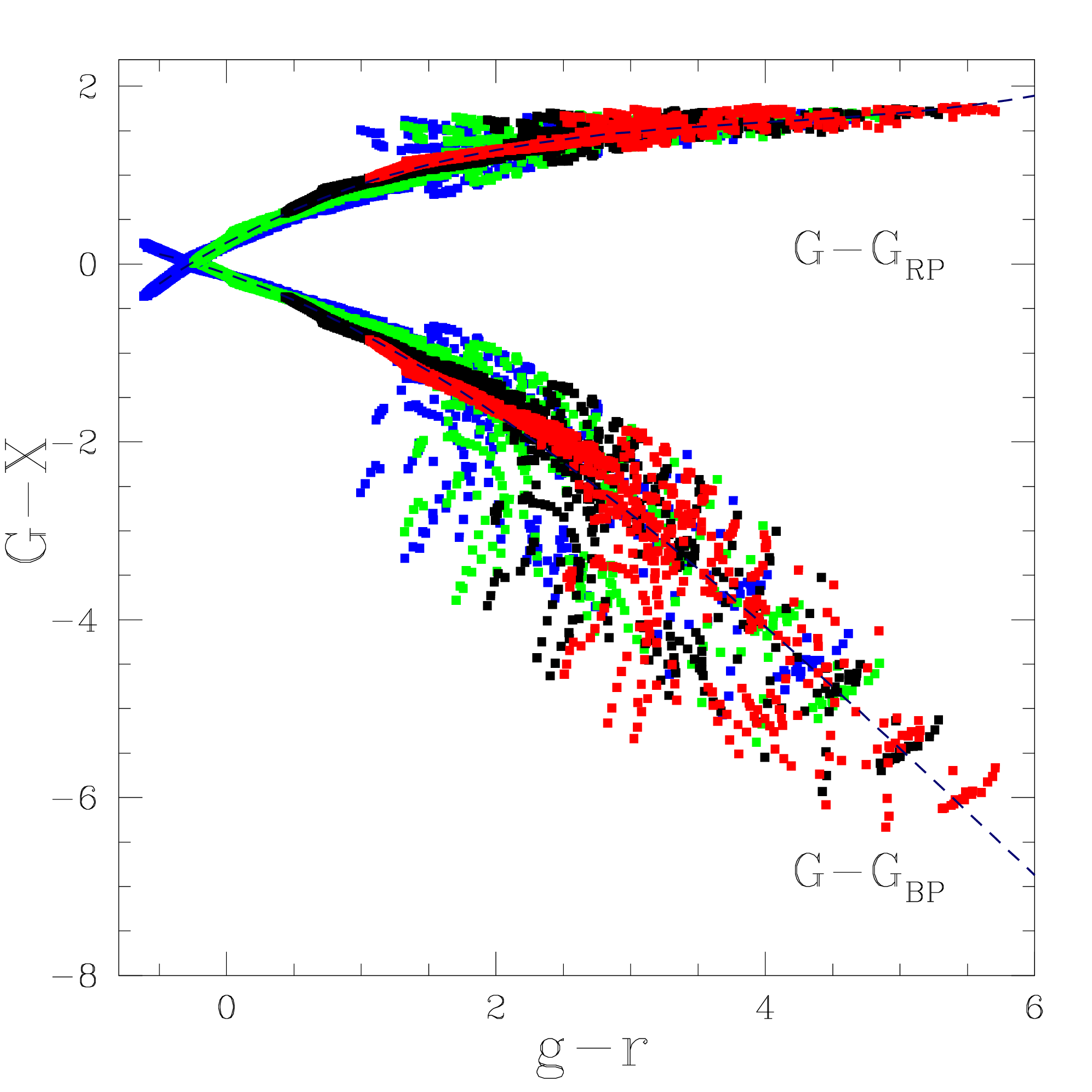}
 \includegraphics[scale=0.2]{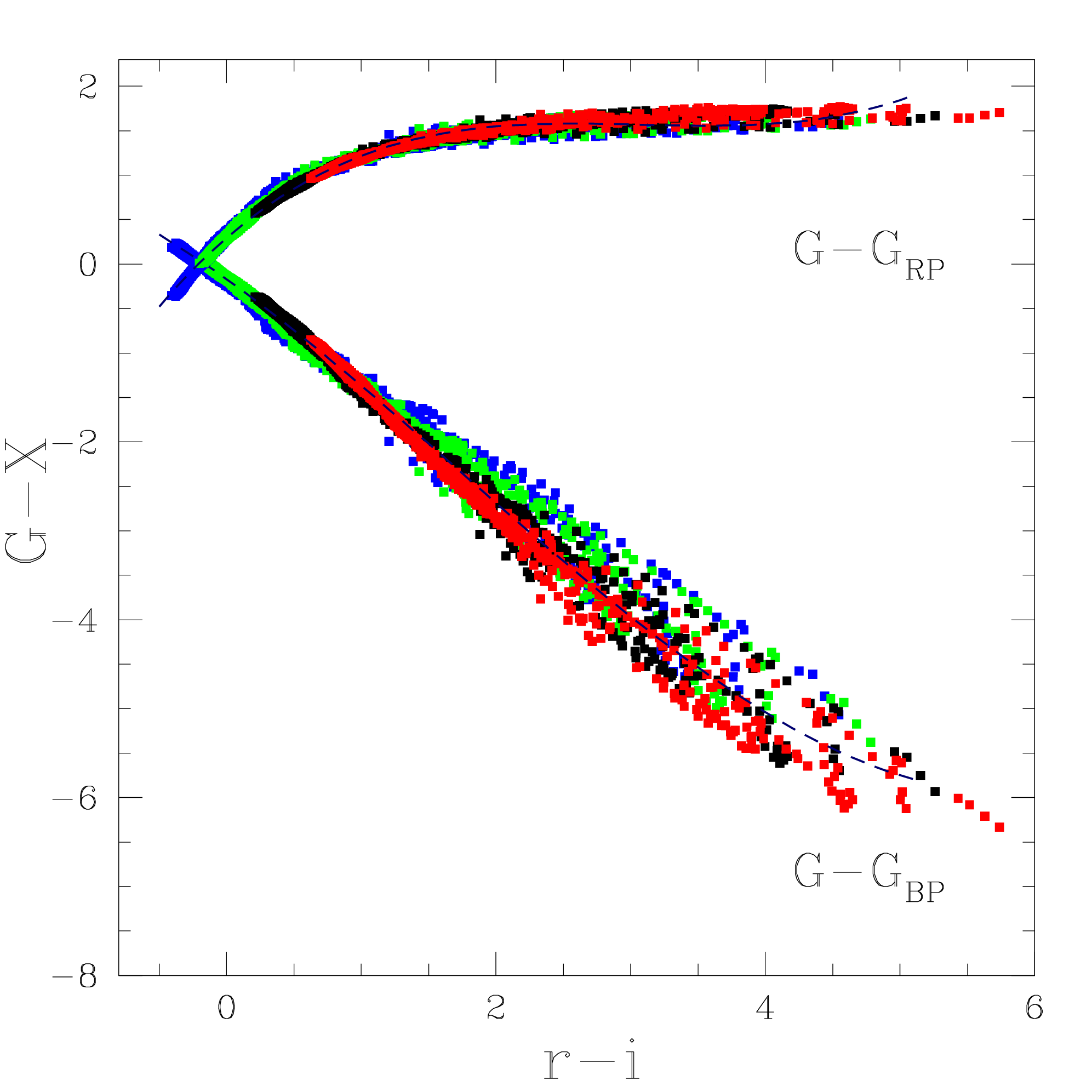}
\caption{Colour-colour diagrams involving the three broad {\Gaia} passbands and SDSS passbands. Different colours are used for different absorption values as in Fig.~\ref{fig:transformation-J-C}. Dashed lines correspond to the fitting in Table \ref{table:coeficients_SDSS}.}
 \label{fig:transformation-SDSS}
 \end{figure}

\begin{figure}[htbp!]

 \centering
 \includegraphics[scale=0.2]{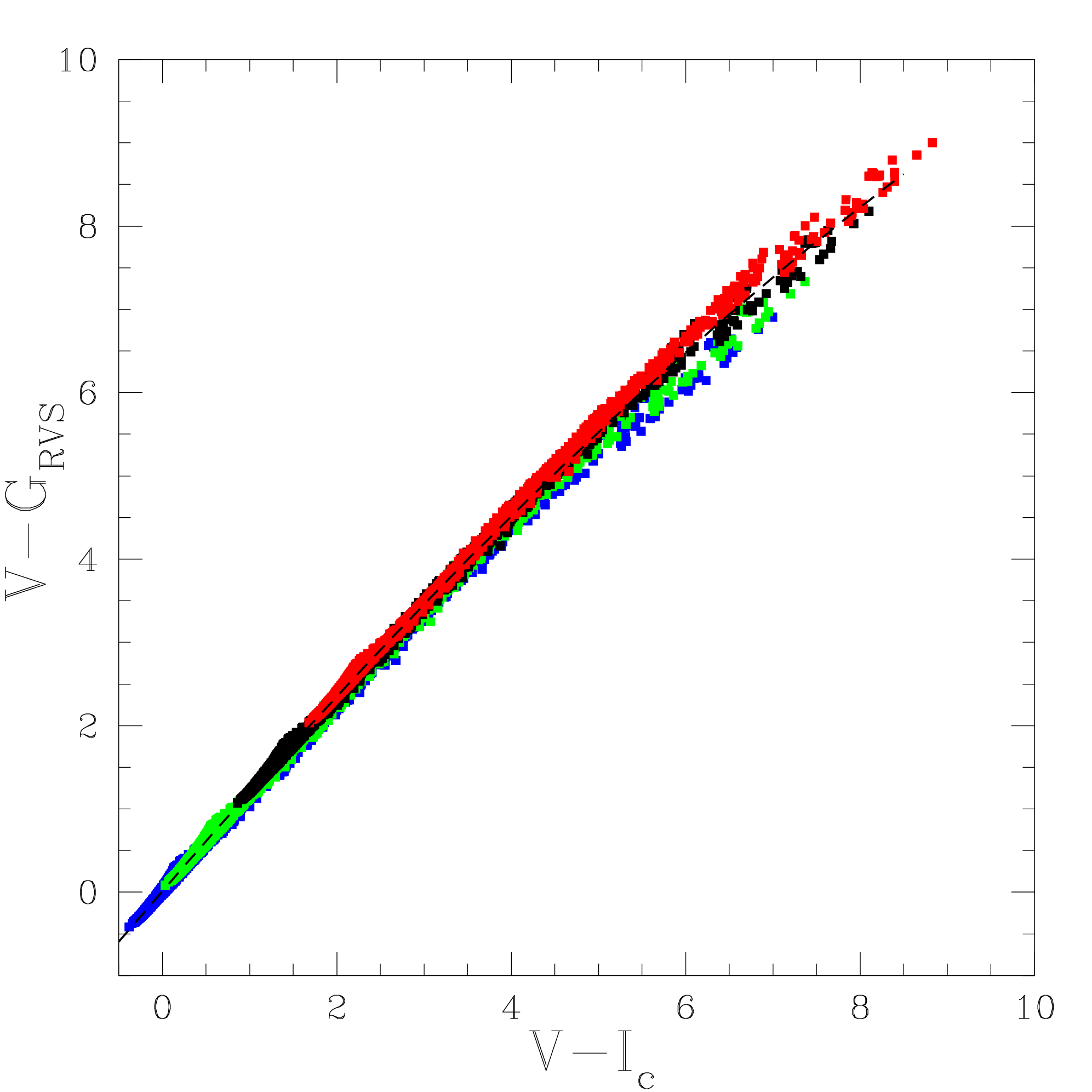}
  \includegraphics[scale=0.2]{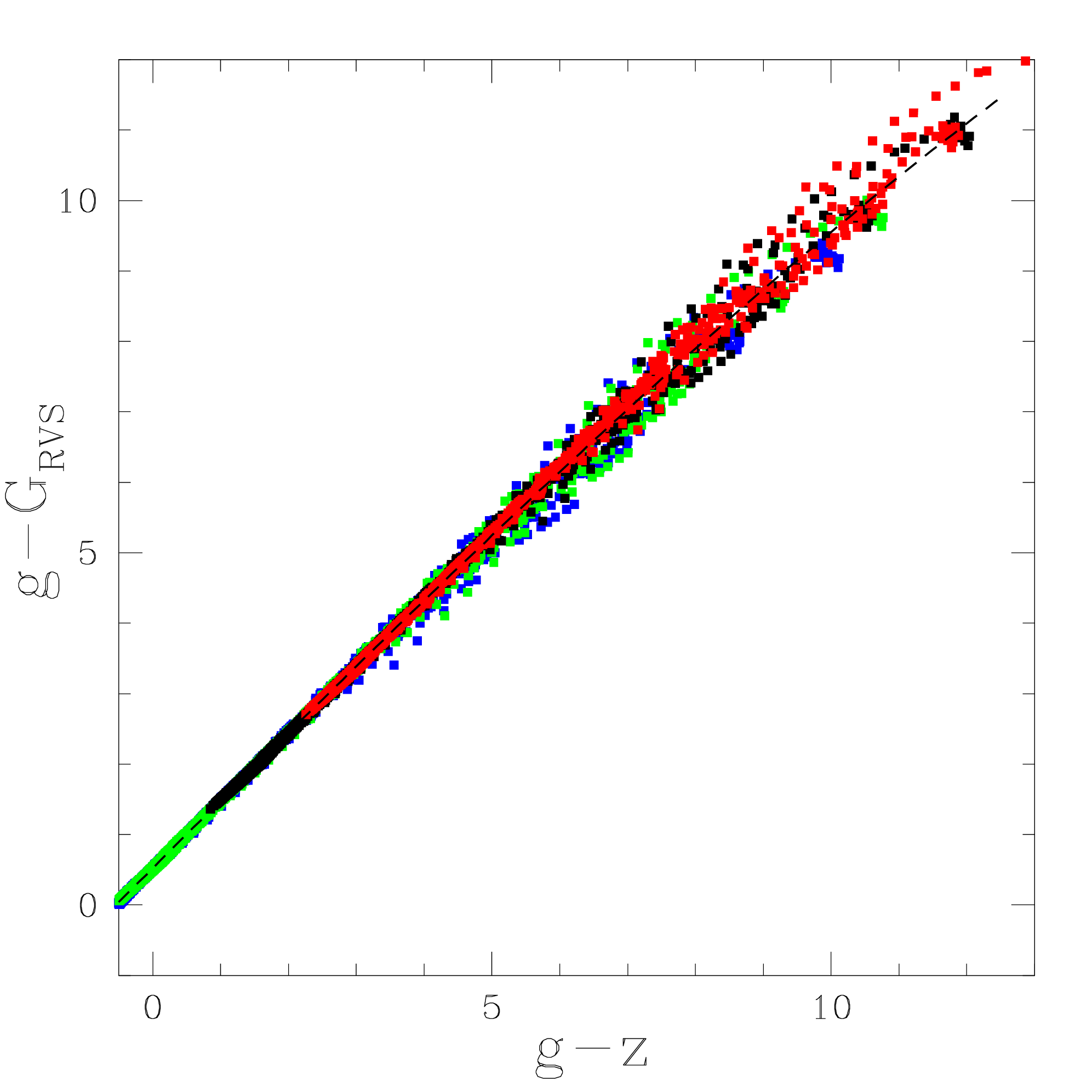}
  \caption{Colour-colour diagrams involving {\Gaia} $G_{\rm{RVS}}$ and Johnson-Cousins, {\it Hipparcos} and SDSS colours. Different colours are used for different absorption values as in Fig.~\ref{fig:transformation-J-C}. Dashed lines correspond to the fitting in Tables \ref{table:coeficients_Johnson} and \ref{table:coeficients_SDSS}.}
 \label{fig:RVS_diag}
\end{figure}

Several of the fits are presented in colour--colour diagrams in Figs.~\ref{fig:Tycho_Teff} 
to \ref{fig:RVS_diag}. For the transformations that involve Johnson-Cousins colours (Fig.~\ref{fig:transformation-J-C}), the relation with $V-I_{c}$ is the one that has the lowest residuals. One can notice an increase in dispersion starting at $V-I_c \gtrsim$ 4.5. This is due to the metallicity. As an example, we mention the upper left panel displaying $G-V$ with respect to $V-I_c$ where for a fixed $V-I_c$ value, metal poor stars have higher $G-V$ values than solar metallicity stars. This effect is the same for every extinction value. The relationships with $V-R_{C}$ or $R_{C}-I_{C}$ have also low residuals, but we only display those with $V-I_{C}$ as an example of the fitting.
The diagrams with the $B-V$ colour show large scatter, especially for $G-V$, $G-$\BP, 
$V-$\RP\ and \BP$-$\RP. The same effects appear with respect to $B_{T}-V_{T}$ as seen in Fig. \ref{fig:Tycho_Teff}. The residuals increase from $B_{T}-V_{T}\sim 1$ and $G-V_{T}<-0.5$, which is mainly due to cool stars ($T_{\rm{eff}}< 4500$~K). Among these cool stars, the scatter is due to the surface gravity and metallicity. It is preferable not to use the transformation with $B-V$ or $B_{T}-V_{T}$ for the cool stars. In Table~\ref{table:coeficients_Hippa} we present a relationship between $G-V_{T}$ and $B_{T}-V_{T}$ for stars with an effective temperature higher than 4500~K (black dots in Fig.~\ref{fig:Tycho_Teff}).

\begin{table*}[t]
\begin{center}
\caption{Coefficients of the colour-colour polynomial fittings using {\it Hipparcos}, {\it Tycho}, and Johnson-Cousins passbands.}
\label{table:coeficients_Hippa}
\tiny
\begin{tabular}{lrrrrcc}
 \hline
& & $(H_{p}-I_{C})$& $(H_{p}-I_{C})^2$&$(H_{p}-I_{C})^3$ &$\sigma$\\
 $G-V$&   -0.0447 &  -0.1634 &   0.0331 &  -0.0371 & 0.10\\
    $G-G_{\rm RVS}$& 0.0430 &   0.6959 &   0.0115 &  -0.0147 & 0.08\\
  $G-$\BP& 0.0142 &  -0.4149 &   0.0702 &  -0.0331 & 0.08\\
  $G-$\RP&    0.0096 &   0.5638 &  -0.0553 &   0.0016 & 0.02\\
   $V-G_{\rm RVS}$&   0.0877 &   0.8593 &  -0.0217 &   0.0224 & 0.11\\
$V-$\BP&     0.0589 &  -0.2515 &   0.0371 &   0.0040 & 0.05\\
    $V-$\RP&   0.0542 &   0.7271 &  -0.0884 &   0.0387 & 0.11\\
  \BP$-$\RP&  -0.0047 &   0.9787 &  -0.1254 &   0.0347 & 0.08\\
\hline
For $T_{\rm{eff}}\leq 2500$ K and [M/H]$<$-1.5 dex & & (\BP-\RP)& (\BP-\RP)$^2$&(\BP-\RP)$^3$ &$\sigma$\\
$G-H_{p}$&   1.0922 &  -1.3980 &   0.1593 &  -0.0073 & 0.05\\ \hline
For $T_{\rm{eff}}> 2500$ K or [M/H]$\geq$-1.5 dex & & (\BP-\RP)& (\BP-\RP)$^2$&(\BP-\RP)$^3$ &$\sigma$\\
  $G-H_{p}$ &0.0169 &  -0.4556 &  -0.0667 &   0.0075 & 0.04\\ \hline
For all the stars & & (\BP-\RP)& (\BP-\RP)$^2$&(\BP-\RP)$^3$ &$\sigma$\\
  $G-H_{p}$&    0.0172 &  -0.4565 &  -0.0666 &   0.0076 & 0.04\\
  $H_{p}-G_{\rm RVS}$&   0.0399 &   1.3790 &  -0.0412 &  -0.0061 & 0.07\\
  $H_{p}-$\BP&  0.0023 &   0.1253 &  -0.0279 &  -0.0029 & 0.05\\
   $H_{p}-$\RP&  0.0023 &   1.1253 &  -0.0279 &  -0.0029 & 0.05\\
\hline
For $T_{\rm{eff}}\geq 4500$ K and \Av$=0$ & & $(B_{T}-V_{T})$& $(B_{T}-V_{T})^2$&$(B_{T}-V_{T})^3$ &$\sigma$\\
$G-V_{T}$& -0.0260 &  -0.1767 &  -0.2980 &   0.1393 & 0.03\\

\hline
\end{tabular}
\tablefoot{Data computed with four values of extinction (\Av$=$ 0, 1, 3 and 5 mag).}

\end{center}
\end{table*}

For the {\it Hipparcos} passbands we show in Fig.~\ref{fig:Hipp-diag} two plots involving the {\Gaia} passbands and $H_{p}$. In the left panel where we display $G-H_{p}$ with respect to \BP$-$\RP, we notice a  deviation from the main trend for \BP$-$\RP$\gtrsim4$. This deviation is caused by cool metal poor stars with $T_{\rm{eff}}<$ 2500~K and [M/H]$<-1.5$ dex. For this reason, we have computed two distinct relationships involving \BP$-$\RP \ and $G-H_{p}$. These relationships are displayed in Table~\ref{table:coeficients_Hippa} and Fig.~\ref{fig:Hipp-diag}.

For the SDSS passbands, the relationships with $g-i$ colour are slightly more sensitive to 
reddening than with $V-I_{C}$. \BP$-$\RP \ correlates better with $g-z$ than with $g-i$ as shown in Fig.~\ref{fig:transformation-SDSS} and Table \ref{table:coeficients_SDSS}. The transformations from SDSS passbands yield residuals larger than with Johnson passbands. We have also plotted $G-$\BP\ and $G-$\RP\ with respect to $g-r$ and $r-i$ because in the SDSS system the stellar locus is defined mainly from the $g-r$ vs $r-i$ diagram \citep{1996AJ....111.1748F}. For stars with $T_{\rm{eff}}<4500$~K, dispersions exist in gravity and metallicity for each absorption value. This dispersion is more present in $g-r$ than in $r-i$.

\begin{table}[h]
\begin{center}
\caption{Coefficients of the colour-colour polynomial fittings using SDSS passbands.}
\label{table:coeficients_SDSS}
\tiny
\begin{tabular}{lrrrrcc}
\hline
& & $(g-i)$& $(g-i)^2$&$(g-i)^3$ &$\sigma$\\	
$G-g$&			 -0.0940 &  -0.5310 &  -0.0974 &   0.0052 & 0.09\\						
$G-G_{\rm RVS}$&  0.3931 &   0.7250 &  -0.0927 &   0.0032 & 0.10\\ 
$G-$\BP&  -0.1235 &  -0.3289 &  -0.0582 &   0.0033 & 0.14\\  
$G-$\RP&   0.2566 &   0.5086 &  -0.0678 &   0.0032 & 0.04\\  
$g-G_{\rm RVS}$&  0.4871 &   1.2560 &   0.0047 &  -0.0020 & 0.12\\ 
$g-$\BP&  -0.0294 &   0.2021 &   0.0392 &  -0.0019 & 0.11\\  
$g-$\RP&   0.3506 &   1.0397 &   0.0296 &  -0.0020 & 0.11\\  
\BP$-$\RP&   0.3800 &   0.8376 &  -0.0097 &  -0.0001 & 0.17\\
 \hline	       
          & & $(g-r)$& $(g-r)^2$&$(g-r)^3$ &$\sigma$\\
  $G-g$&		  -0.0662 &  -0.7854 &  -0.2859 &   0.0145 & 0.30\\
  $G-G_{\rm RVS}$& 0.3660 &   1.1503 &  -0.2200 &   0.0088 & 0.13\\
  $G-$\BP&  -0.1091 &  -0.5213 &  -0.1505 &   0.0083 & 0.29\\ 
  $G-$\RP&   0.2391 &   0.8250 &  -0.1815 &   0.0150 & 0.07\\ 
  $g-G_{\rm RVS}$& 0.4322 &   1.9358 &   0.0660 &  -0.0057 & 0.36\\
  $g-$\BP&  -0.0429 &   0.2642 &   0.1354 &  -0.0061 & 0.03\\ 
  $g-$\RP&   0.3053 &   1.6104 &   0.1044 &   0.0006 & 0.36\\ 
  \BP$-$\RP& 0.3482 &   1.3463 &  -0.0310 &   0.0067 & 0.35\\ 
 \hline	       
               & & $(r-i)$& $(r-i)^2$&$(r-i)^3$ &$\sigma$\\
   $G-g$&	        -0.1741 &  -1.8240 &  -0.1877 &   0.0365 & 0.28\\
   $G-G_{\rm RVS}$&  0.4469 &   1.9259 &  -0.6724 &   0.0686 & 0.07\\
   $G-$\BP&       -0.1703 &  -1.0813 &  -0.1424 &   0.0271 & 0.10\\
   $G-$\RP&        0.2945 &   1.3156 &  -0.4401 &   0.0478 & 0.04\\
   $g-G_{\rm RVS}$&  0.6210 &   3.7499 &  -0.4847 &   0.0321 & 0.28\\
   $g-$\BP&        0.0038 &   0.7427 &   0.0453 &  -0.0094 & 0.22\\
   $g-$\RP&        0.4686 &   3.1396 &  -0.2523 &   0.0113 & 0.29\\
   \BP$-$\RP&      0.4649 &   2.3969 &  -0.2976 &   0.0207 & 0.12\\
\hline
         & & $(g-z)$& $(g-z)^2$&$(g-z)^3$ &$\sigma$\\	
  $G-g$&	      -0.1154 &  -0.4175 &  -0.0497 &   0.0016 & 0.08\\
  $G-G_{\rm RVS}$&  0.4087 &   0.5474 &  -0.0519 &   0.0012 & 0.07\\
  $G-$\BP&     -0.1350 &  -0.2545 &  -0.0309 &   0.0011 & 0.09\\
  $G-$\RP&      0.2702 &   0.3862 &  -0.0401 &   0.0015 & 0.02\\
  $g-G_{\rm RVS}$&  0.5241 &   0.9649 &  -0.0022 &  -0.0004 & 0.08\\
  $g-$\BP&     -0.0195 &   0.1630 &   0.0188 &  -0.0005 & 0.14\\
  $g-$\RP&      0.3857 &   0.8037 &   0.0096 &  -0.0001 & 0.06\\
  \BP$-$\RP&    0.4052 &   0.6407 &  -0.0091 &   0.0004 & 0.11\\
\hline
\end{tabular}
\tablefoot{Data computed with four values of extinction (\Av$=$ 0, 1, 3 and 5 mag).}

\end{center}
\end{table}

Finally, Fig.~\ref{fig:RVS_diag} displays two plots involving the {\Gaia} $G_{RVS}$ narrow band, Johnson-Cousins, and SDSS passbands. The relationships can be found in Tables \ref{table:coeficients_Johnson} and \ref{table:coeficients_SDSS}.

Transformations using two Johnson or two SDSS colours have also been computed in the form 
$$C_1=a+b\cdot C_2+c\cdot C_2^2+ d\cdot C_2^3+e\cdot C_3+f\cdot C_3^2+ g\cdot C_3^3+h\cdot C_2C_3$$

\noindent
and they are shown in Table~\ref{table:coeficients_2colors}. The residuals are lower 
than using only one colour. For the Johnson-Cousins system, the residuals do not decrease 
much, but for the SDSS system the improvement is substantial and the residuals are of the same order as those derived with $V-I_{C}$. Thus, for Sloan, transformations 
with two colours are preferred.

The residuals can still be decreased if different transformations are considered for 
different ranges of colours, reddening values, luminosity classes, and metallicities.
As an example, for unreddened stars (nearby stars or stars above the galactic plane), the fittings are those in Table \ref{table:coeficients-unred}.

\begin{table}[h]
\begin{center}
\caption{Coefficients of the unreddened colour-colour polynomial fittings using Johnson-Cousins and SDSS passbands.}
\tiny
\begin{tabular}{lrrrrcc}
 \hline
& & $(V-I_{C})$& $(V-I_{C})^2$&$(V-I_{C})^3$ &$\sigma$\\
 $G-V$&  -0.0354 &  -0.0561 &  -0.1767 &   0.0108 & 0.04\\
  $G-G_{\rm RVS}$&  -0.0215 &   1.0786 &  -0.1713 &   0.0068 & 0.06\\
 $G-$\BP&   0.0379 &  -0.4697 &  -0.0450 &   0.0008 & 0.05\\
 $G-$\RP&   -0.0360 &   0.8279 &  -0.1549 &   0.0105 & 0.03\\
  $V-G_{\rm RVS}$ & 0.0139 &   1.1347 &   0.0054 &  -0.0040 & 0.05\\
  $V-$\BP&  0.0733 &  -0.4136 &   0.1316 &  -0.0100 & 0.05\\
  $V-$\RP&  -0.0007 &   0.8840 &   0.0218 &  -0.0003 & 0.05\\
 \BP$-$\RP & -0.0740 &   1.2976 &  -0.1099 &   0.0097 & 0.07\\
\hline
 & & $(g-i)$& $(g-i)^2$&$(g-i)^3$ &$\sigma$\\
   $G-g$&    -0.0912 &  -0.5310 &  -0.1042 &   0.0068 & 0.07\\
  $G-G_{\rm RVS}$&   0.3356 &   0.6528 &  -0.0655 &   0.0004 & 0.08\\
    $G-$\BP& -0.1092 &  -0.3298 &  -0.0642 &   0.0045 & 0.12\\
    $G-$\RP&     0.2369 &   0.4942 &  -0.0612 &   0.0027 & 0.04\\
   $g-G_{\rm RVS}$&    0.4268 &   1.1838 &   0.0387 &  -0.0063 & 0.09\\
   $g-$\BP&    -0.0180 &   0.2013 &   0.0399 &  -0.0023 & 0.10\\
    $g-$\RP&   0.3281 &   1.0252 &   0.0429 &  -0.0041 & 0.09\\
   \BP$-$\RP &    0.3461 &   0.8240 &   0.0030 &  -0.0018 & 0.16\\
\hline
\end{tabular}
\label{table:coeficients-unred}
\end{center}
\end{table}

\begin{table*}[t]
\begin{center}
\caption{Coefficients of the colour-colour polynomial fittings using two colours.}
\label{table:coeficients_2colors}
\tiny
\begin{tabular}{lrrrrrrrcc} 
 \hline
 & &        $(V-I_{C})$&$(V-I_{C})^2$&$(V-I_{C})^3$&$(B-V)$&$(B-V)^2$&$(B-V)^3$&$(V-I_{C})(B-V)$&$\sigma$ \\ 
 $G-V$  &  -0.0099 &  -0.2116 &  -0.1387 &   0.0060 &   0.1485 &  -0.0895 &   0.0094 &   0.0327 & 0.04\\
$G-G_{\rm RVS}$  &  -0.0287 &   1.2419 &  -0.2260 &   0.0097 &  -0.1493 &  -0.0194 &  -0.0026 &   0.0710 & 0.07\\
 $G-$\BP &   0.0088 &  -0.1612 &  -0.1569 &   0.0080 &  -0.3243 &   0.0692 &  -0.0082 &   0.0582 & 0.04\\
  $G-$\RP&  -0.0075 &   0.6212 &  -0.0816 &   0.0040 &   0.2206 &  -0.1020 &   0.0131 &  -0.0053 & 0.02\\
 $V-G_{\rm RVS}$ &  -0.0188 &   1.4535 &  -0.0873 &   0.0038 &  -0.2977 &   0.0701 &  -0.0120 &   0.0383 & 0.06\\
$V-$\BP &   0.0188 &   0.0504 &  -0.0181 &   0.0021 &  -0.4728 &   0.1586 &  -0.0176 &   0.0255 & 0.03\\
 $V-$\RP &   0.0024 &   0.8328 &   0.0572 &  -0.0020 &   0.0722 &  -0.0126 &   0.0037 &  -0.0380 & 0.06\\
\BP$-$\RP &  -0.0163 &   0.7825 &   0.0753 &  -0.0040 &   0.5450 &  -0.1712 &   0.0213 &  -0.0635 & 0.06\\
\hline
         & &                  $(g-i)$&$(g-i)^2$&$(g-i)^3$&$(g-r)$&$(g-r)^2$&$(g-r)^3$&$(g-r)(g-i)$&$\sigma$\\
 $G-g$ &  -0.1005 &  -0.5358 &  -0.1207 &   0.0082 &  -0.0272 &   0.1270 &  -0.0205 &  -0.0176 & 0.08\\
 $G-G_{\rm RVS}$&   0.4341 &   1.7705 &  -0.4126 &   0.0130 &  -1.6834 &  -0.0036 &  -0.0463 &   0.5242 & 0.07\\
$G-$\BP &  -0.1433 &  -0.5200 &  -0.1086 &   0.0087 &   0.2110 &   0.3126 &  -0.0301 &  -0.0669 & 0.08\\
 $G-$\RP &   0.2726 &   0.7978 &  -0.1449 &   0.0035 &  -0.4358 &  -0.1458 &   0.0025 &   0.1990 & 0.02\\
 $g-G_{\rm RVS}$ &   0.5345 &   2.3063 &  -0.2919 &   0.0049 &  -1.6562 &  -0.1306 &  -0.0257 &   0.5419 & 0.09\\
 $g-$\BP &  -0.0428 &   0.0158 &   0.0122 &   0.0005 &   0.2382 &   0.1855 &  -0.0096 &  -0.0493 & 0.02\\
$g-$\RP &   0.3731 &   1.3335 &  -0.0242 &  -0.0047 &  -0.4086 &  -0.2729 &   0.0230 &   0.2166 & 0.10\\
  \BP$-$\RP&   0.4159 &   1.3177 &  -0.0364 &  -0.0052 &  -0.6468 &  -0.4584 &   0.0326 &   0.2659 & 0.10\\
\hline
         & &                  $(r-i)$&$(r-i)^2$&$(r-i)^3$&$(g-r)$&$(g-r)^2$&$(g-r)^3$&$(g-r)(r-i)$&$\sigma$\\
$G-g$ &  -0.0992 &  -0.5749 &  -0.2427 &   0.0365 &  -0.5277 &  -0.1158 &   0.0086 &  -0.0337 & 0.08\\
  $G-G_{\rm RVS}$&   0.4362 &   1.7138 &  -0.6096 &   0.0585 &   0.1399 &  -0.0572 &   0.0000 &   0.0578 & 0.07\\
   $G-$\BP &  -0.1430 &  -0.5925 &  -0.2220 &   0.0366 &  -0.2521 &   0.0198 &   0.0015 &  -0.0452 & 0.08\\
$G-$\RP &   0.2733 &   0.7851 &  -0.1993 &   0.0159 &   0.3745 &  -0.1357 &   0.0149 &   0.0057 & 0.02\\
 $g-G_{\rm RVS}$ &   0.5355 &   2.2887 &  -0.3669 &   0.0220 &   0.6676 &   0.0587 &  -0.0086 &   0.0915 & 0.09\\
$g-$\BP  &  -0.0438 &  -0.0176 &   0.0207 &   0.0001 &   0.2755 &   0.1356 &  -0.0071 &  -0.0114 & 0.02\\
  $g-$\RP&   0.3725 &   1.3600 &   0.0434 &  -0.0206 &   0.9022 &  -0.0199 &   0.0062 &   0.0394 & 0.10\\
  \BP$-$\RP &   0.4163 &   1.3776 &   0.0227 &  -0.0206 &   0.6267 &  -0.1555 &   0.0133 &   0.0509 & 0.10\\
\hline
\end{tabular}
\tablefoot{Data computed with four values of extinction (\Av$=$ 0, 1, 3 and 5 mag).}

\end{center}
\end{table*}

\section {Bolometric correction}
\label{sec:Mbol}
Luminosity is a fundamental stellar parameter that is essential for testing stellar structure and evolutionary models. Luminosity is derived by computing the integrated energy flux over the entire wavelength range (bolometric magnitude). The relation between the absolute magnitude in a specific passband and the bolometric one is done through the bolometric correction ($BC$).

For a given filter transmission curve, $S_X(\lambda)$, the bolometric correction is defined by
\begin{equation}
BC_{S_X}=M_{\rm{bol}}-M_{S_X}.
\label{bolo_corre1}
\end{equation}
This correction can be derived for each star of known $T_{\rm{eff}}$ and $ \log g$, using the following equation from \citet{Girardi2002}
\begin{eqnarray*}
BC_{S_X}&=&M_{\rm{bol},\odot}-2.5 \log [4\pi(10\rm{pc})^{2} F_{\rm{bol}}/L_{\odot}] 
\end{eqnarray*}
\begin{eqnarray}
 \  \ \ \ \ \ \ &\ & +2.5 \log \left(\frac{\int_{\lambda_{1}}^{\lambda_{2}}  F_{\lambda} S_X(\lambda) d\lambda}{\int_{\lambda_{1}}^{\lambda_{2}}  f^{0}_{\lambda} S_X(\lambda) d\lambda}\right) - m^{0}_{S_X},
\label{bolo_corre2}
\end{eqnarray}\\

\noindent where $M_{\rm{bol,\odot}}=4.75$ \citep{1999Obs...119..289A} is the bolometric magnitude of the Sun and
$L_{\odot}=3.856\times10^{26}$ W is its luminosity\footnote{http://www.spenvis.oma.be/spenvis/ecss/ecss06/ecss06.html}. $f^{0}_{\lambda}$ stands for the reference spectrum (e.g. Vega) at the Earth with its apparent magnitude 
$m^{0}_{S_X(\lambda)}$. $F_{\rm{bol}}$ is the total flux at the surface of the star 
($F_{\rm{bol}}=\int_{0}^{\infty} F_{\lambda} d\lambda = \sigma T_{\rm{eff}}^{4}$).

Substituting $F_{\rm{bol}}$ by $ \sigma T_{\rm{eff}}^{4}$, Eq.~\ref{bolo_corre2} can be
rewritten as

\begin{equation} 
BC_{S_X}=-\textsf{M}_{S_X}-2.5 \log (T_{\rm{eff}}^{4})-0.8637,
\label{bolo_corre3}
\end{equation}
where $\textsf{M}_{S_X}=-2.5 \log \left(\frac{\int_{\lambda_{1}}^{\lambda_{2}}F_{\lambda}  S_X(\lambda) d\lambda}{\int_{\lambda_{1}}^{\lambda_{2}} f^{0}_{\lambda} S_X(\lambda) d\lambda}\right) + m^{0}_{S_X}$  
is computed using the SED of the star at its surface. Equation \ref{bolo_corre3} 
is similar to the one derived in \cite{Bessel98} and will be used to compute the 
bolometric correction in the {\Gaia} photometric bands. Once we compute the bolometric correction $BC_{S_X}$, the bolometric absolute magnitude $M_{\rm{bol}}$ can be derived from Eq.~\ref{bolo_corre1}. 

\begin{figure}[htbp!]
 \centering
 \includegraphics[scale=0.45]{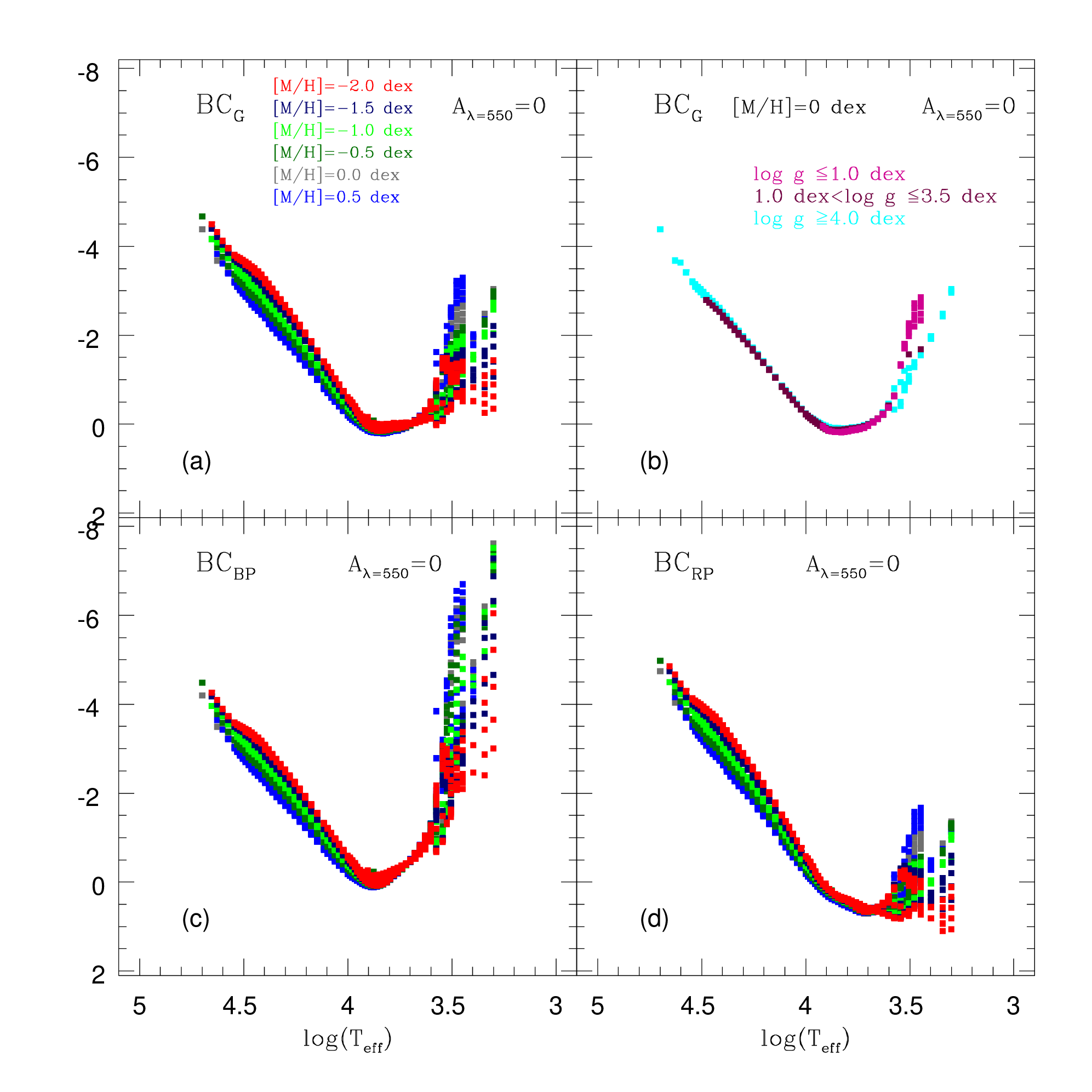}
  \caption{Panels (a), (c) and (d) show the bolometric correction in $G$, \BP, and \RP\ with respect to the effective 
temperature and metallicity. Panels (a), (c) and (d) have the same legends. Panel (b) displays the variation of the bolometric correction in $G$ with respect to surface gravity for solar metallicity. }
 \label{fig:Bolom-plot}
\end{figure}

Figure \ref{fig:Bolom-plot} and Table\footnote{Table 12 is only available in electronic form.}~12 display the bolometric correction for the $G$, \BP, and \RP\ bands for different metallicities and surface gravities. Panels (a), (c) and (d) of Fig.~\ref{fig:Bolom-plot} display the bolometric correction and its dependence with effective temperature and metallicity. 
Panel (b) shows the variation of the bolometric correction in $G$ with respect to surface gravity and for solar 
metallicity. The bolometric corrections in $G$ and \BP\ are near zero for F-type stars and for the entire metallicity range. For \RP, the maximum around $BC_{RP}=$0.75 is found to be related to stars with $T_{\rm{eff}}$ around $4500-5000$~K. For cool stars ($\log T_{\rm{eff}}\leq3.6$ dex) and for each temperature, there is a large dispersion in bolometric correction values with respect to surface gravity and metallicity.

\section{Interstellar absorption}
\label{sec:Av}

\begin{figure}[th!]
 \centering
  \includegraphics[scale=0.2]{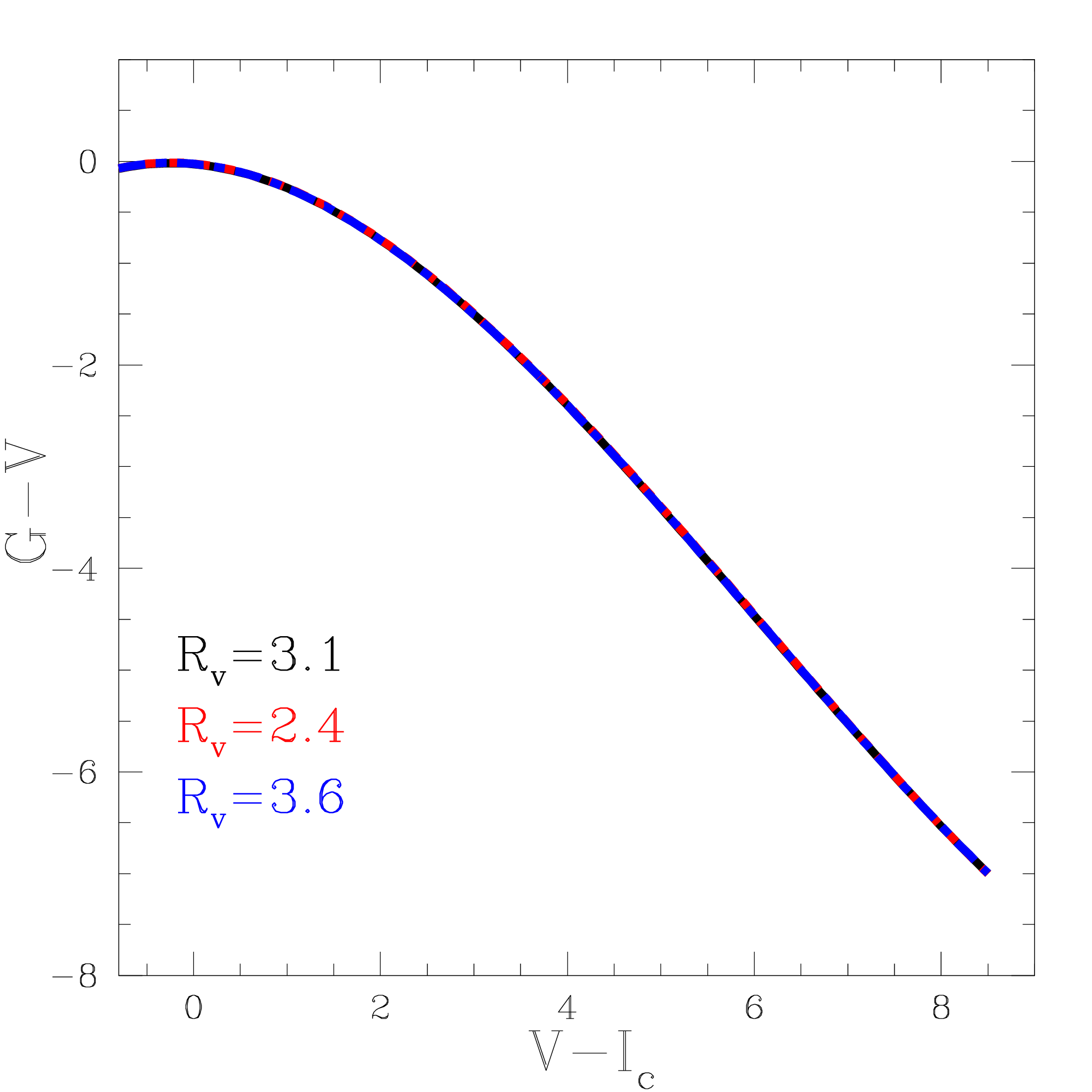}
    \includegraphics[scale=0.2]{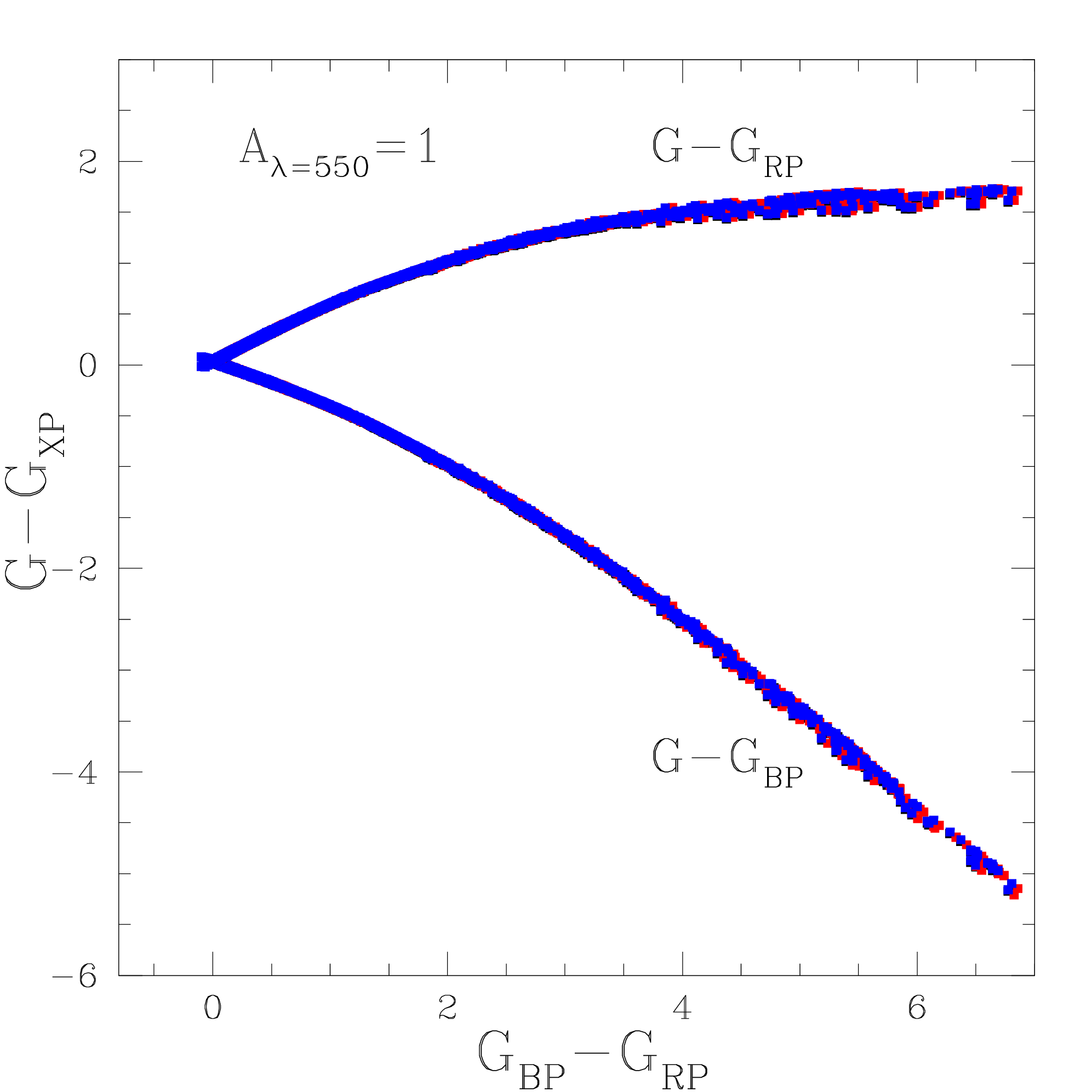}
  \caption{Left: polynomial fitting of different transformations for $G-V$ with respect to $V-I_{c}$ using Cardelli et al. (1989) law with different $R_{V}$ values ($R_{V}=2.4$, 3.1 and 3.6). In each case the four absorption values 
have been considered (\Av$=0$, 1, 3 and 5 mag). Right: different transformations for $G-G_{XP}$ with respect to $G_{BP}-G_{RP}$ using different $R_{V}$ values ($R_{V}=2.4$, 3.1 and 3.6). Only \Av$=1$ is displayed.}
 \label{fig:RV_var}
\end{figure}

The extinction curve used in the previous section was taken from
\cite{1989ApJ...345..245C} assuming an average galactic value of $R_V=3.1$. 
This curve agrees with \cite{1999PASP..111...63F} and \cite{2007ApJ...663..320F} in the
wavelength range of {\Gaia}'s passbands, 330--1000~nm. In \cite{2007ApJ...663..320F}, 
which contains the most updated discussion on the absorption law, extinction curves 
with $R_V$ values in the range 2.4--3.6 are considered for a sample 243 stars in
sight lines with diffuse interstellar medium. {\Gaia}~magnitudes have been
recomputed for all spectra of the BaSeL3.1 library and \Av$=0$, 1, 3 and 5 mag with 
$R_V=2.4$ and $R_V=3.6$. The left panel of Fig.~\ref{fig:RV_var} shows the $G-V$ vs $V-I_c$ 
polynomial relationships for each value of $R_V$. No differences are noticeable: the polynomials overlap in the three cases. In the right panel of Fig.~\ref{fig:RV_var}, we display the effect of the variation of $R_V$ on the {\Gaia} colour-colour diagram. We have computed $G-G_{BP}$ and $G-G_{RP}$ with respect to \BP$-$\RP \ colour for an absorption value \Av$=1$~mag, and using the three different values of $R_V$. The effect of modifying $R_V$ is also negligible in this case. 

\begin{table*}[th!]
\begin{center}
\caption{Coefficients of the polynomial fittings of $A_{G}/A_{V}$ with respect to the unreddened $(V-I_{C})_{0}$.}
\begin{tabular}{llrrrrcc}
 \hline
& & & $(V-I_{C})_{0}$& $(V-I_{C})_{0}^2$&$(V-I_{C})_{0}^3$ &$\sigma$\\
For \Av = 1& $A_{G}/A_{V}$  & 0.9500 &  -0.1569 &   0.0210 &  -0.0010 & 0.01\\
For \Av = 3& $A_{G}/A_{V}$  &  0.8929 &  -0.1378 &   0.0186 &  -0.0008 & 0.01\\
For \Av = 5& $A_{G}/A_{V}$  & 0.8426 &  -0.1187 &   0.0157 &  -0.0007 & 0.01\\
\hline
\end{tabular}
\label{Tab:Ag/Av}
\end{center}
\end{table*}

\begin{figure}[htbp!]
\centering
\includegraphics[scale=0.22]{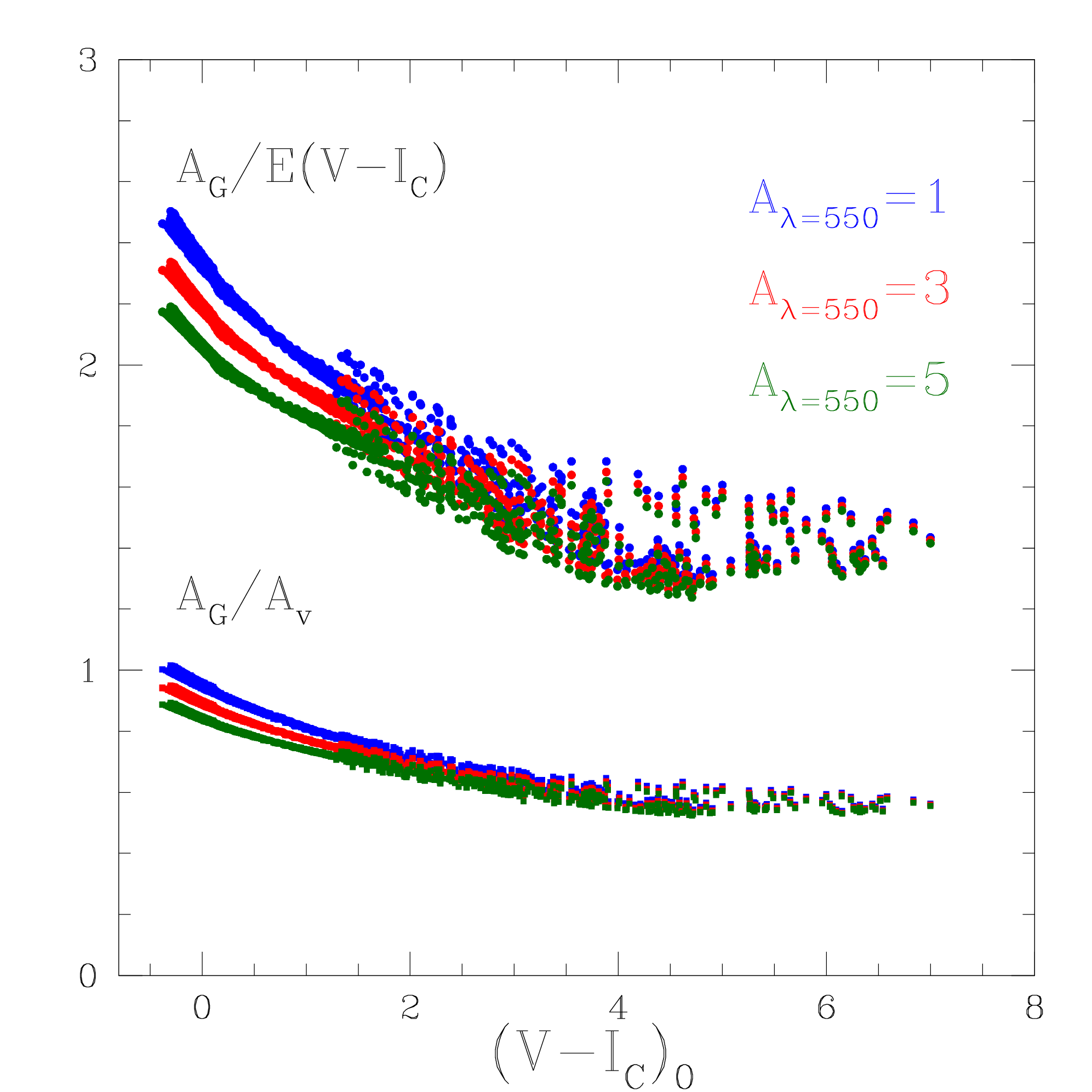}
\includegraphics[scale=0.22]{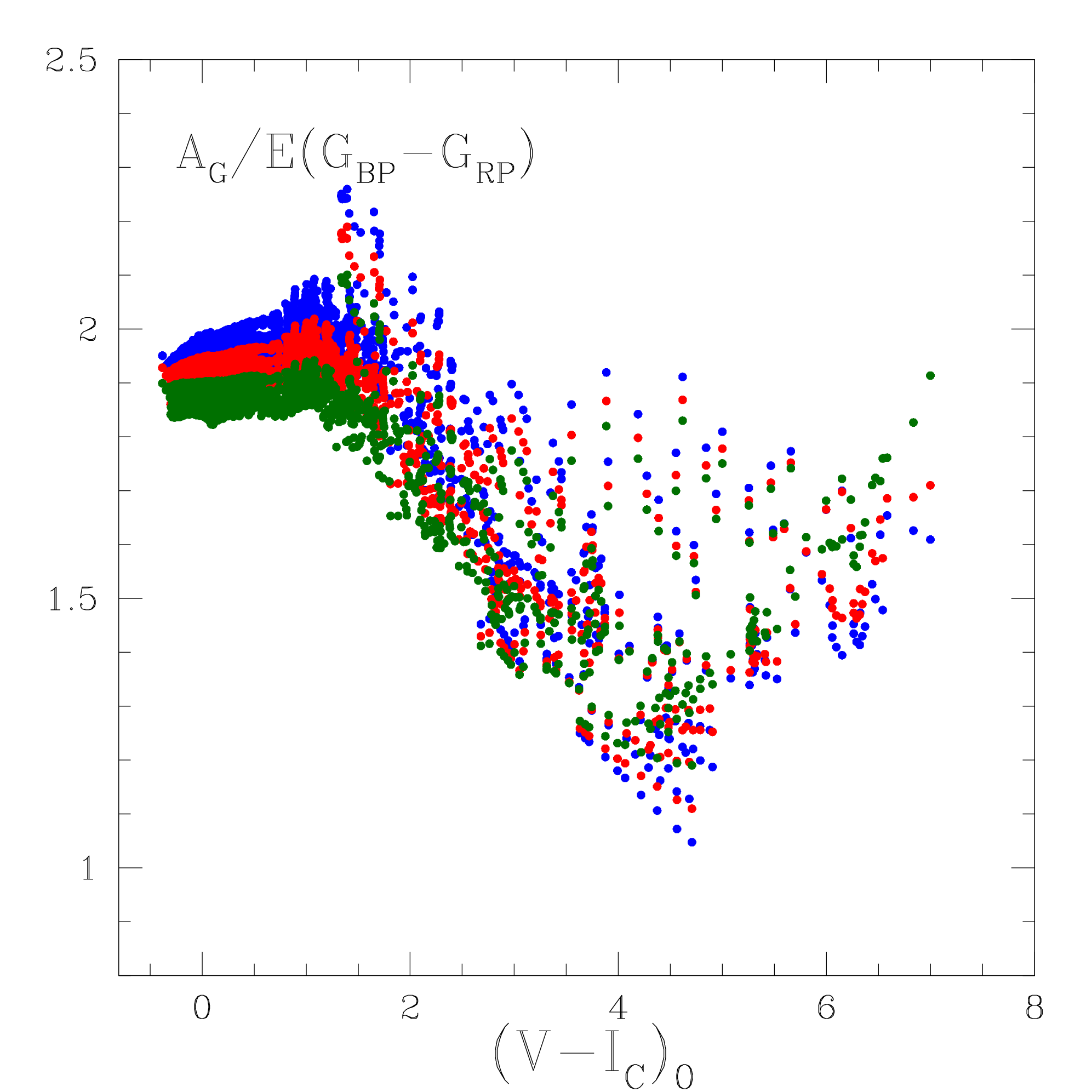}
\includegraphics[scale=0.22]{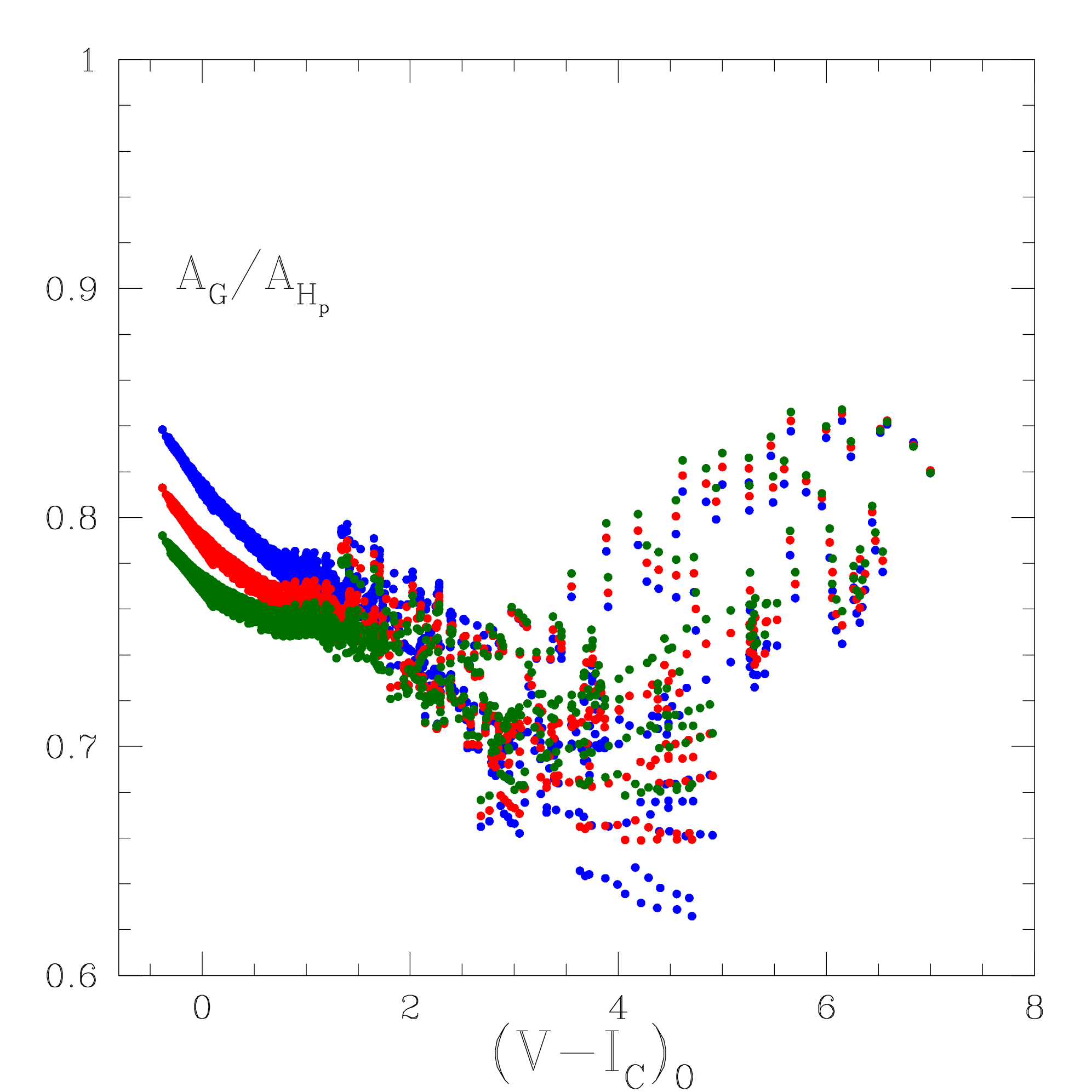}
\includegraphics[scale=0.22]{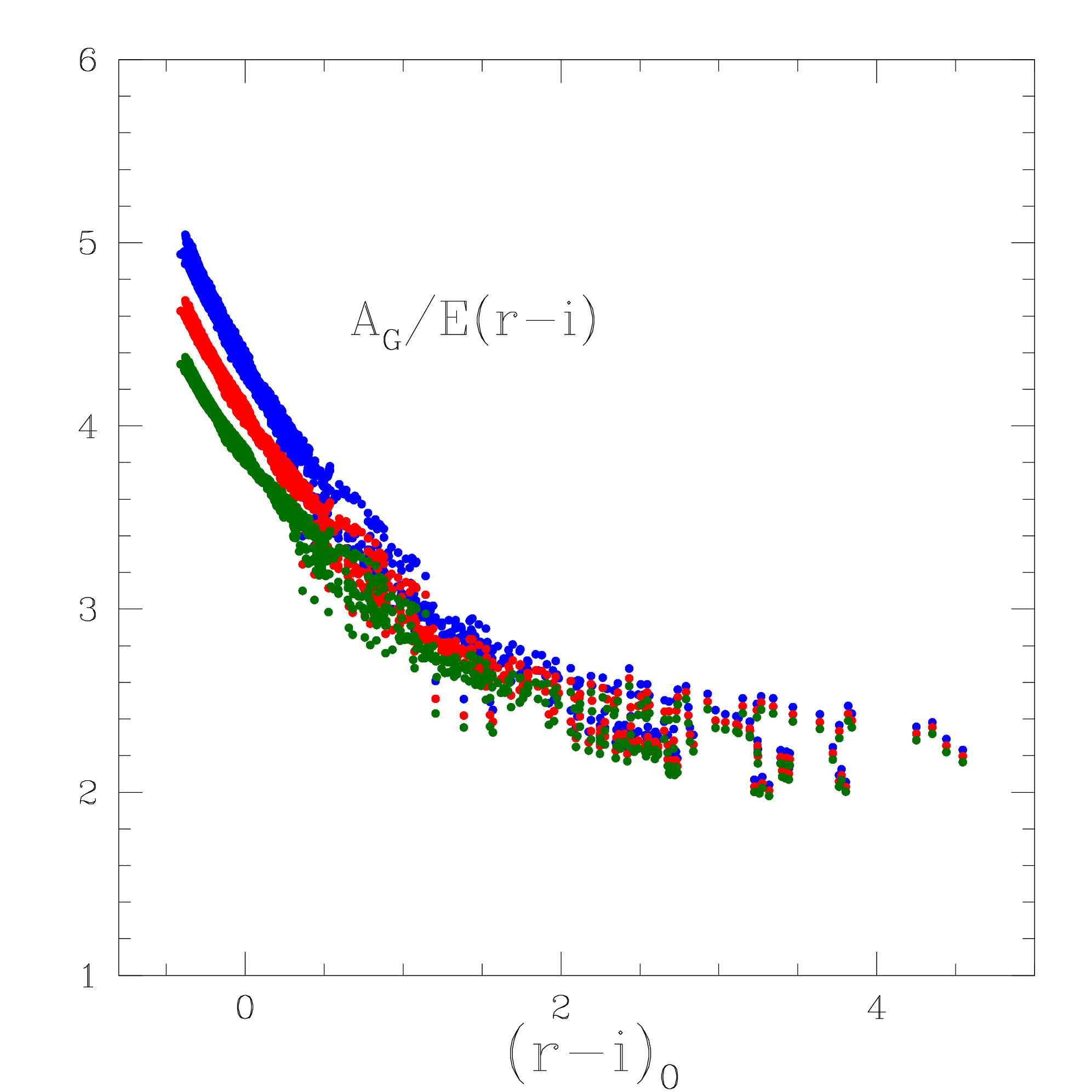}
\caption{Examples of ratios of total-to-selective absorption $A_G$ in the 
$G$ band and colour excess in \BP--\RP, $E$(\BP$-$\RP).}
 \label{fig:absorcio1}
\end{figure}

Figure~\ref{fig:absorcio1} and Table\footnote{Table 13 is only available in electronic form.}~13 show several ratios of total-to-selective absorption including absorption $A_G$ in the $G$ band and colour excess $E($\BP--\RP$)$. 
The ratios in all the bands depend on the stellar effective temperature, and less on surface gravity and metallicity \citep{Grebel1995}. There is also a dependence on $A_{V}$ itself. The scattering that appears for $(V-I_{C})_{0} \gtrsim 1.5$ or $(r-i)_{0} \gtrsim 0.3$ (i.e. $T_{\rm{eff}}\lesssim 4500$~K) is due to the dependence on metallicity and gravity. Table~\ref{Tab:Ag/Av} displays a third degree polynomial relationship involving $A_{G}/A_{V}$ and $(V-I_{C})_{0}$.

\section{Stellar isochrones}
\label{sec:tracks}

The {\Gaia} parallaxes will allow users to locate stars in Hertzsprung-Russell (HR) diagrams with unprecedented precision, and as a consequence the colour-magnitude $M_G$ vs (\BP$-$\RP)$_{0}$ diagram will provide lots of astrophysical information. Tracks and isochrones in that \Gaia-HR diagram are needed.

Several sets of isochrones are available in the literature:

\begin{enumerate}
\item {\sl \bf Padova isochrones}

Several sets of stellar tracks and isochrones have been calculated by the Padova group in the past 20 years
(Bressan et al. 1993; \cite{1994A&AS..104..365F,1994A&AS..105...29F}, \cite{1994A&AS..106..275B}, Girardi et al. 2000). \cite{2007A&A...469..239M} and Marigo et al. (2008) included  an updated modelling of the AGB phase. The evolution from the first thermal pulse up to the complete ejection of the stellar envelope is followed, including the  transition to the C-star phase due to the third dredge-up event, and the proper effective temperatures  for carbon stars, and suitable mass-loss rates for the M and C-type stars. Those AGB models are not calibrated. Recently, new stellar evolution models appeared in the Padova database for low-mass stars and high-mass stars in a large region of the $Z$-$Y$ plane \citep{2008A&A...484..815B} including a 
grid of abundances in the He content $Y$. The initial chemical composition is in the range $0.0001\le Z \le 0.070$ for the metal content and for the helium content in the range $0.23 \le Y \le 0.40$.  For each value of $Z$, the fractions of different metals follow a scaled solar distribution, as compiled by \cite{1993PhST...47..133G} and adopted in the OPAL opacity tables.  The radiative opacities for scaled solar mixtures are from the OPAL group \citep{1996ApJ...464..943I} for temperatures higher than $\log T =4$, and the molecular opacities from \cite{1994ApJ...437..879A} for $\log T < 4.0$ as in \cite{2000yCat..33611023S}. For very high temperatures ($\log T \ge 8.7$) the opacities by Weiss et al. (1990) are used.  The stellar models are computed for initial masses from $0.15$  to $20 M_{\odot}$, for stellar phases going from the ZAMS to the end of helium burning. These tracks include the convective overshoot in the core of the stars, the hydrogen semiconvection during the core H-burning phase of massive stars; and the helium semiconvection  in the convective core of low-mass stars during the early stages of the horizontal branch. 

\item {\sl \bf Teramo isochrones: BASTI data base}

\cite{2000ApJ...543..955B} presented intermediate-mass standard models with different
helium and metal content ($ 3 \le M \le 15 M_{\odot}$) and the \cite{2004ApJ...612..168P,Teramo06} extended database makes available stellar models and isochrones
for scaled-solar and $\alpha$-enhanced metal distribution for the mass range 
between $0.5$ and $10 M_{\odot}$ for a standard evolutionary scenario (no atomic diffusion, no overshooting) and including overshooting. The stellar evolution models and isochrones are extended along the AGB stage to cover the full thermal pulse phase, using the synthetic AGB technique \citep{1978ApJ...220..980I}.  All models have been computed using a scaled solar distribution (Grevesse \& Noels 1993) for the heavy elements.
The mass range goes from 0.5 $M_{\odot}$ to 10.0 $M_{\odot}$. The metallicity ranges from Z=0.0001 to 0.04, the He content from Y=0.245 to 0.30 following a $\Delta Y/\Delta Z$ relation. 

\end{enumerate}

\begin{figure}[ht!]
\centering
 \begin{tabular}{cc}
\includegraphics[scale=0.21]{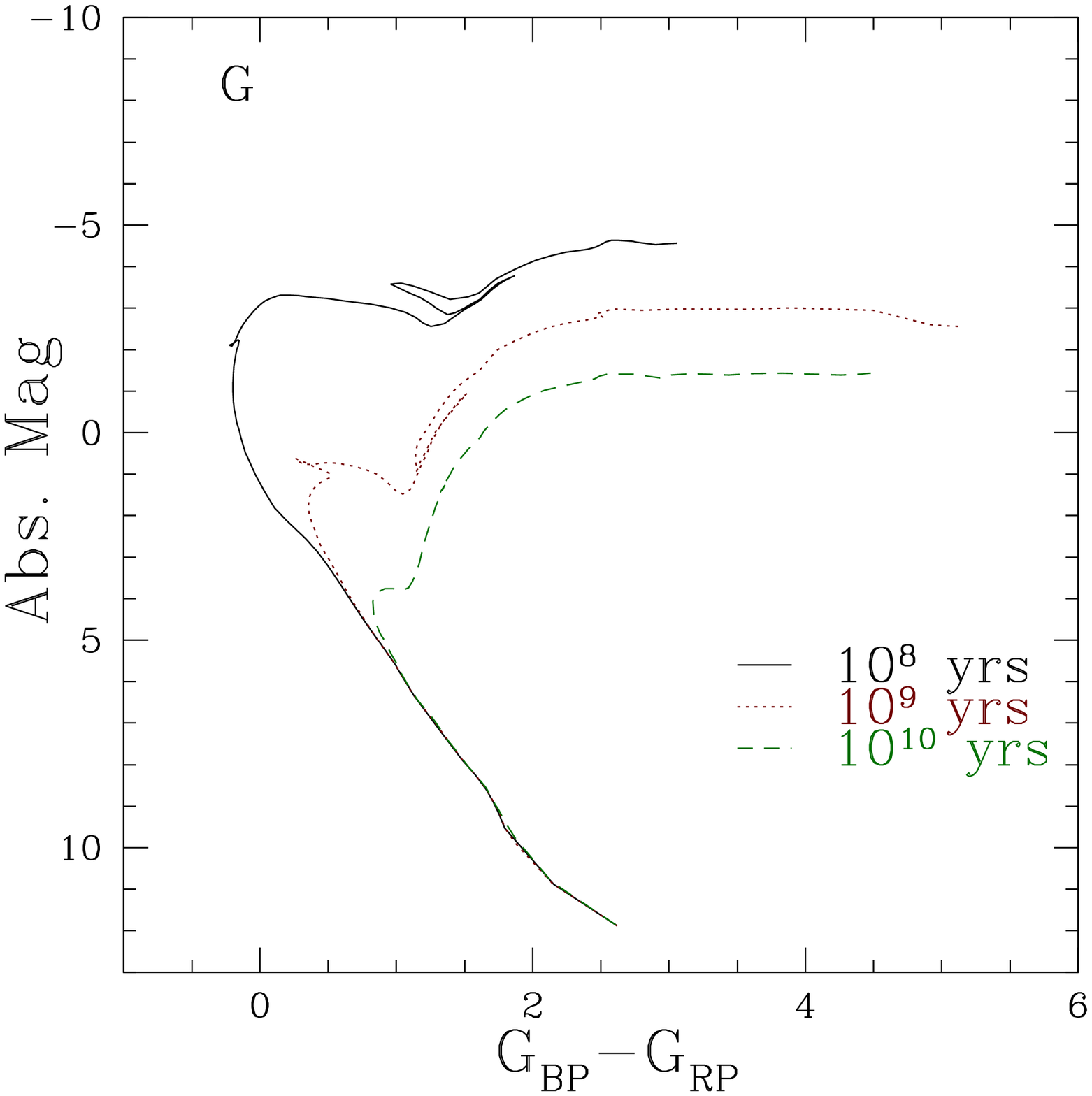}&
\includegraphics[scale=0.21]{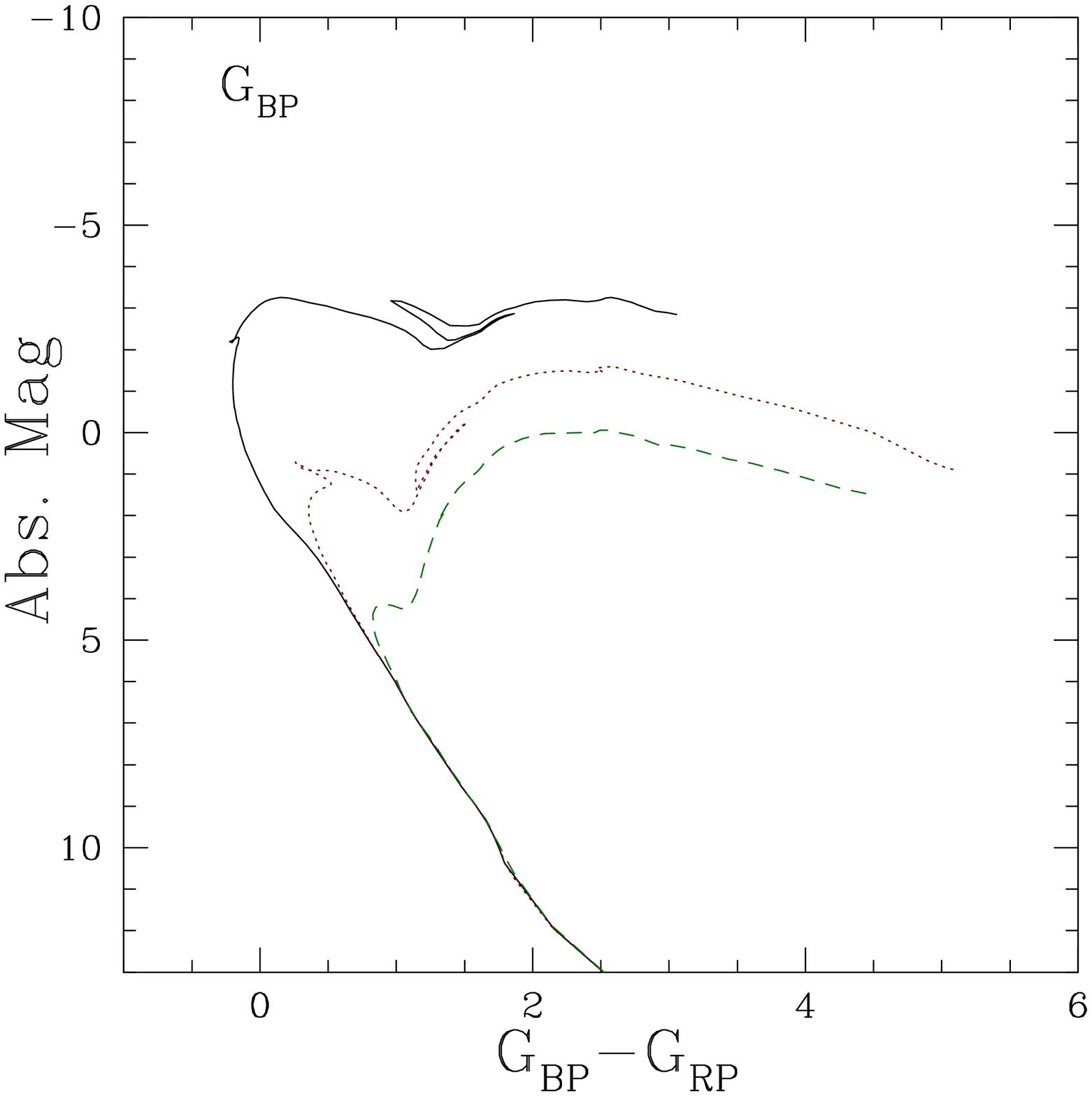}\\
\includegraphics[scale=0.21]{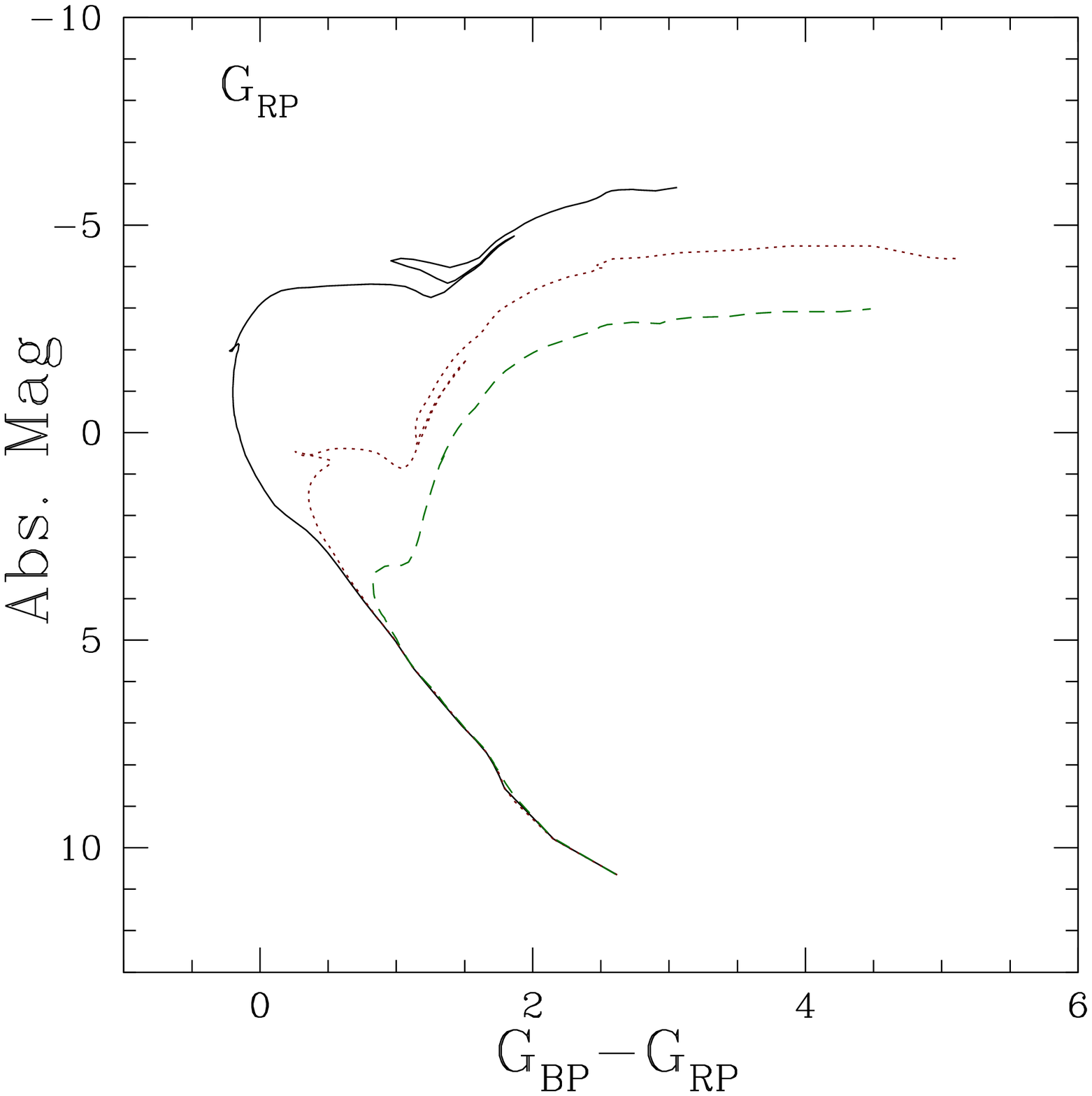}
\end{tabular}
\caption{Isochrones computed at different ages, for $G$, \BP, and \RP\ using the luminosities of \cite{Marigo2008}. }
\label{fig:iso}
\end{figure}

The {\Gaia} passbands have been implemented on the two web sites of Padova (http://stev.oapd.inaf.it/) and BASTI (http://albione.oa-teramo.inaf.it/). This way stellar tracks and isochrones can be computed and downloaded.
Figure~\ref{fig:iso} shows the Padova isochrones (Marigo et al. 2008) in the {\Gaia} passbands for solar metallicity and for different ages, just as an example.

\section{Performances}
\label{sec:performances}

As explained in Sect. 6.2 of \cite{2006MNRAS.367..290J}, during the observational process
only the pixels in the area immediately surrounding the target source are sent
to the ground in the form of a 'window'. In most cases the pixels in the window
are binned in the across-scan direction so that the resulting data consist of 
a one-dimensional set of number counts per sample (a set of pixels). 
The images in the one- or two- dimensional windows will be fitted with line-spread or point-spread functions to estimate the fluxes of the objects.
The estimated associated error of the derived flux is related to the signal within the window as in the case of an 'aperture photometry'
approach. It is assumed that the object flux $f_X$ within a given passband $X$
is measured in a rectangular 'aperture' of $n_s$ samples within the window. Some
light loss is produced because of the finite extent of the
'aperture'. Hence the actual flux in the window will be $g_{\rm aper} \cdot f_X$,
where $g_\mathrm{aper} \le 1$.

While scanning the sky, {\Gaia} will observe the sources transiting the focal plane. In each transit, the same source will be observed nine times in AF\footnote{Actually, when transiting along the central row, the source will be observed only eight times in AF CCDs. Therefore the number of AF observations per transit is 8.86 on average.} and one time in BP and RP CCDs (see Fig.~\ref{fig:focalplane}).
The magnitude error for a transit ($\sigma _{X}$) is computed taking into account (1) the photon noise, (2) the total detection noise per sample $r$, which includes the detector read-out noise, (3) the sky background contribution $b_X$ assumed to be derived from $n_b$ background samples, (4) the contribution
of the calibration error per observation $\sigma_{\rm cal}$, and (5) the averaged total number of columns in each band $n_{\rm strips}$\footnote{8.86 for $G$, 1 for {\BP} and 1 for \RP.}.
 
\begin{eqnarray} 
\label{eq:eqerror} 
\sigma _{X} [{\rm mag}] &= &\frac{m}{\sqrt{n_{\rm strips}}} \cdot \bigg[\sigma^{2}_{\rm cal} +\Big(
2.5 \cdot {\rm log}_{\rm 10}e \cdot \nonumber\\
 & &  \frac{[g_{\rm aper} \cdot f_X + (b_{X} + r^{2}) \cdot n_{s} \cdot
(1+ n_{s}/n_{b})]^{1/2}}{g_{\rm aper} \cdot f_X} \Big)^{2} \bigg]^{1/2} 
\end{eqnarray} 

The magnitude errors are artificially increased by  20 per cent ($m=1.2$). This safety margin accounts for sources of error not considered here such as the dependence of the calibration error on the sky density, complex background, etc. For the calculations here we have assumed $\sigma^{2}_{\rm cal}=0$, i.e. negligible compared to the poissonian and read-out noise. In reality this might not be the case because the complexity of the instrumental effects is rather challenging. At present it is not completely understood to which level of perfectness effects like saturation, non-linearity, radiation damage, 
and charge transfer inefficiencies on the data can be calibrated. Therefore a general calibration error of a few mmag at the end of the mission cannot be ruled out at the moment. A calibration error of this level would
mainly affect the quality of the bright sources. Furthermore it cannot be ruled out that sources with very extreme colours might have larger final errors. 

\begin{figure*}[htbp!]
 
 \centering

 \begin{tabular}{cc}
 \includegraphics[scale=0.22]{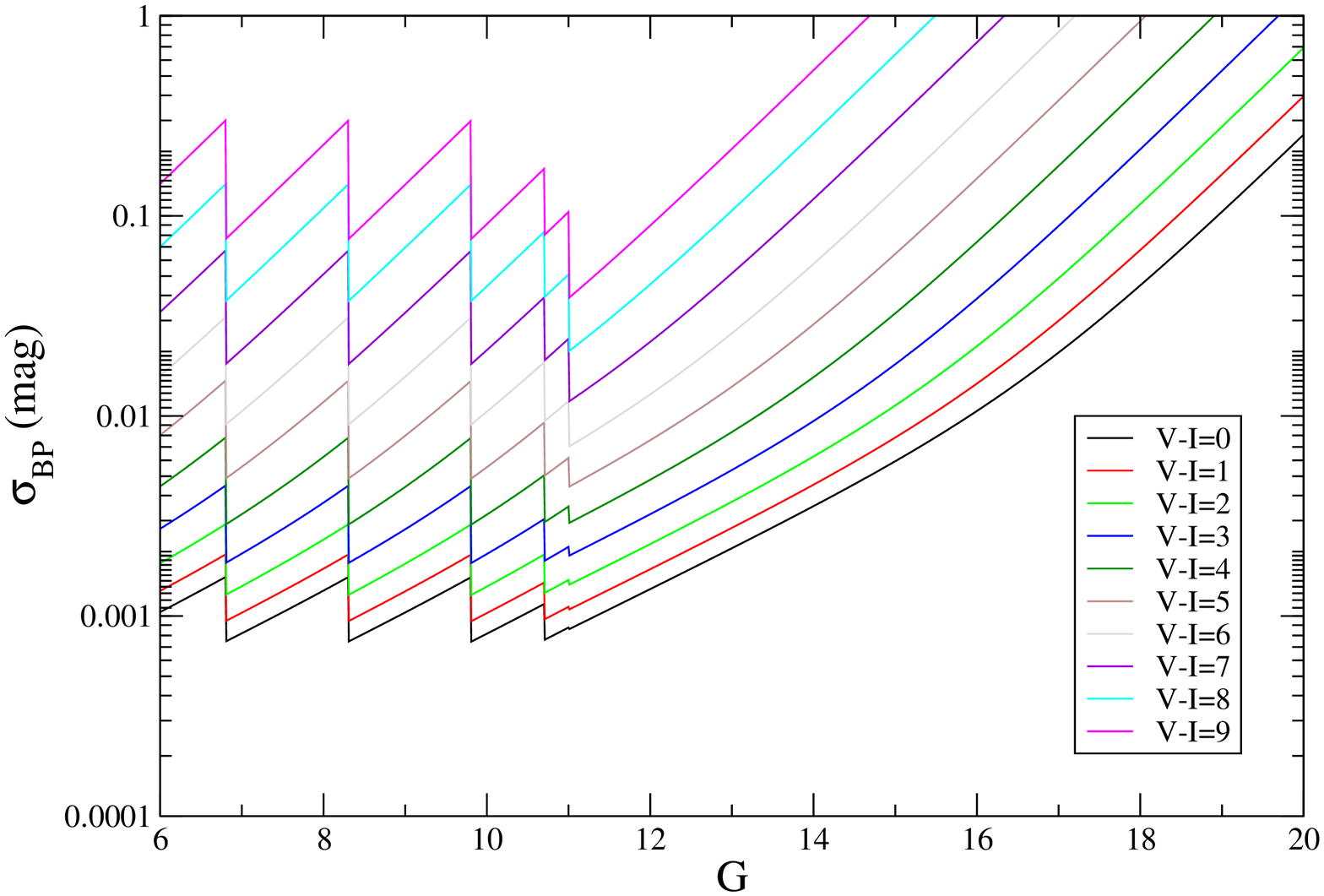}&
 \includegraphics[scale=0.22]{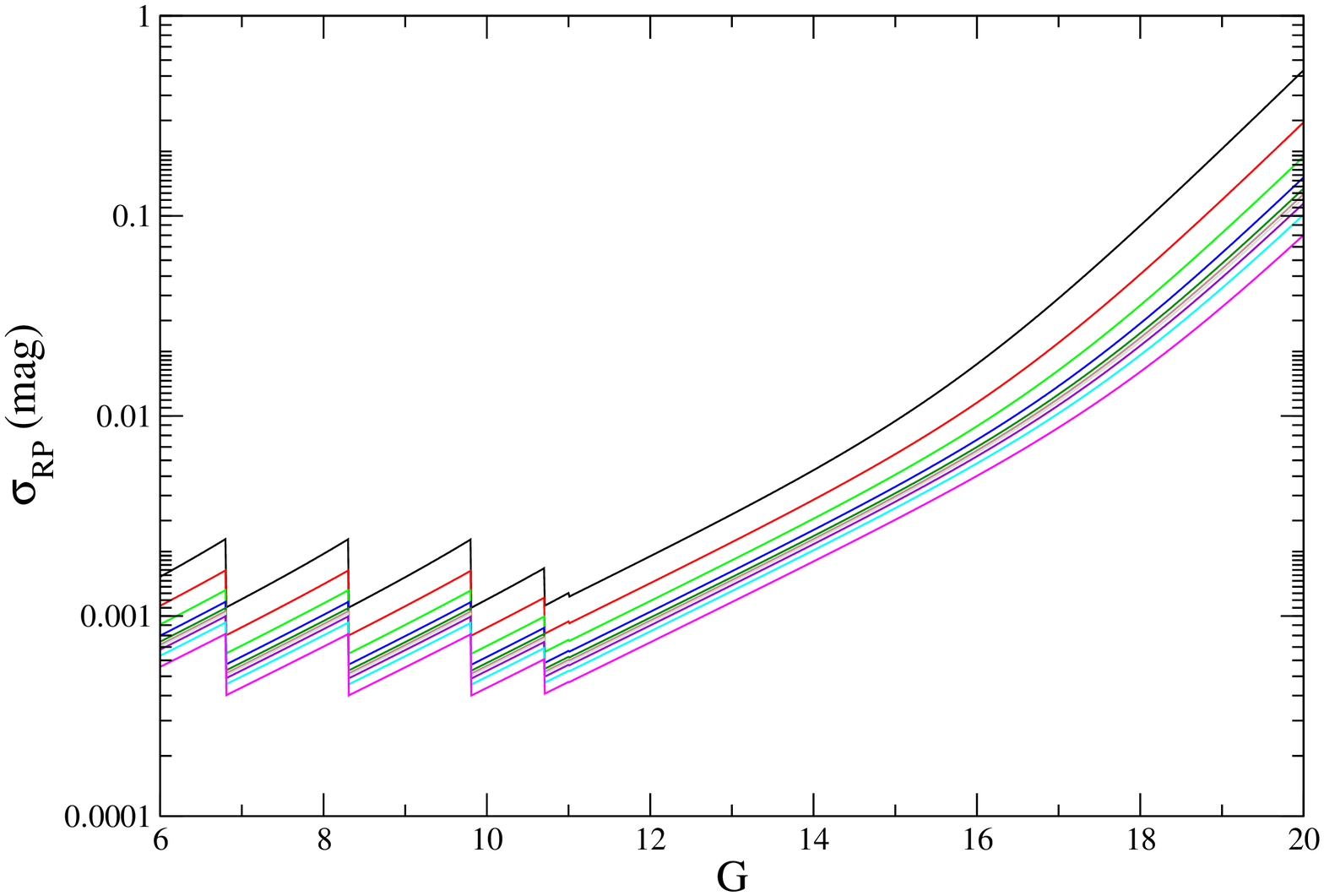} \\
  \includegraphics[scale=0.22]{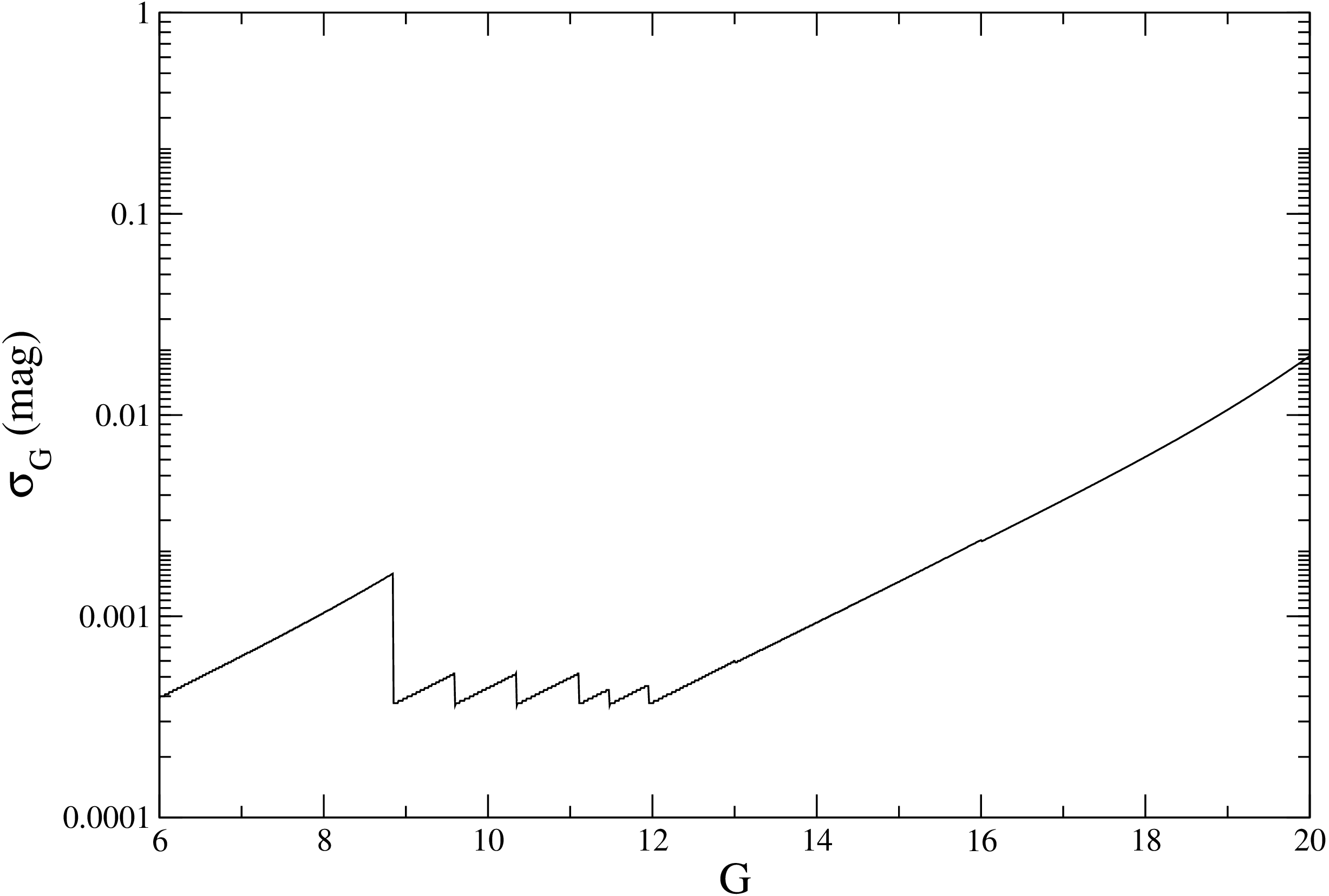}
\end{tabular}
  \caption{Estimated precisions for one transit on the focal plane and for $G$, {\BP}, and \RP \ with respect to the $G$ magnitude. The shape for bright stars is due
to the decrease of the exposure time to avoid saturation of the pixels.}
\label{fig:errors}
\end{figure*}

The attainable precisions in \BP \ and \RP \ are shown in Fig.~\ref{fig:errors} where the estimated $\sigma _{X}$ are plotted for one single transit along the focal plane as a function of the $G$ magnitude for different ($V-I_{C}$) colours, respectively. The precision in $G$, which depends only on the $G$ value, is also plotted. The discontinuities in Fig.~\ref{fig:errors} for bright stars are owing to different integration times at different magnitude intervals to avoid saturation of the pixels. 

The end-of-mission error should consider the true number of observations $N_{obs}$ and is given by $\sigma_{X}^{EOM}=\sigma_{X}/\sqrt{N_{obs}\times DP_{G}}$. $DP_{G}$ is a factor that takes into account the detection probability. It gives the probability that a star is detected and selected on-board for observation, as function of the apparent magnitude $G$. Table~\ref{detection_prob} displays the values of $DP_{G}$ in different ranges of magnitudes. As a result of the scanning law, the number of observations per star is related to the ecliptic latitude as given in Table~\ref{tab:beta}. The data in this table are taken from the ESA\footnote{http://www.rssd.esa.int/index.php?project$=$GAIA$\&$page$=$Science$\_$ Performance} Science Performance information. The ecliptic latitude can be calculated from the equatorial coordinates ($\alpha$,$\delta$) or galactic coordinates $(l,b)$ according to Eq.~\ref{relation-beta-alpha}:

 \begin{eqnarray}
\sin \beta &=&0.9175\cdot\sin \delta -0.3978\cdot\cos \delta \cdot\sin \alpha \\
&=& 0.4971\cdot\sin b + 0.8677 \cos b \sin(l - 6^{\rm o}.38). \nonumber
\label{relation-beta-alpha}
\end{eqnarray}

According to this formulation, and for a specific object, one can estimate the end-of-mission precision by knowing the celestial coordinates, the $G$ magnitude of the source and its colour.

\begin{table}
\center
\caption{Number of focal plane transits after 5 years of mission as a function of the ecliptic latitude $\beta$.}
\label{tab:beta}
\begin{tabular}{cccc}
\hline
$|\sin(\beta)|$ &    $\beta_{min}$ (deg)  &   $\beta_{max}$ (deg)  &   $N_{obs}$ \\ \hline    
0.025   &               0.0         &            2.9         &      61     \\ 
0.075    &                2.9      &               5.7       &        61   \\   
0.125   &                 5.7      &               8.6        &       62    \\  
0.175   &                 8.6      &              11.5        &       62    \\  
0.225   &                11.5      &              14.5        &       63    \\  
0.275   &                14.5      &              17.5        &       65    \\   
0.325    &               17.5      &              20.5          &     66    \\  
0.375    &               20.5      &              23.6          &     68    \\  
0.425     &              23.6       &             26.7           &    71    \\  
0.475     &              26.7       &             30.0           &    75    \\ 
0.525     &              30.0        &            33.4           &    80    \\  
0.575    &               33.4          &          36.9           &    87    \\  
0.625     &              36.9         &             40.5      &         98 \\     
0.675    &               40.5         &           44.4           &   122     \\ 
0.725    &               44.4        &            48.6          &    144     \\ 
0.775    &               48.6           &         53.1           &   106     \\ 
0.825    &               53.1          &          58.2           &    93     \\ 
0.875    &               58.2           &         64.2           &    85     \\ 
0.925    &               64.2           &         71.8           &    80     \\ 
0.975    &               71.8          &          90.0           &    75     \\  \hline
Mean        &             0.0           &         90.0           &    81     \\  \hline
\end{tabular}
\end{table}

\begin{table}[h!]
 \center
 \caption{Observation probability in percent as function of the apparent magnitude $G$.}
 \label{detection_prob}
 \begin{tabular}{cccc}
 \hline
 $G_{min}$&$G_{max}$ & AF (\%)& BP/RP (\%)\\ \hline
 6&16&100&100 \\
 16&17&98.7&98.7 \\
 17&18&98.2&97.3 \\
 18&19&98.1&98.1 \\
 19&19.9&96.5&92.5 \\
 19.9&20&95.1&92.5 \\
\hline
 \end{tabular}
 \end{table}

\section{Conclusions}
\label{sec:conclusions}

We have presented the characterisation of the {\Gaia} passbands ($G$, {\BP}, {\RP}, and $G_{\rm RVS}$) based on the most up-to-date information from the industrial
partners. Not all data are yet real measurements of flight hardware, but close enough for the scientific exploitation preparation. The results of this paper will not be severely affected since no drastic changes are expected.

{\Gaia} magnitudes and colours have been computed for all spectral energy distributions in the BaSeL3.1 stellar library and for four reddening values.
In addition, colours in the most commonly used photometric systems (Johnson-Cousins, {\it Hipparcos-Tycho}, and SDSS) have also been derived. All the computed colours are provided in an online table. Based on this table, colour-colour transformations have been calculated. {\BP}$-${\RP} colour correlates very well with $V-I_{C}$ and a bit less well with $g-z$. 
Therefore, {\BP}$-${\RP} is a raw indicator of effective temperature, especially for $T_{\rm{eff}}\geq4500$~K (i.e. {\BP}$-${\RP}$< 1.5$), yielding residuals of about 5\% in temperature. 

Relationships involving $V-I_{C}$ colour are the ones with the lowest residuals among Johnson-Cousins colours and are of the order of $0.03-0.08$ mag depending on the colour-colour pair considered. For the SDSS system, $g-z$ colour is the one providing tighter transformations with residuals in the range $0.02-0.14$ mag. For the {\it Hipparcos-Tycho} photometric system, the residuals are between 0.02 and 0.12 mag. 
The use of two colours decreases the residuals of the transformations, as can be seen in Table~\ref{table:coeficients_2colors}.
Independently of the choice of the passbands, scattering exists due to metallicities and gravities, especially for $T_{\rm{eff}} < 4500$~K. The level of the scattering varies among the transformations.

Relationships to predict the {\RVS} magnitude have been derived and permit one to know the conditions in which a given star can be observed by the high-resolution radial velocity spectrometer.

The choice of the spectral library is not critical in this work, although other libraries could provide slightly different relationships. Either the relationships or the online table allow the prediction of {\Gaia} magnitudes and colours from knowledge of $T_{\rm{eff}}$, gravity, and metallicity, or from existing photometry. The online table even allows users to compute their own relationships based on their needs (specific objects, colours, reddening,...). We remind users that the computed polynomials are only valid in the range of BaSeL3.1 astrophysical-parameters space, and no extrapolation is recommended. 

Bolometric corrections have been computed in {\Gaia}'s passbands and are also provided in an online table, which allows the correspondence between absolute {\Gaia} magnitudes and luminosity. In addition, the passbands are included in the Padova and BASTI isochrones web sites, allowing the computation of any track and isochrone in the {\Gaia} colour-magnitude diagram, an essential tool for the derivation of ages or the analysis of clusters. The paper presents some examples of the Padova isochrones in $G$, \BP, and \RP\ for solar metallicity and for different ages. 

Absorption and colour excess in {\Gaia} passbands have been computed as well as the ratios with respect to $A_V$, $A_{H_p}$ and several colour excesses have been provided for the whole spectral energy distributions in BaSeL3.1 and for three absorption values. A polynomial fitting of $A_G/A_V$ ratio is provided for the three absorption values considered.

Finally, we have provided the photometric performances by computing the estimated errors on the $G$, \BP, and \RP \ magnitudes for one observation. End-of-mission precision can be derived knowing the number of observations, which depends on the celestial coordinates due to the scanning law of the satellite. 

Therefore, the paper provides the ingredients and tools to predict in an easy manner the {\Gaia} magnitudes and associated precisions for all kind of stars, either from theoretical stellar parameters or from existing photometry, plus interstellar absorptions and isochrones. All can be combined and used to predict how the {\Gaia} sky will look, in which conditions a known object will be observed, etc. Besides the $G$, \BP, \RP \ and \RVS \ photometry discussed here, low-resolution BP and RP spectra will be available. We wish to emphasize that BP and RP spectra are the best suited elements to derive astrophysical parameters of the observed objects because they have been designed for exactly this goal. However, a lot of research can be done with only a colour-magnitude diagram, and this is what pushed us to perform the present work. Furthermore, for very faint objects or in the early releases of the mission (when few observations are available per object), broadband photometry may be the only available product. The present work can be used also to plan ground-based complementary observations and to build catalogues with auxiliary data for the {\Gaia} data processing and validation.

\begin{acknowledgements}
This programme was supported by Ministerio de Ciencia y Tecnolog\'\i a under contract AYA2009-14648-C02-01.
\end{acknowledgements}

\bibliographystyle{aa} 
\bibliography{bibliography} 

\begin{thebibliography}{58}
\expandafter\ifx\csname natexlab\endcsname\relax\def\natexlab#1{#1}\fi

\bibitem[{{Alexander} \& {Ferguson}(1994)}]{1994ApJ...437..879A}
{Alexander}, D.~R. \& {Ferguson}, J.~W. 1994, \apj, 437, 879

\bibitem[{{Allard} {et~al.}(2000){Allard}, {Hauschildt}, \&
  {Schweitzer}}]{Allard00}
{Allard}, F., {Hauschildt}, P.~H., \& {Schweitzer}, A. 2000, \apj, 539, 366

\bibitem[{{Alvarez} \& {Plez}(1998)}]{Alva98}
{Alvarez}, R. \& {Plez}, B. 1998, \aap, 330, 1109

\bibitem[{Andersen(1999)}]{1999Obs...119..289A}
Andersen, J. 1999, The Observatory, 119, 289

\bibitem[{{Bertelli} {et~al.}(1994){Bertelli}, {Bressan}, {Chiosi}, {Fagotto},
  \& {Nasi}}]{1994A&AS..106..275B}
{Bertelli}, G., {Bressan}, A., {Chiosi}, C., {Fagotto}, F., \& {Nasi}, E. 1994,
  \aaps, 106, 275

\bibitem[{{Bertelli} {et~al.}(2008){Bertelli}, {Girardi}, {Marigo}, \&
  {Nasi}}]{2008A&A...484..815B}
{Bertelli}, G., {Girardi}, L., {Marigo}, P., \& {Nasi}, E. 2008, \aap, 484, 815

\bibitem[{{Bessell}(1990)}]{1990PASP..102.1181B}
{Bessell}, M.~S. 1990, \pasp, 102, 1181

\bibitem[{{Bessell}(2000)}]{2000PASP..112..961B}
{Bessell}, M.~S. 2000, \pasp, 112, 961

\bibitem[{{Bessell} {et~al.}(1991){Bessell}, {Brett}, {Scholz}, \&
  {Wood}}]{Bessell2}
{Bessell}, M.~S., {Brett}, J.~M., {Scholz}, M., \& {Wood}, P.~R. 1991, \aaps,
  89, 335

\bibitem[{{Bessell} {et~al.}(1989){Bessell}, {Brett}, {Wood}, \&
  {Scholz}}]{Bessell1}
{Bessell}, M.~S., {Brett}, J.~M., {Wood}, P.~R., \& {Scholz}, M. 1989, \aaps,
  77, 1

\bibitem[{{Bessell} {et~al.}(1998){Bessell}, {Castelli}, \& {Plez}}]{Bessel98}
{Bessell}, M.~S., {Castelli}, F., \& {Plez}, B. 1998, \aap, 333, 231

\bibitem[{{Bohlin}(2007)}]{2007ASPC..364..315B}
{Bohlin}, R.~C. 2007, in Astronomical Society of the Pacific Conference Series,
  Vol. 364, The Future of Photometric, Spectrophotometric and Polarimetric
  Standardization, ed. {C.~Sterken}, 315

\bibitem[{{Bohlin} \& {Gilliland}(2004)}]{2004AJ....127.3508B}
{Bohlin}, R.~C. \& {Gilliland}, R.~L. 2004, \aj, 127, 3508

\bibitem[{{Bono} {et~al.}(2000){Bono}, {Caputo}, {Cassisi}, {Marconi},
  {Piersanti}, \& {Tornamb{\`e}}}]{2000ApJ...543..955B}
{Bono}, G., {Caputo}, F., {Cassisi}, S., {et~al.} 2000, \apj, 543, 955

\bibitem[{{Bouret} {et~al.}(2008){Bouret}, {Lanz}, {Fr{\'e}mat}, {Martins},
  {Lefever}, {Blomme}, {Martayan}, {Neiner}, {Quinet}, \& {Zorec}}]{Bouret08}
{Bouret}, J., {Lanz}, T., {Fr{\'e}mat}, Y., {et~al.} 2008, in Revista Mexicana
  de Astronomia y Astrofisica Conference Series, Vol.~33, Revista Mexicana de
  Astronomia y Astrofisica Conference Series, 50--50

\bibitem[{{Brott} \& {Hauschildt}(2005)}]{Brott05}
{Brott}, I. \& {Hauschildt}, P.~H. 2005, in ESA Special Publication, Vol. 576,
  The Three-Dimensional Universe with Gaia, ed. {C.~Turon, K.~S.~O'Flaherty, \&
  M.~A.~C.~Perryman}, 565

\bibitem[{{Cardelli} {et~al.}(1989){Cardelli}, {Clayton}, \&
  {Mathis}}]{1989ApJ...345..245C}
{Cardelli}, J.~A., {Clayton}, G.~C., \& {Mathis}, J.~S. 1989, \apj, 345, 245

\bibitem[{{Cousins}(1976)}]{1976MmRAS..81...25C}
{Cousins}, A.~W.~J. 1976, \memras, 81, 25

\bibitem[{{ESA}(1997)}]{1997yCat.1239....0E}
{ESA}. 1997, VizieR Online Data Catalog, 1239, 0

\bibitem[{ESA(2000)}]{2000ESA}
ESA. 2000, Technical Report ESA-SCI(2000)4, (scientific case on-line at
  http://www.rssd.esa.int/index.php?project=Gaia), 4

\bibitem[{{Fagotto} {et~al.}(1994{\natexlab{a}}){Fagotto}, {Bressan},
  {Bertelli}, \& {Chiosi}}]{1994A&AS..104..365F}
{Fagotto}, F., {Bressan}, A., {Bertelli}, G., \& {Chiosi}, C.
  1994{\natexlab{a}}, \aaps, 104, 365

\bibitem[{{Fagotto} {et~al.}(1994{\natexlab{b}}){Fagotto}, {Bressan},
  {Bertelli}, \& {Chiosi}}]{1994A&AS..105...29F}
{Fagotto}, F., {Bressan}, A., {Bertelli}, G., \& {Chiosi}, C.
  1994{\natexlab{b}}, \aaps, 105, 29

\bibitem[{{Fitzpatrick}(1999)}]{1999PASP..111...63F}
{Fitzpatrick}, E.~L. 1999, \pasp, 111, 63

\bibitem[{{Fitzpatrick} \& {Massa}(2007)}]{2007ApJ...663..320F}
{Fitzpatrick}, E.~L. \& {Massa}, D. 2007, \apj, 663, 320

\bibitem[{{Fluks} {et~al.}(1994){Fluks}, {Plez}, {The}, {de Winter},
  {Westerlund}, \& {Steenman}}]{Fluks}
{Fluks}, M.~A., {Plez}, B., {The}, P.~S., {et~al.} 1994, \aaps, 105, 311

\bibitem[{{Fukugita} {et~al.}(1996){Fukugita}, {Ichikawa}, {Gunn}, {Doi},
  {Shimasaku}, \& {Schneider}}]{1996AJ....111.1748F}
{Fukugita}, M., {Ichikawa}, T., {Gunn}, J.~E., {et~al.} 1996, \aj, 111, 1748

\bibitem[{{Girardi} {et~al.}(2002){Girardi}, {Bertelli}, {Bressan}, {Chiosi},
  {Groenewegen}, {Marigo}, {Salasnich}, \& {Weiss}}]{Girardi2002}
{Girardi}, L., {Bertelli}, G., {Bressan}, A., {et~al.} 2002, \aap, 391, 195

\bibitem[{{Grebel} \& {Roberts}(1995)}]{Grebel1995}
{Grebel}, E.~K. \& {Roberts}, W.~J. 1995, \aaps, 109, 293

\bibitem[{{Grevesse} \& {Noels}(1993)}]{1993PhST...47..133G}
{Grevesse}, N. \& {Noels}, A. 1993, Physica Scripta Volume T, 47, 133

\bibitem[{{Gustafsson} {et~al.}(2008){Gustafsson}, {Edvardsson}, {Eriksson},
  {J{\o}rgensen}, {Nordlund}, \& {Plez}}]{Gust08}
{Gustafsson}, B., {Edvardsson}, B., {Eriksson}, K., {et~al.} 2008, \aap, 486,
  951

\bibitem[{{Hauschildt} {et~al.}(1999){Hauschildt}, {Allard}, \&
  {Baron}}]{1999ApJ...512..377H}
{Hauschildt}, P.~H., {Allard}, F., \& {Baron}, E. 1999, \apj, 512, 377

\bibitem[{{Iben} \& {Truran}(1978)}]{1978ApJ...220..980I}
{Iben}, Jr., I. \& {Truran}, J.~W. 1978, \apj, 220, 980

\bibitem[{{Iglesias} \& {Rogers}(1996)}]{1996ApJ...464..943I}
{Iglesias}, C.~A. \& {Rogers}, F.~J. 1996, \apj, 464, 943

\bibitem[{{Ivezi{\'c}} {et~al.}(2007){Ivezi{\'c}}, {Smith}, {Miknaitis}, {Lin},
  {Tucker}, {Lupton}, {Gunn}, {Knapp}, {Strauss}, {Sesar}, {Doi}, {Tanaka},
  {Fukugita}, {Holtzman}, {Kent}, {Yanny}, {Schlegel}, {Finkbeiner},
  {Padmanabhan}, {Rockosi}, {Juri{\'c}}, {Bond}, {Lee}, {Stoughton}, {Jester},
  {Harris}, {Harding}, {Morrison}, {Brinkmann}, {Schneider}, \&
  {York}}]{2007AJ....134..973I}
{Ivezi{\'c}}, {\v Z}., {Smith}, J.~A., {Miknaitis}, G., {et~al.} 2007, \aj,
  134, 973

\bibitem[{{Johnson}(1963)}]{1963bad..book..204J}
{Johnson}, H.~L. 1963, {Photometric Systems}, ed. {Strand, K.~A.} (the
  University of Chicago Press), 204

\bibitem[{{Jordi} {et~al.}(2006){Jordi}, {H{\o}g}, {Brown}, {Lindegren},
  {Bailer-Jones}, {Carrasco}, {Knude}, {Strai{\v z}ys}, {de Bruijne},
  {Claeskens}, {Drimmel}, {Figueras}, {Grenon}, {Kolka}, {Perryman},
  {Tautvai{\v s}ien{\.e}}, {Vansevi{\v c}ius}, {Willemsen}, {Brid{\v z}ius},
  {Evans}, {Fabricius}, {Fiorucci}, {Heiter}, {Kaempf}, {Kazlauskas}, {Ku{\v
  c}inskas}, {Malyuto}, {Munari}, {Reyl{\'e}}, {Torra}, {Vallenari},
  {Zdanavi{\v c}ius}, {Korakitis}, {Malkov}, \& {Smette}}]{2006MNRAS.367..290J}
{Jordi}, C., {H{\o}g}, E., {Brown}, A.~G.~A., {et~al.} 2006, \mnras, 367, 290

\bibitem[{{Katz} {et~al.}(2004){Katz}, {Munari}, {Cropper}, {Zwitter},
  {Th{\'e}venin}, {David}, {Viala}, {Crifo}, {Gomboc}, {Royer}, {Arenou},
  {Marrese}, {Sordo}, {Wilkinson}, {Vallenari}, {Turon}, {Helmi}, {Bono},
  {Perryman}, {G{\'o}mez}, {Tomasella}, {Boschi}, {Morin}, {Haywood},
  {Soubiran}, {Castelli}, {Bijaoui}, {Bertelli}, {Prsa}, {Mignot}, {Sellier},
  {Baylac}, {Lebreton}, {Jauregi}, {Siviero}, {Bingham}, {Chemla}, {Coker},
  {Dibbens}, {Hancock}, {Holland}, {Horville}, {Huet}, {Laporte}, {Melse},
  {Say{\`e}de}, {Stevenson}, {Vola}, {Walton}, \&
  {Winter}}]{2004MNRAS.354.1223K}
{Katz}, D., {Munari}, U., {Cropper}, M., {et~al.} 2004, \mnras, 354, 1223

\bibitem[{{Kochukhov} \& {Shulyak}(2008)}]{Koch08}
{Kochukhov}, O. \& {Shulyak}, D. 2008, Contributions of the Astronomical
  Observatory Skalnate Pleso, 38, 419

\bibitem[{{Kurucz}(1992)}]{1992IAUS..149..225K}
{Kurucz}, R.~L. 1992, in IAU Symposium, Vol. 149, The Stellar Populations of
  Galaxies, ed. {B.~Barbuy \& A.~Renzini}, 225

\bibitem[{{Lejeune} {et~al.}(1997){Lejeune}, {Cuisinier}, \&
  {Buser}}]{Lejeune1}
{Lejeune}, T., {Cuisinier}, F., \& {Buser}, R. 1997, \aaps, 125, 229

\bibitem[{{Lejeune} {et~al.}(1998){Lejeune}, {Cuisinier}, \&
  {Buser}}]{Lejeune2}
{Lejeune}, T., {Cuisinier}, F., \& {Buser}, R. 1998, \aaps, 130, 65

\bibitem[{{Lindegren}(2010)}]{2010IAUS..261..296L}
{Lindegren}, L. 2010, in IAU Symposium, Vol. 261, IAU Symposium, ed.
  {S.~A.~Klioner, P.~K.~Seidelmann, \& M.~H.~Soffel}, 296--305

\bibitem[{{Marigo} \& {Girardi}(2007)}]{2007A&A...469..239M}
{Marigo}, P. \& {Girardi}, L. 2007, \aap, 469, 239

\bibitem[{{Marigo} {et~al.}(2008){Marigo}, {Girardi}, {Bressan}, {Groenewegen},
  {Silva}, \& {Granato}}]{Marigo2008}
{Marigo}, P., {Girardi}, L., {Bressan}, A., {et~al.} 2008, \aap, 482, 883

\bibitem[{{Martayan} {et~al.}(2008){Martayan}, {Fr{\'e}mat}, {Blomme},
  {Jonckheere}, {Borges}, {de Batz}, {Leroy}, {Sordo}, {Bouret}, {Martins},
  {Zorec}, {Neiner}, {Naz{\'e}}, {Alecian}, {Floquet}, {Hubert}, {Briot},
  {Miroshnichenko}, {Kolka}, {Stee}, {Lanz}, \& {Meynet}}]{Martayan08}
{Martayan}, C., {Fr{\'e}mat}, Y., {Blomme}, R., {et~al.} 2008, in SF2A-2008,
  ed. {C.~Charbonnel, F.~Combes, \& R.~Samadi}, 499

\bibitem[{{Megessier}(1995)}]{1995A&A...296..771M}
{Megessier}, C. 1995, \aap, 296, 771

\bibitem[{{Oke} \& {Gunn}(1983)}]{1983ApJ...266..713O}
{Oke}, J.~B. \& {Gunn}, J.~E. 1983, \apj, 266, 713

\bibitem[{{Perryman} {et~al.}(2001){Perryman}, {de Boer}, {Gilmore}, {H{\o}g},
  {Lattanzi}, {Lindegren}, {Luri}, {Mignard}, {Pace}, \& {de
  Zeeuw}}]{2001A&A...369..339P}
{Perryman}, M.~A.~C., {de Boer}, K.~S., {Gilmore}, G., {et~al.} 2001, \aap,
  369, 339

\bibitem[{{Perryman} {et~al.}(1997){Perryman}, {Lindegren}, {Kovalevsky},
  {Hoeg}, {Bastian}, {Bernacca}, {Cr{\'e}z{\'e}}, {Donati}, {Grenon}, {van
  Leeuwen}, {van der Marel}, {Mignard}, {Murray}, {Le Poole}, {Schrijver},
  {Turon}, {Arenou}, {Froeschl{\'e}}, \& {Petersen}}]{1997A&A...323L..49P}
{Perryman}, M.~A.~C., {Lindegren}, L., {Kovalevsky}, J., {et~al.} 1997, \aap,
  323, L49

\bibitem[{{Pietrinferni} {et~al.}(2004){Pietrinferni}, {Cassisi}, {Salaris}, \&
  {Castelli}}]{2004ApJ...612..168P}
{Pietrinferni}, A., {Cassisi}, S., {Salaris}, M., \& {Castelli}, F. 2004, \apj,
  612, 168

\bibitem[{{Pietrinferni} {et~al.}(2006){Pietrinferni}, {Cassisi}, {Salaris}, \&
  {Castelli}}]{Teramo06}
{Pietrinferni}, A., {Cassisi}, S., {Salaris}, M., \& {Castelli}, F. 2006, \apj,
  642, 797

\bibitem[{{Salasnich} {et~al.}(2000){Salasnich}, {Girardi}, {Weiss}, \&
  {Chiosi}}]{2000yCat..33611023S}
{Salasnich}, B., {Girardi}, L., {Weiss}, A., \& {Chiosi}, C. 2000, VizieR
  Online Data Catalog, 336, 11023

\bibitem[{{Scholz}(1997)}]{Scholz1997}
{Scholz}, M. 1997, {private communication to the authors of Westera et al.
  (2002)}

\bibitem[{{Shulyak} {et~al.}(2004){Shulyak}, {Tsymbal}, {Ryabchikova},
  {St{\"u}tz}, \& {Weiss}}]{2004A&A...428..993S}
{Shulyak}, D., {Tsymbal}, V., {Ryabchikova}, T., {St{\"u}tz}, C., \& {Weiss},
  W.~W. 2004, \aap, 428, 993

\bibitem[{{van Leeuwen} {et~al.}(1997){van Leeuwen}, {Evans}, {Grenon},
  {Grossmann}, {Mignard}, \& {Perryman}}]{1997A&A...323L..61V}
{van Leeuwen}, F., {Evans}, D.~W., {Grenon}, M., {et~al.} 1997, \aap, 323, L61

\bibitem[{{Westera} {et~al.}(2002){Westera}, {Lejeune}, {Buser}, {Cuisinier},
  \& {Bruzual}}]{Lejeune3}
{Westera}, P., {Lejeune}, T., {Buser}, R., {Cuisinier}, F., \& {Bruzual}, G.
  2002, \aap, 381, 524

\bibitem[{{Wilkinson} {et~al.}(2005){Wilkinson}, {Vallenari}, {Turon},
  {Munari}, {Katz}, {Bono}, {Cropper}, {Helmi}, {Robichon}, {Th{\'e}venin},
  {Vidrih}, {Zwitter}, {Arenou}, {Baylac}, {Bertelli}, {Bijaoui}, {Boschi},
  {Castelli}, {Crifo}, {David}, {Gomboc}, {G{\'o}mez}, {Haywood}, {Jauregi},
  {de Laverny}, {Lebreton}, {Marrese}, {Marsh}, {Mignot}, {Morin}, {Pasetto},
  {Perryman}, {Pr{\v s}a}, {Recio-Blanco}, {Royer}, {Sellier}, {Siviero},
  {Sordo}, {Soubiran}, {Tomasella}, \& {Viala}}]{2005MNRAS.359.1306W}
{Wilkinson}, M.~I., {Vallenari}, A., {Turon}, C., {et~al.} 2005, \mnras, 359,
  1306

\bibitem[{{Worthey} \& {Lee}(2006)}]{Worthey06}
{Worthey}, G. \& {Lee}, H. 2006, ArXiv Astrophysics e-prints

\end{thebibliography}




\end{document}